\documentclass[aps,superscriptaddress,twocolumn,showpacs,floatfix,notitlepage]{revtex4-1}

\usepackage{amsmath}
\usepackage[makeroom]{cancel}
\usepackage{amsfonts}
\usepackage{amssymb}
\usepackage{graphicx}
\usepackage{color}
\usepackage[svgnames]{xcolor}
\usepackage{xspace}
\usepackage{dsfont}
\usepackage{physics}
\usepackage[mathscr]{euscript}
\usepackage{xr}

\DeclareFontFamily{OT1}{pzc}{}
\DeclareFontShape{OT1}{pzc}{m}{it}%
{<-> s * [1.15] pzcmi7t}{}
\DeclareMathAlphabet{\mathpzc}{OT1}{pzc}{m}{it}

\newcommand{\ud}[1]{{#1^{\dagger}}}

\newcommand{\Imamoglu}{\u Imamo\=glu\xspace}
\newcommand{\Vuckovic}{Vu\u{c}kovi\'c\xspace}
\newcommand{\g}[1]{g^{(#1)}}
\newcommand{\av}[1]{\langle  #1 \rangle}
\newcommand{\pop}[1]{\langle \ud{#1} #1 \rangle}
\newcommand{\coh}[1]{\langle  #1 \rangle}

\newcommand{\corr}[3]{\langle #1^{\dagger #2} #1^{#3} \rangle}

\usepackage[colorlinks=true, linkcolor=blue, citecolor=magenta]{hyperref}

\newcommand{\mean}[1]{\langle#1\rangle}
\allowdisplaybreaks

\begin{document} 
\title{Tuning photon statistics with coherent fields}

\author{Eduardo {Zubizarreta Casalengua}}
\email{e.zubizarreta.casalengua@gmail.com}
\affiliation{Faculty of Science and Engineering,
  University of Wolverhampton, Wulfruna St, Wolverhampton WV1 1LY, UK}

\author{{Juan Camilo} {L\'{o}pez~Carre\~no}}
\email{juclopezca@gmail.com}
\affiliation{Faculty of Science and Engineering, University of
  Wolverhampton, Wulfruna St, Wolverhampton WV1 1LY, UK}

\author{Fabrice P. Laussy}
\email{fabrice.laussy@gmail.com}
\affiliation{Faculty of Science and Engineering,
  University of Wolverhampton, Wulfruna St, Wolverhampton WV1 1LY, UK}
\affiliation{Russian Quantum Center, Novaya 100, 143025 Skolkovo,
  Moscow Region, Russia}

\author{Elena {del Valle}}
\email{elena.delvalle.reboul@gmail.com}
\affiliation{Faculty of Science and Engineering,
  University of Wolverhampton, Wulfruna St, Wolverhampton WV1 1LY, UK}
\affiliation{Departamento de F\'isica Te\'orica de la Materia
Condensada, Universidad Aut\'onoma de Madrid, 28049 Madrid,
Spain}

\date{\today}

\begin{abstract}
  Photon correlations, as measured by Glauber's $n$-th order coherence
  functions~$\g{n}$, are highly sought to be minimized and/or
  maximized. In systems that are coherently driven, so-called
  blockades can give rise to strong correlations according to two
  scenarios based on level-repulsion (conventional blockade) or interferences (unconventional
  blockade). Here we show how these two approaches relate to the admixing of a coherent state with a quantum state such as a squeezed state for the simplest and most recurrent case.
  The emission from a
  variety of systems, such as resonance fluorescence, the
  Jaynes--Cummings model or microcavity polaritons, as a few examples
  of a large family of quantum optical sources, are shown to be
  particular cases of such admixtures, that can further be doctored-up
  externally by adding an amplitude- and phase-controlled coherent
  field with the effect of tuning the photon statistics from exactly
  zero to infinity. We show how such an understanding also allows to
  classify photon statistics throughout platforms according to
  conventional and unconventional features, with the effect of
  optimizing the correlations and with possible spectroscopic
  applications. In particular, we show how configurations that can
  realize simultaneously conventional and unconventional antibunching
  bring the best of both worlds: huge antibunching (unconventional)
  with large populations and being robust to dephasing (conventional).
\end{abstract}
\maketitle

\section{Introduction}

That the sum (or superposition) of two fields does not simply add
their respective intensities but introduce an interference term is the
basic principle of optics and other wave
theories~\cite{young_book1807a}. In this text, we study the impact
that such interferences have on the quantum-optical aspect of
light. Since quantum optics typically deals with correlations between
photons, our concern is primarily with the photon statistics of
interfering fields~\cite{ficek_book04a}. The effect of interferences
on photon statistics has been previously studied in the literature,
both past and
recent~\cite{ritze79a,paul82a,buzek95a,prakash05a,majumdar12a,boddeda19a}
but its ubiquity in a wealth of physical systems has been greatly
overlooked. In particular, it appears to be central in
coherently-driven systems, even when one is not directly controlling
the phase and amplitude of the coherent field with the aim of
interfering it with its quantum counterpart.  The problem was first
considered in this form by Vogel~\cite{vogel91a, vogel95a} to bring to
the quantum realm the signal-engineering technique of homodyning,
namely, to extract squeezing (rather than phase modulation in the
classical case). In fact, a related case when the coherent field
interferes with a quantum field that it generates as the result of
driving a quantum emitter, had been previously introduced by
Carmichael under the apt denomination of
``self-homodyning''~\cite{carmichael85a}. The technique has been
lately championed in a recent series of works from the \Vuckovic
group~\cite{fischer16a, muller16a, dory17a, fischer17a, fischer18a}
and a recent work controlled the interference to tune the photon
statistics~\cite{foster19a}. The problem is so widespread, often in
disguise, that even a brief overview of its occurrences and its
identification throughout platforms and realizations, takes the
character of a self-contained Review~\cite{zubizarretacasalengua20a},
to which we refer for further discussion of the literature. Here, we
consider mainly the case of mixing a squeezed state with a coherent
field, as this is the most common configuration, although the general
case is clearly also of fundamental
interest~\cite{lvovsky02a,windhager11a,xu12a,mehringer18a}.

\begin{figure}
  \centering \includegraphics[width=.78\linewidth]{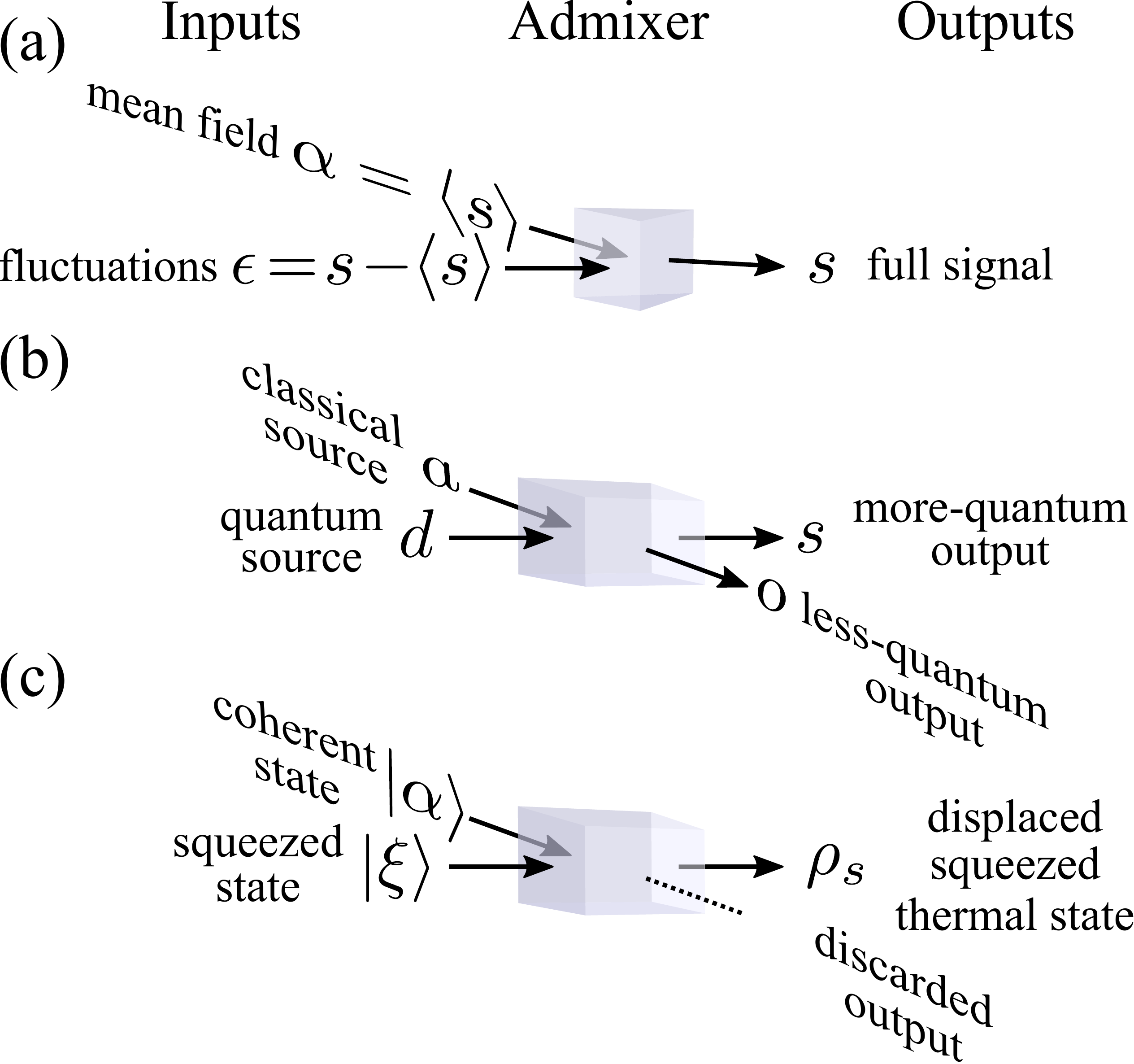}
  \caption{Schemes considered in this text. (a) A quantum field~$s$
    can be seen as composed of a classical component (its mean
    field)~$\langle s\rangle$ and quantum fluctuations~$\epsilon$. (b)
    By using a beam splitter, a classical source~$a$ can be used to
    remove the classical component from a quantum source~$d$ resulting
    in a more quantum output~$s$. The excess of classical signal is
    redirected towards the other branch of the interferometer
    in~$o$. (c) A particular case of great popularity in the
    literature admixes a coherent state~$\ket{\alpha}$ to a squeezed
    state~$\ket{\xi}$, producing a displaced squeezed thermal state in
    a density matrix form~$\rho_s$ as the other output is discarded
    (traced over).}
  \label{fig:Fri27Mar172913CET2020}
\end{figure}

\section{Homodyning a quantum field.}

A simple but powerful decomposition in quantum field theory separates
the mean field~$\alpha\equiv\langle s\rangle$ from the quantum
fluctuations~$\epsilon\equiv s-\langle s\rangle$ of a quantum
field~$s$, which is recovered by bringing these two components
together~\cite{blaizot_book85a}:
\begin{equation}
  \label{eq:Wed8Apr181920CEST2020}
  s=\alpha+\epsilon\,.
\end{equation}
Note that~$\epsilon$, like~$s$, is an operator, and thus describes a
quantum field. In contrast, $\alpha$ is a $c$ number, and describes a
classical field.  This decomposition is sketched in
Fig.~\ref{fig:Fri27Mar172913CET2020}(a). This trivial mathematical
fact, to which theorists recourse to treat the two components
separately, can be in essence realized in a physical setting for
instance by using a beam splitter, i.e., a two-inputs ($a$, $d$),
two-outputs ($o$, $s$) linear and unitary transformation which, for
optical fields, and assuming a transmittance $\mathrm{T}^2$ and
reflectance $\mathrm{R}^2$, reads as~\cite{gerry_book05a}:
\begin{equation}
\label{eq:Thu14Nov141149GMT2019}
    \begin{pmatrix}
      o\\s
    \end{pmatrix}
=
\begin{pmatrix}
  i\mathrm{R} & \mathrm{T}\\
  \mathrm{T} & i\mathrm{R}
\end{pmatrix}
\begin{pmatrix}
  d\\a
\end{pmatrix}
\end{equation}
with $0 \leq \mathrm{T}, \mathrm{R} \leq 1$ and
$\mathrm{T}^2 + \mathrm{R}^2 = 1$. The $i$ term comes from Stokes'
relations and can be corrected with a half-wave plate, so that,
assuming a balanced beam-splitter $\mathrm{R}=\mathrm{T}=1/\sqrt2$, we
can have $o=(i/\sqrt{2})(-a+d)$ and~$s={1/\sqrt{2}}(a+d)$, with the
outputs providing an attenuated fraction of the difference and sum of
the two input fields. The phase shift in the difference could also be
removed with another half-wave plate. The attenuation comes from
energy-conservation and the impossibility to amplify faithfully a
quantum field.  Since one can, however, amplify faithfully a classical
field, then an amplification by a factor $\mathrm{T}/\mathrm{R}$,
together with the half-wave plate phase shift of~$-i$ and passing
through an unbalanced beam splitter, returns on the transmitted output
the signal $s=T(a+d)$ and on the reflected one
$o=i[\mathrm{R}d-(\mathrm{T}^2/\mathrm{R})a]$. Therefore, in the
limit~$\mathrm{T}\rightarrow 1$, which is theoretical only since this
requires infinite amplification, one can recover the sum of the two
fields without attenuation, $s=a+d$. The other field then collects the
diverging (due to amplification) classical field alone
$-i(\mathrm{T}/\mathrm{R})\langle\sigma\rangle$. As this classical
field can be discarded without affecting the quantum field, this
hypothetical setup realizes the scheme sketched in
Fig.~\ref{fig:Fri27Mar172913CET2020}(a) by feeding it with the quantum
fluctuations $d=\epsilon$ and the mean-field $a=\langle s\rangle$. In
the following, we will exploit such decompositions as
Eq.~(\ref{eq:Wed8Apr181920CEST2020}) as well as the possibility to
interfere in a beam splitter a classical source~$a$ with a quantum
source~$d$, to similarly collect on the one hand the magnified quantum
character of the quantum source, that is, devoid of its mean field,
that is re-routed, on the other hand, in another channel which can be
traced over. This is sketched in
Fig.~\ref{fig:Fri27Mar172913CET2020}(b). We have shown for instance
how one can collect in this way the pure quantum emission from a
coherently driven two-level system by destructively interfering its
mean-field~\cite{lopezcarreno18b}.  We now generalize this scheme to
other quantum sources.

Being interested in the~$s$ component alone, and in particular in its
quantum averages, to describe the beam-splitter situation, we perform
a binomial expansion on $s=\mathrm{T}d+i\mathrm{R}a$ and its conjugate
to compute
%
\begin{multline}
\label{eq:2LScorrBS}
\corr{s}{n}{m} = T^{n+m}\\\sum_{p=0}^n \sum_{q=0}^m \binom{n}{p} \binom{m}{q}
i^{q-p}
(\mathrm{R}/\mathrm{T})^{p+q}\langle a^{\dagger p}a^qd^{\dagger(n-p)}d^{m-p}\rangle
\end{multline}
which includes the beam-splitting signal reduction as well as the
$\pi$ phase-shift for reflection. While those will typically be
present in an actual experiment, they are not so important for the
actual process of interfering (or mixing) the fields and can be
absorbed in an overall leading factor which will, in fact, cancel in
all normalized correlators. This reflects the fact that photon
statistics is not affected by linear processes. In this case, one can
consider directly the correlators that arise from the simple addition
of the fields $s=a+d$:
\begin{equation}
  \label{eq:Fri10Apr194032CEST2020}
  \corr{s}{n}{m} = \sum_{p=0}^{n} \sum_{q=0}^{m}
  \binom{n}{p}\binom{m}{q}\langle a^{\dagger p}a^qd^{\dagger(n-p)}d^{m-p}\rangle\,.
\end{equation}
Besides, one can assume no correlations between the two inputs, so
that
$\langle a^{\dagger p}a^qd^{\dagger(n-p)}d^{m-p}\rangle=\langle
a^{\dagger p}a^q\rangle\langle d^{\dagger(n-p)}d^{m-p}\rangle$ and
since we presently concern ourselves with the case where~$a$ is
classical, i.e., with~$\alpha=\langle a\rangle$, we can further
simplify Eq.~(\ref{eq:Fri10Apr194032CEST2020}) into
\begin{equation}
  \label{eq:homodynecorrelators}
  \corr{s}{n}{m} = \sum_{p=0}^{n} \sum_{q=0}^{m}
  \binom{n}{p}\binom{m}{q} \alpha^{* p} \alpha^{q} \corr{d}{(n-p)}{m-q} \,.
\end{equation}
This yields, as the simplest ($n=m=1$) case, the classical version of
interfering fields, that holds at the one-photon level (in the sense
of first-order correlations). Namely, the output field~$s$ sums the
intensities of the two fields, plus or minus an interfering term:
\begin{equation}
  \label{eq:Thu14Nov150835GMT2019}
  \mean{n_s} \equiv \pop{s} = |\alpha|^2 + \mean{n_d} + 2 \mathrm{Re}[ \alpha^* \coh{d} ] \,.
\end{equation}
So far this is just a sophisticated way to write one of the most basic
and best known features of optics: interferences. One departs from the
classical realm in the next orders, with~$n=m\ge 2$.  Correlations are
typically normalized, which yields the $n$th-order photon
correlation~$g^{(n)}\equiv {\mean{s^{\dagger
      n}s^n}}/{\mean{n_s}^n}$~\cite{glauber63a} that one can write in
increasing powers of~$\alpha$, namely,
\begin{subequations}
\label{eq:Mon20Apr190800CEST2020}
\begin{align}
  \g{2}_{s} &= 1 + \mathcal{I}_0 + \mathcal{I}_1 + \mathcal{I}_2\,,\\
  \g{3}_s &= 1 + \mathcal{J}_0 + \mathcal{J}_1+ \mathcal{J}_2+ \mathcal{J}_3+ \mathcal{J}_4\,,\\
  \g{k}_s&=1+\sum_{m=0}^{2n-2}\mathcal{K}_m\,,
\end{align}
\end{subequations}
with $\mathcal{I}_m\langle n_s\rangle^2$,
$\mathcal{J}_m\langle n_s\rangle^3$,
$\mathcal{K}_m\langle n_s\rangle^4 \propto |\alpha|^m$. Note that
there are no explicit terms $\mathcal{K}_{2n-1}$, $\mathcal{K}_{2n}$
(e.g., $\mathcal{I}_3$, $\mathcal{I}_4$ for~$\g{2}$ and
$\mathcal{J}_5, \mathcal{J}_6$ for~$\g{3}$) because, through
simplifications, these get absorbed in the unit term~1, which
otherwise comes from the coherent field. The non-coherent (quantum)
contributions read as, for
$\g{2}$~\cite{mandel82a,carmichael85a,vogel91a,vogel95a},
\begin{subequations}
  \label{eq:ThuFeb22121240CET2018}
  \begin{align}
      \label{eq:ThuFeb22121240CET2018a}
	\mathcal{I}_0 &= \frac{\corr{d}{2}{2} -
                        \pop{d}^2}{\mean{n_s}^2}\,,\\
      \label{eq:ThuFeb22121240CET2018b}
	\mathcal{I}_1 &=4\frac{ \mathrm{Re}[\alpha^{*} (\av{\ud{d}  d^2}-
                        \pop{d} \av{d})]}{\mean{n_s}^2}\,,\\
      \label{eq:ThuFeb22121240CET2018c}
      \mathcal{I}_2 &= 2\frac{\mathrm{Re}[ {\alpha^*}^{2}
          \av{d^2}] - 2\mathrm{Re}[\alpha^*
                      \av{d}]^2  + |\alpha|^2 \pop{d}}{\mean{n_s}^2}
  \end{align}
\end{subequations}
and, for~$\g{3}$,
\begin{subequations}
\begin{align}
  \mathcal{J}_0 &=\frac{\corr{d}{3}{3} - \pop{d}^3 }{\av{n_s}^3} \, ,\\
  \mathcal{J}_1 &= 6 \frac{ \mathrm{Re}[ \alpha^* \left( \corr{d}{2}{3} - \coh{d} \pop{d} \right)] }{\av{n_s}^3} \, ,\\
  \mathcal{J}_2 &= 3\Big[ 2\mathrm{Re}[{\alpha^*}^2 \av{\ud{d} d^3} ] +
                  |\alpha|^2 \left(3\corr{d}{2}{2} +\pop{d}^2 \right)\nonumber\\
&\hskip3.5cm- 4 \pop{d} \mathrm{Re}[ \alpha^* \coh{d}]^2 \Big]/ \av{n_s}^3 \,,\\ 
  \mathcal{J}_3 &= 2 \Big[ \mathrm{Re}[ {\alpha^*}^3 \av{d^3} ] +  |\alpha|^2 \mathrm{Re}[
                  \alpha^* \left( 9 \av{\ud{d} d^2} - 6 \coh{d} \pop{d} \right)]\nonumber\\
                 &\hskip3.5cm - 4 \mathrm{Re}[ \alpha^* \coh{d}]^3 \Big]/ \av{n_s}^3 \, ,\\
  \mathcal{J}_4 &= 6|\alpha|^2 \frac{ \mathrm{Re}[ {\alpha^*}^2 \av{d^2} ] + |\alpha|^2 \pop{d} - 2 \mathrm{Re}[\alpha^*  \coh{d}]^2}{\av{n_s}^3}\,.
\end{align}
\end{subequations}
It is possible, though not necessary presently, to provide the
higher-order-correlators terms~$\mathcal{K}_m$. In connection of our
coming discussion on squeezing, note that~$\mathcal{I}_2$ can be
rewritten in terms of the field quadratures:
\begin{multline}
  \mathcal{I}_2 =4 \Big[ |\alpha|^2 \left( \cos^2 \phi \
    \av{{:}X_{d}^2{:}} + \sin^2 \phi
    \ \av{{:}Y_{d}^2{:}} +{} \right. \\
  \quad \left. {}+ \cos \phi \sin \phi \ \av{\lbrace X_d , Y_d
      \rbrace} \right) - \mathrm{Re}[ \alpha^* \av{d}]^2 \Big]/\mean{n_s}^2\,.
\end{multline}
Here, the notation ``$::$'' indicates normal ordering,
$\lbrace X_d,Y_d \rbrace \equiv X_dY_d+Y_dX_d$, and
$X_d \equiv \frac{1}{2} \left(\ud{d}+d\right) $,
$Y_d \equiv \frac{i}{2} \left(\ud{d}- d \right)$ are the quadratures of the
field described with the annihilation operator~$d$. Such
decompositions of the photon statistics,
[Eqs.~(\ref{eq:Mon20Apr190800CEST2020})], pinpoint which mechanism is at
play in accounting for the photon correlations. The coefficients
$\mathcal{I}_0$ and~$\mathcal{J}_0$, for instance, quantify the
statistics of the quantum part of the signal, and through the first
term in the numerator, are basically the Glauber correlators
themselves.  $\mathcal{I}_0$ also bears some resemblance to
Mandel's~$Q$ parameter~\cite{mandel79a}. Indeed, when there is no
coherence involved and the full signal is quantum, i.e., when
$s\approx d$, then $\mathcal{I}_0=g^{(2)}-1$ and~$\mathcal{J}_0=g^{(3)}-1$
(and similarly at still higher orders) with all other~$\mathcal{I}_k$,
$\mathcal{J}_k$ cancelling. In this case, the photon statistics can be
fully attributed to the quantum dynamics of the naked emitter. In
other cases, it includes interferences with the coherent contribution,
which are the multi-photon counterpart for the photon
statistics of the usual field (single-photon) interference,
[Eq.~\ref{eq:Thu14Nov150835GMT2019}], for intensities. At the two-photon
level, $\mathcal{I}_2$ shows that the interference can be well
described through squeezing of the quantum signal, and we will study
exactly this configuration in the next section. At the three-photon
level, the situation becomes more complex with departures from the
exact squeezed and coherent-states interference
scenario~\cite{zubizarretacasalengua20a}, and we will therefore focus
on the simpler, and so far more popular, case of two-photon
statistics. Nevertheless, the same underlying idea of multiphoton
interferences prevails.

We conclude this section with the version of
Eqs.~(\ref{eq:ThuFeb22121240CET2018}) which will be used in the rest
of the text, where we shall consider both the cases where the coherent
state is brought externally to the emitter (homodyning) and the case
where the coherent state is the classical, or mean-field, component of
the emitter itself (self-homodyning). In the latter case, coming back
to Eq.~(\ref{eq:Wed8Apr181920CEST2020}),
with~$d=\epsilon=s-\langle s\rangle$ and~$a=\langle s\rangle$, one
finds for the~$\mathcal{I}$ coefficients (the same could be done
for~$\mathcal{J}$, etc.):
\begin{subequations}
  \label{eq:decompositiontermswhole}
 \begin{align}
   \mathcal{I}_0 &= \Big[{\corr{s}{2}{2} - \av{\ud{s}s}^2 - 4
                   |\mean{s}|^4 + 6 |\mean{s}|^2 
                   \pop{s}}+{}\nonumber\\
                 &\kern1cm{2\mathrm{Re}[   {\mean{\ud{s}}}^2 \av{s^2} - 2
                   \mean{\ud{s}} \av{\ud{s} s^2} ]
                   }\Big]/{\pop{s}^2}\, ,\\ 
   \mathcal{I}_1 &=4 \frac{\mathrm{Re}[ \mean{\ud{s}} \av{\ud{s} s^2} -
                   \mean{\ud{s}}^2 \av{s^2}] + 2
                   |\mean{s}|^2 \left(|\mean{s}|^2
                   -\mean{\ud{s}s}\right) }{\pop{s}^2}\, , \\
   \mathcal{I}_2 &=
                   2 \frac{ \mathrm{Re}[\mean{\ud{s}}^{2}  \av{s^2}] +  |\mean{s}|^2 \pop{s} - 2|\mean{s}|^4  }{\pop{s}^2}\, .
 \end{align}
\end{subequations}

\section{Interfering a squeezed and a coherent state}
\label{sec:Tue14Apr124542CEST2020}

We now focus on one case of considerable interest, as it is possibly
the most recurring case, although not always identified as such,
namely, when the quantum state is a squeezed
state. Note that a squeezed state has a zero mean,
  $\langle d\rangle=0$, so it technically qualifies as a
  fluctuation term in the sense of
  Eq.~(\ref{eq:Wed8Apr181920CEST2020}). However, since it can be the
  dominant term of the total population, we will also refer to it as
  the quantum part of the signal.  The operators~$a$ and~$d$ are now
both annihilation operators for bosonic fields, and we define
$D_a(\alpha) \equiv \exp(\alpha \ud{a}-\alpha^\ast a)$ the displacement
operator for the coherent state~$\ket{\alpha}=D_a(\alpha)\ket{0}$,
with~$\alpha = |\alpha|e^{i\phi}$ and
$S_d(\xi) \equiv \exp({1\over2}[\xi d^{\dagger\,2}-\xi^\ast d^2])$ the squeezing
operator for the squeezed state~$\ket{\xi}=S_d(\xi)\ket{0}$,
with~$\xi=re^{i\theta}$ the squeezing parameter.  The total input
state is then
\begin{equation}
\ket{\psi_{\mathrm{in}}} = {D}_a \left(\alpha\right) {S}_d \left(\xi\right)
\ket{0}_{da},
\end{equation}
where the state subscript indicates the input and output subspaces where
operators are acting upon, namely, here, in the input basis. Now,
applying the transformation Eq.~\eqref{eq:Thu14Nov141149GMT2019} and
rearranging terms, we obtain, first for the displacement operator,
${D}_a \left(\alpha\right) = \exp \left( \alpha_o \ud{o} - \alpha_o^*
  o + \alpha_s \ud{s} - \alpha_s^* s \right)$, where
$\alpha_s = i \mathrm{R} \alpha$ and $\alpha_o = \mathrm{T} \alpha$.
Exponentials of operators can be factorized since both outputs are
independent from each other and commute. This leads to
$D_a \left(\alpha \right) = D_o \left(\alpha_o \right) D_s
\left(\alpha_s\right) = D_s \left(\alpha_s\right) D_o \left(\alpha_o
\right)$, where each displacement operator $D_j \left(\alpha_j\right)$
($j=o,s$) only acts over its assigned output. Second, the squeezing
operator in the output basis reads as
\begin{multline}
{S}_d \left(\xi \right) = \exp \left[ \frac{1}{2}
  \left(\xi_o^* o^2 - \xi_o \ud{o} \right) + \frac{1}{2}
  \left(\xi_s^* s^2 - \xi_s \ud{s} \right) + \right.\\
  \left. \left(\xi_{os}^* o s - \xi_{os} \ud{o} \ud{s} \right)
  \right] =\exp(S_o+S_s+S_{os})\,,
\end{multline}
where $\xi_s = \mathrm{T}^2 \xi$, $\xi_o = - \mathrm{R}^2 \xi$ and
$\xi_{os} = i \mathrm{R} \mathrm{T} \, \xi$. This exponential can be
split into a product, only if $\lbrack S_o + S_s , S_{os} \rbrack =0$,
which is fulfilled in the particular case of a balanced BS,
$\mathrm{T} = \mathrm{R}$.  This restriction is however not very
stringent since its first-order correction grows proportionally to
$r^2 \mathrm{T} \mathrm{R} \left(\mathrm{T}^2-\mathrm{R}^2
\right)$. Thus, for either low squeezing signal ($r \ll 1$) or almost
symmetrical BS ($\mathrm{T}-\mathrm{R} \approx 0$), the output signal
can still be described as follows. Since the commutator
$\lbrack S_o , S_s \rbrack$ vanishes for all possible values, the
exponential simplifies into
$S_d \left(\xi \right) = S_o \left(\xi_o \right) S_s \left(\xi_s
\right) S_{os} \left(\xi_{os} \right)$. Therefore, the output state
can be written as
%
\begin{multline}
\ket{\psi_{\mathrm{out}}} = D_o \left(\alpha_o\right)
S_o \left(\xi_o\right) D_s \left(\alpha_s\right)
S_s \left(\xi_s\right) S_{os}
\left(\xi_{os}\right) \ket{0}_{os} \\ = D_o
\left(\alpha_o\right) S_o \left(\xi_o\right) D_s
\left(\alpha_s\right) S_s \left(\xi_s\right)
\ket{\xi_{os}}_{os},
\end{multline}
%
where $\ket{\xi_{os}}$ is a two-mode squeezed state, which in the Fock
basis reads as~\cite{gerry_book05a}:
\begin{equation}
\ket{\xi_{os}} = \frac{1}{\cosh r_{os}} \sum_{n=0}^{\infty} \left(\tanh r_{os}\right)^n
\ket{n, n}_{os},
\end{equation}
where $r_{os} = |\xi_{os}| = \mathrm{R} \mathrm{T} \, r$. The
corresponding density matrix for this pure state reads as
$\rho_{\mathrm{out}} = \ket{\psi_{\mathrm{out}}}
\bra{\psi_{\mathrm{out}}}$. Tracing out output $o$, we obtain the
density matrix for output $s$ only (our signal of interest):
$\rho_s = \Tr_o\{ \rho_{\mathrm{out}}\}$.  With the cyclic properties
of the trace, we move operators clockwise to act over the output
subspace~$o$ and use
$\ud{D_o} \left(\alpha_o\right) D_o \left(\alpha_o\right) = \ud{S_o}
\left(\xi_o\right) S_o \left(\xi_o\right) = {\mathds{1}}_o $, where
${\mathds{1}}_o $ is the identity. Furthermore, any operator that only
acts on the $s$-subspace can be taken out of the trace. This brings us
to an expression for the quantum state of the signal:
\begin{equation}
\rho_s= D_s \left(\alpha_s\right) S_s
\left(\xi_s\right) \big(\Tr_o \lbrace \ket{\xi_{os}} \bra{\xi_{os}}
\rbrace \big) \ud{S_s} \left(\xi_s\right) \ud{D_s}
\left(\alpha_s\right) \, .
\end{equation}
The partial trace $\Tr_o \lbrace \ket{\xi_{os}} \bra{\xi_{os}} \rbrace$
has the form of a thermal state
\begin{equation}
  \rho_\mathrm{th}=\Tr_o \lbrace \ket{\xi_{os}} \bra{\xi_{os}} \rbrace =
        \, \frac{1}{\cosh^2 r_{os}} \sum_{n}^{\infty} \left(\tanh
        r_{os}\right)^{2n} \ket{n}_s\!\!\bra{n}_s
\end{equation}
with mean population
$\mathrm{p_{th}} \equiv \pop{s} = \sinh^2 r_{os}$. To sum up, admixing
a coherent and a squeezed state as shown in
Fig.~\ref{fig:Fri27Mar172913CET2020}(c) produces on one arm of a
beam-splitter a \textit{displaced squeezed thermal}
state~\cite{lemonde14a} where the displacement and squeezing are both
in terms of~$s=a+d$:
\begin{equation}
  \label{eq:Sun3May200019CEST2020}
  \rho_s = D_s \left(\alpha_s\right) S_s
  \left(\xi_s\right) \rho_{\mathrm{th}} \ud{S_s}
  \left(\xi_s\right) \ud{D_s} \left(\alpha_s\right),
\end{equation}
with parameters $\alpha_s = i\mathrm{R}|\alpha| e^{i \phi}$,
$\xi_s = r_s e^{i \theta_s} = \mathrm{R}^2 e^{i \left( \theta + \pi
  \right)}$, and
$ \mathrm{p_{th}} = \sinh[2](\mathrm{R}\mathrm{T} \, r)$. Even though
$\mathrm{T}$ and $\mathrm{R}$ appear as free parameters, we remind
that Eq.~(\ref{eq:Sun3May200019CEST2020}) is valid for
$\mathrm{R} \approx \mathrm{T}$ (and is exact
for~$\mathrm{R}=\mathrm{T}$). We now restrict ourselves to the case of
a 50:50 beam splitter ($\mathrm{T}^2 = \mathrm{R}^2 = 1/2$).  The
thermal population reads as, in terms of the squeezed population of the
input signal $\av{n_d} = \sinh^2 r$,
\begin{equation}
\mathrm{p_\mathrm{th}} = \frac{1}{2}\left(\sqrt{1+ \av{n_d}}-1\right)\,.
\end{equation}
From $\rho_s$ we can compute the observables for the mixed signal:
\begin{subequations}
  \label{eq:Fri17Apr123043CEST2020}
  \begin{equation}
    \av{n_s} = \frac{|\alpha|^2}{2} + \frac{\av{n_d}}{2} \, , \quad |\av{s^2}| = \left(\mathrm{p_{th}} + \frac{1}{2}\right) \sinh(r)\,,
  \end{equation}
  \begin{multline}
    \label{eq:DST_g2}
    \g{2}_s =  1 + \av{n_s}^{-2} \sinh^2 r
    \left[\cosh 2r + {}\right.\\\left. {}+ 2 |\alpha|^2 \left(1- \cos(\theta-2\phi) \coth r\right) \right]  \, ,
  \end{multline}
  \begin{multline}
    \g{3}_s =  1 + \av{n_s}^{-3} \sinh^2 r \, \Big\lbrace \sinh^2 2r + 5 \sinh^2 r \cosh 2r +{}\\
    {}+ 6 |\alpha|^4 \left(1- \cos(\theta-2\phi) \coth r \right) + \\
    \: 
    3 |\alpha|^2 \left[3 \coth^2 r - 1 + 6 \left(1- \cos(\theta-2\phi) \coth r\right) \right]
    \Big\rbrace  \,.
  \end{multline}
\end{subequations}


\begin{figure}[t]
  \centering \includegraphics[width=.9\linewidth]{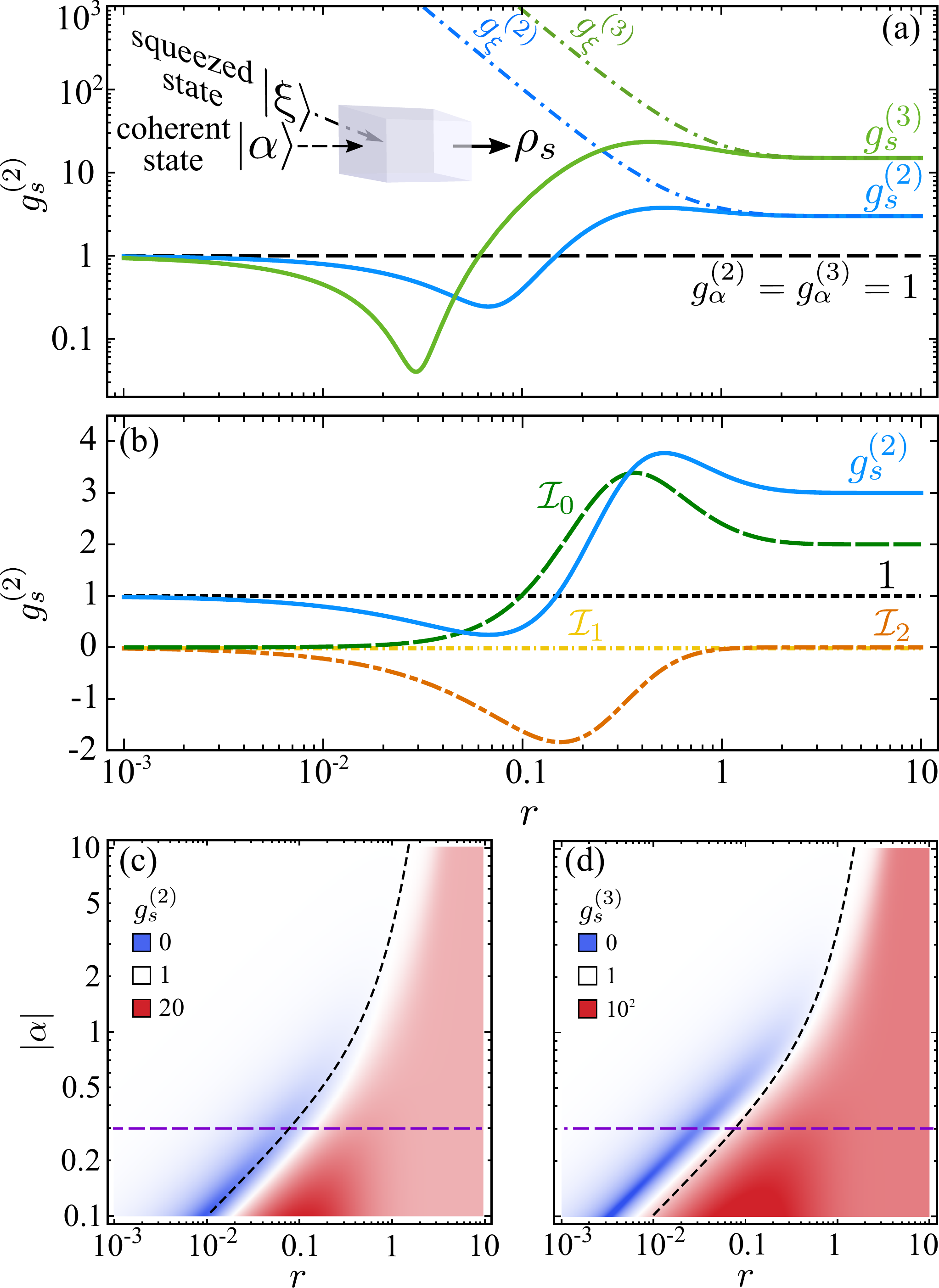}
  \caption{(Color online) (a) Admixing a squeezed state with
    statistics~$g_\xi^{(n)}$ (dash-dotted lines, $n=2,3$) with a
    coherent state with $g_\alpha^{(n)}=1$ (dashed) yields a displaced
    thermal state whose statistics~$g_s^{(2)}$ ranges between
    antibunching and superbunching, here shown as a function of
    squeezing~$r$. (b) Decomposition of~$g_s^{(2)}$ in terms of
    the~$\mathcal{I}$ coefficients, with~$\mathcal{I}_1=0$ for a
    squeezed-coherent states admixture. (c) Map of the~$g_s^{(2)}$
    realized for arbitrary admixtures of coherent~$\ket{\alpha}$ and
    $r$-squeezed states. The cases in~(a) and~(b) correspond to the
    purple-dashed cut shown at~$|\alpha|=0.3$ and $\theta=2\phi$. (d)
    Same as panel~(c) but for~$g^{(3)}_s$. In both cases, the
    black-dashed line that optimizes two-photon antibunching shows the
    mismatch to different photon-orders.}
  \label{fig:Tue14Apr124339CEST2020}
\end{figure}

The second- and third-order correlators,
[Eqs.~(\ref{eq:Fri17Apr123043CEST2020}b--c)], are shown in
Fig.~\ref{fig:Tue14Apr124339CEST2020}(a) for~$|\alpha|=0.3$ and as a
function of the squeezing parameter~$r$, with~$\theta=2\phi$ (blue and
green solid lines), along with the corresponding correlators for the
coherent state alone, $g_\alpha^{(n)}=1$ for all~$n$ (black dashed),
and for the squeezed state alone, $g_\xi^{(2)}=2+\coth(r)^2$ and
$g_\xi^{(3)}=6+9\coth(r)^2$, these being particular cases of
Eqs.~(\ref{eq:Fri17Apr123043CEST2020}b--c) when~$\alpha=0$ (blue and
green dashed-dotted line). Counter-intuitively, $g^{(2,3)}_s\ll 1$ is
obtained when the squeezed light itself is, on the opposite,
super-Poissonian, $g_\xi^{(2)}, g_\xi^{(3)}\gg1$. This is the
multiphoton counterpart of Eq.~(\ref{eq:Thu14Nov150835GMT2019}), that
shows how adding two maxima can yield a minimum (when the amplitudes
have opposite phases). Here, an interference at the $n\ge2$ level
shows how adding two fields with Poissonian or super-Poissonian
fluctuations can yield a sub-Poissonian field.

The second-order correlation~(\ref{eq:DST_g2}) is decomposed according
to Eq.~(\ref{eq:ThuFeb22121240CET2018}) into~$\mathcal{I}$
coefficients that read as
\begin{subequations}
    \label{eq:FriOct20184657CEST2017}
    \begin{align}
        \label{eq:FriOct20184657CEST2017a}
       \mathcal{I}_0&= \frac{ \sinh^4(r)}{\av{n_s}^2}
               [1+\coth(r)^2]\,,\\
        \label{eq:FriOct20184657CEST2017b}
       \mathcal{I}_1&=0\,,\\
        \label{eq:FriOct20184657CEST2017c}
       \mathcal{I}_2&=\frac{2 |\alpha|^2 \sinh^2(r)}{\av{n_s}^2} \left[ 1 - \cos(\theta -
           2 \phi) \coth(r) \right] \,,
  \end{align}
\end{subequations}
where~$\mean{n_s}=|\alpha|^2 + \sinh^2(r)$. They are shown in
Fig.~\ref{fig:Tue14Apr124339CEST2020}(b).  From
Eqs.~(\ref{eq:FriOct20184657CEST2017}), $\g{2}_s<1$ only if
$\mathcal{I}_2$ is negative, so this is related to squeezing, which can
also be small as long as it is nonzero. Note that while this is a
necessary condition, it is not a sufficient one, i.e., $\mathcal{I}_2$
can be negative without~$\g{2}_s$ being less than one. The phase
between the squeezed and the coherent sates must
satisfy~$|\theta-2\phi|<\pi/2$, being optimum when $\theta=2\phi$,
i.e., when coherent and squeezed states have the same phase, since the
phase of a squeezed state is $\theta/2$.  The optimum sub-Poissonian
character is obtained for a small amount of squeezing~$r$, in which
case the coherent state that minimizes two-photon antibunching has
amplitude
\begin{equation}
  \label{eq:Thu16Apr192441CEST2020}
  |\alpha|_\mathrm{min}=e^{r}\sqrt{\cosh(r)\sinh(r)}\,,
\end{equation}
which yields the optimum two-photon antibunching
\begin{equation}
  \label{eq:FriOct20172135CEST2017}
  g_{s,\,\mathrm{min}}^{(2)}=1-\frac{e^{-2r}}{1+\sinh(2r)}
\end{equation}
which is always~$\le 1$ and goes to zero as both~$r$ and~$\alpha$
vanish, in the proportion of Eq.~(\ref{eq:Thu16Apr192441CEST2020}).
One can find the counterpart $|\alpha|_\mathrm{max}$ which yields the
maximum bunching~$g_{s,\,\mathrm{max}}^{(2)}$ but the expressions are
too bulky to be given in closed-form. The correlations obtained in
this way are strong when the fields are weak, which will be a
recurring theme in the following sections. We will show how they
indeed become exactly zero and infinite to first-order in the driving,
regardless of the other parameters in the system, which has made a
lasting impression for the case of
antibunching~\cite{liew10a,bamba11a,flayac17b}.  The possible photon
statistics as a function of the coherent and squeezed states
admixtures, for optimum phase matching, is shown in
Figs.~\ref{fig:Tue14Apr124339CEST2020}(c--d).  There is, therefore, a
great tunability from such a simple admixture, since $g^{(2)}_s$ takes
all the values from $g_{s,\mathrm{min}}^{(2)}$,
Eq.~(\ref{eq:FriOct20172135CEST2017}) (which is zero with vanishing
signal) to $g_{s,\mathrm{max}}^{(2)}$ (which is $\infty$ with
vanishing signal), simply by adjusting the magnitudes of the coherent
field and the squeezing parameter. This is shown in
Fig.~\ref{fig:Tue14Apr124339CEST2020}(b) for the fixed coherent
amplitude~$|\alpha|=0.3$ by changing the amount of squeezing, which
alters the~$\mathcal{I}$ coefficients with effect of tuning from
antibunching to bunching, with $g_{s,\mathrm{min}}^{(2)}\approx0.24$
at~$r\approx0.07$ and $g_{s,\mathrm{max}}^{(2)}\approx3.77$
at~$r\approx0.52$. Such a controlled tuning has been recently
experimentally implemented by Foster \emph{et al.}~\cite{foster19a}
with a quantum dot in a waveguide cavity, in which case the
transmittance~$\mathrm{T}$ and the detuning between the external laser
and the quantum dot served as the control parameters to vary the
$\mathcal{I}_0$ and~$\mathcal{I}_2$ coefficients, which these Authors
interpreted as a two-photon bound state and an interference term,
respectively.  Similarly, at the three-photon level, and still for the
fixed~$|\alpha|=0.3$ of Fig.~\ref{fig:Tue14Apr124339CEST2020}, one has
$g_{s,\mathrm{min}}^{(3)}\approx0.04$ at~$r\approx0.03$ and
$g_{s,\mathrm{max}}^{(3)}\approx23.45$ at~$r\approx0.44$. The curve
optimizing two-photon antibunching,
Eq.~(\ref{eq:Thu16Apr192441CEST2020}), in these correlation-spaces is
shown dashed black in
Figs.~\ref{fig:Tue14Apr124339CEST2020}(c--d). Importantly, and this
will be another recurrent theme in what follows, the admixture that
optimizes~$g_s^{(2)}$ antibunching is not the one that
optimizes~$g_s^{(3)}$ antibunching, and vice versa, as shown in
Fig.~2(d) where the optimum two-photon antibunching falls in a region
of three-photon bunching. This suggests that such antibunching is not
suitable for single-photon emission. In contrast, superbunching tends
to be degenerate at all photon numbers, which is only approximately
realized here due to the broad maximum.

\section{Dynamics}
\label{sec:Fri17Apr144500CEST2020}

The above considerations are general, and apply equally well to the
case just discussed of interfering two pure states admixed as an
initial condition, or as a result of some self-consistent dynamics
whereby a coherent field (typically, a laser driving a system)
interferes with by-products of its excitation which, if squeezed, will
be decomposed essentially as described above, thus producing the same
type of photon statistics ranging from sub-Poissonian to superbunched
depending on the coherent vs squeezed states relationship.  

As the simplest case, consider feeding a beam splitter with the output
of two cavities with Hamiltonians
\begin{subequations}
\begin{align}
  \label{eq:Mon20Apr121211CEST2020}
  H_a&=\Delta_a\ud{a}a + \Omega_a (e^{i \phi} \ud{a}+ e^{-i \phi} a)\,,\\
  H_d&=\Delta_d \ud{d}d + i \lambda/2 (d^{\dagger 2} - d^2)\,,
\end{align}
\end{subequations}
driven by a laser of intensity~$\Omega_{a}$ (for the~$a$ cavity) and
$\lambda$ (for the~$d$ cavity) with respective
detunings~$\Delta_{a,d}$ in their steady-states of a coherent state
(for~$a$) and a squeezed thermal state (for~$d$) through the master
equations ($c = a, d$)
\begin{equation}
\partial_t \rho = i[\rho,H_c]+\frac{\gamma_c}{2}\mathcal{L}_c \rho
\end{equation}
with~$\mathcal{L}_c=(2c\rho \ud{c} - \ud{c}c\rho - \rho\ud{c}c)$ to
include dissipation. This provides a self-consistent, fully dynamical
model which establishes the correspondence with the displacement
$\alpha$, squeezing parameters $\xi=re^{i\theta}$, and the thermal
population $\mathrm{p_{th}}$, that are given by:
\begin{subequations}
	\begin{align}
	\alpha&=\mean{a}\,,\\
	|\mean{d^{2}}|& =\sinh(r) \cosh(r) \left(1 + 2 \mathrm{p_{th}} \right)\,,\\
	\mean{\ud{d}d} & = \sinh[2](r) + \mathrm{p_{th}} \cosh(2r) \,,\\
	\theta&= \arg [\mean{d^{2}} - \mean{d}^2])\,.
	\end{align}
\end{subequations}
The two systems can each be solved exactly, yielding the steady-state
solutions for the parameters defined above as
\begin{equation}
\label{eq:FriOct20213722CEST2017a}
\alpha = - \frac{2 i \, \Omega_a e^{i \phi}}{\gamma_a + 2 i  \Delta_a} ,
\end{equation}
for the coherent state and
\begin{subequations}
\label{eq:FriOct20213722CEST2017b}
\begin{align}
  \mathrm{p_{th}} &= \sinh^2(r) = \frac{1}{2}
\left(\sqrt{\frac{\Gamma_d^2}{\Gamma_d^2 - 4 \lambda^2 }} - 1 \right)\,,\\
\theta &= \arg \left(\gamma_d - 2 i \Delta_d \right)\,,
\end{align}
\end{subequations}
for the squeezed thermal state, where
$\Gamma_i^2 \equiv \gamma_i^2 + 4\Delta_i^2$.  Tuning these easily
accessible parameters (laser detunings and intensities) one can thus
produce an output field with the desired photon statistics, from
$g^{(2)}=0$ to~$\infty$ according to
Eqs.~(\ref{eq:Fri17Apr123043CEST2020}). The population for the mixed
signal is given by:
\begin{equation}	
n_s  = \frac{1}{2} \left( \frac{4 \Omega_a^2}{\Gamma_a^2} + \frac{2 \lambda^2}{\Gamma_d^2 - 4 \lambda^2} \right) \,,
\end{equation}
while the two-photon statistics is
\begin{equation}
\begin{split}
\g{2}_s = & \,\Big\lbrace
\lambda^2 \Gamma_a^4 \big(\Gamma_d^2 + 8 \lambda^2 \big) + 8 \Omega_a^2 \lambda \big(\Gamma_d^2
- 4 \lambda^2 \big) \times \\ 
&  \Big[ 4 \Gamma_a^2 \lambda - \cos (2 \phi) \big(\gamma_a^2 \gamma_d
- 4 \gamma_d \Delta_a^2 + 8 \gamma_a \Delta_a \Delta_d \big)  + \\ & 2  \sin (2 \phi)  
\big(-2 \gamma_a \gamma_d \Delta_a + \gamma_a^2 \Delta_d - 4 \Delta_a^2 \Delta_d \big) \Big] + \\
& 16 \Omega_a^4 \big(\Gamma_d^2 - 4 \lambda^2 \big)^2 \Big\rbrace  \Big/ 
\Big\lbrace 4 \big[ \Gamma_a^2 \lambda^2  + 2 \Omega_a^2 \big(\Gamma_d^2 - 4 \lambda^2 \big)\big]^2 \Big\rbrace \,.
\end{split}
\end{equation}
The decomposition of $\g{2}_s$ in terms of the~$\mathcal{I}$
coefficients reads as
\begin{subequations}
\begin{align}
\mathcal{I}_0 & = {\lambda^2\over n_s^2}
\frac{  \Gamma_d^2 + 4 \lambda^2 }{\big(\Gamma_d^2 - 4 \lambda^2 \big)^2} \,, \\
\mathcal{I}_1 & = 0 \,,  \\
\mathcal{I}_2 & = {8\over n_s^2}\frac{\lambda \Omega_a^2}{\Gamma_a^4 \big(\Gamma_d^2 - 4 \lambda^2 \big) }  \times {}\nonumber \\
& \quad \big[2 \lambda \Gamma_a^2 - \cos(2 \phi) \big(\gamma_a^2 \gamma_d  - 4 \gamma_d \Delta_a^2
+ 8 \gamma_a \Delta_a \Delta_d \big) -{} \nonumber \\
\ & \quad \sin(2 \phi) \big(4 \gamma_a \gamma_d \Delta_a
- 2 \gamma_a^2 \Delta_d + 8 \Delta_a^2 \Delta_d \big) \big]
\,,
\end{align}
\end{subequations}
which landscape of correlations, one can easily check, bears a close
resemblance to the results shown in
Fig.~\ref{fig:Tue14Apr124339CEST2020}, with also identical features
such as $\mathcal{I}_1$ being identically zero.  This confirms that
the results obtained with admixing pure states transpose directly into
a dynamical setting with steady states of open quantum systems. This
particular case could be further investigated, which will no doubt be
the case following its experimental implementation. For now, we turn
in the remainder of the text to similar dynamical systems which
describe important and a significant fraction of currently studied
quantum optical sources.

\section{Resonance fluorescence statistics}

The mixing of a coherent and squeezed state occurs at a fundamental
level in the problem of resonance fluorescence. It is, in fact, in
this particular case that we have ourselves first observed the results
which we now generalize~\cite{lopezcarreno18b}.  In the low-driving
limit, the so-called Heitler regime, of a two-level system with
annihilation operator~$\sigma$, the output of the system is, to first
order, $\alpha\equiv\langle\sigma\rangle$ which, as a complex number
(with a modulus and phase), can be assimilated to a coherent
state. Indeed, this contributes to what is referred to as the
``coherent (or elastic) scattering'' fraction of the emission. The
incoherent emission $\epsilon \equiv \sigma -\langle\sigma\rangle$ 
completes the total emission according to
Eq.~(\ref{eq:Wed8Apr181920CEST2020}):
\begin{equation}
  \label{eq:Thu27Feb160136GMT2020}
  \sigma=\alpha+\epsilon .
\end{equation}
This effectively describes the
original emission~$\sigma$ as a self-homodyning whereby a pure quantum
signal $\epsilon$ is admixed internally to a coherent fraction
$\alpha$.  In the following sections, we shall remain at this level of
the description. In the present case, however, since this is the
simplest configuration, we will take control of the homodyning by
bringing in ourselves an additional coherent field~$\beta$ to tamper
with the fraction~$\alpha$ naturally present in the original emission.
This is easily achieved in principle with a laser, and since we are
considering the emission from coherently driven systems in the first
place, in line with how homodyning is typically performed for reasons
of practicality and stability, the same laser that drives the system
can have a fraction of its beam diverted upstream of the emitter to
provide a phase- and amplitude-controlled beam to be admixed with the
emitter's output. We will also use this self-homodyning picture to
characterize in more details the fluctuations $\epsilon$, that we will
show correspond to a squeezed thermal state, thereby indeed making the
analysis of this section another dynamical version of admixing
squeezed and coherent states as described in
Section~\ref{sec:Tue14Apr124542CEST2020}, although this time not a
contrived dynamics like in Section~\ref{sec:Fri17Apr144500CEST2020},
but one at the core of light-matter interactions, namely, with
Hamiltonian ($\hbar=1$)
\begin{equation}
\label{eq:2LShamiltonian}
H_\mathrm{rf} = (\omega_\sigma-\omega_\mathrm{L}) \ud{\sigma} \sigma
+ \Omega_{\sigma} \left(\ud{\sigma} + \sigma \right)\,,
\end{equation}
with~$\sigma$ the annihilation operator for a two-level system (2LS)
and~$\Omega_\sigma$ the strength of (classical) driving. The formalism
to include dissipation and obtain correlators are as in the previous
Section~\ref{sec:Fri17Apr144500CEST2020}, with obvious notations,
$\Delta_\sigma\equiv\omega_\sigma-\omega_\mathrm{L}$
and~$\gamma_\sigma$ the decay rate of the 2LS.  By applying
Eq.~(\ref{eq:2LScorrBS}) with $a=\beta$ and~$d=\sigma$, since
$\corr{\sigma}{p}{q} =0$ for $p,q >1$, we find
\begin{multline}
  \corr{s}{n}{m} = (-i \mathrm{R}\beta^*)^{n}
  (i \mathrm{R}\beta)^{m}\\
  {}-i \mathrm{R} \mathrm{T} \ (-i \mathrm{R}\beta^*)^{n-1} (i
  \mathrm{R}\beta)^{m-1} \left(m \beta \coh{\sigma} - n \beta^*
    \coh{\sigma}^*  \right)\\
  {}+n m \ (-i \mathrm{R}\beta^*)^{n-1} (i \mathrm{R}\beta)^{m-1}
  \mathrm{T}^2 \av{n_{\sigma}}\,.
\end{multline}
In this case, the coherent fraction and total population of the output
field are found to be
\begin{subequations}
  \label{eq:FriFeb23210546CET2018}
  \begin{align}
    \label{eq:FriFeb23210546CET2018a}
    \coh{s} &= i \mathrm{R} \beta + \mathrm{T} \coh{\sigma}\,,\\
    \label{eq:FriFeb23210546CET2018b}
    \pop{s} &= \mathrm{R}|\beta|^2 + \mathrm{T}^2 \av{n_\sigma} +2 \mathrm{R} \mathrm{T}
    \mathrm{Re} [ i  \beta^* \coh{\sigma}]\,.
  \end{align}
\end{subequations}
Clearly, one can choose the coherent field to compensate exactly the
coherent component of the 2LS $\alpha=\mean{\sigma}$ in a
beam-splitter, namely,
with~$\beta= i \frac{\mathrm{T}}{\mathrm{R}}\mean{\sigma}$ so that
only the transmitted fluctuations~$s=T\epsilon$ are retained. This is a
way to extract the ``pure'' quantum emission of the two-level
system. In such a case, the correlators simplify even further, to:
\begin{multline}
\label{eq:2LSincohCorrs}
\corr{\epsilon}{n}{m}=\corr{s}{n}{m}/ \mathrm{T}^{n+m} ={}\\ (-1)^{m+n}
\alpha^{m-1} \alpha^{* (n-1)} [ nm \ \av{n_\sigma} -
  \left(n+m-1 \right) \big |\alpha \big |^2]\,.
\end{multline}
%


With the general result Eq.~(\ref{eq:2LSincohCorrs}), one then easily
finds the general $N$th-order correlation function for the
fluctuations, that are now directly available on the
  output port of the beam splitter:
\begin{equation}
  g^{(N)}_\epsilon =\frac{|\alpha|^{2(N-1)}
	\big(N^2\,\mean{n_\sigma}+(1-2N)|\alpha|^2\big)}{(\mean{n_\sigma}- |\alpha|^2)^N}\,,
\end{equation}
where $\langle n_\sigma\rangle$ and~$\alpha=\langle\sigma\rangle$ are
found from the steady-state solution for the 2LS
\begin{equation}
\label{eq:2LSrhoSS}
\rho =
\begin{pmatrix}
1 - \mean{n_\sigma} & \alpha^\ast \\ \alpha & \mean{n_\sigma}
\end{pmatrix},
\end{equation}
as
\begin{equation}
\label{eq:2LS_observables}
\mean{n_\sigma} = \frac{4 \Omega_{\sigma}^2}{\gamma_\sigma^2 + 4
	\Delta_{\sigma}^2 + 8 \Omega_{\sigma}^2}\quad\text{and}\quad
\alpha = \frac{2 \Omega_{\sigma} (2 \Delta_{\sigma} + i
	\gamma_\sigma)}{\gamma_\sigma^2 + 4 \Delta_{\sigma}^2 + 8  
	\Omega_{\sigma}^2}\,.
\end{equation}
In terms of the physical parameters, reads as
%
%
\begin{equation}
\label{eq:FriFeb23174312CET2018}
\g{N}_\epsilon = \frac{ \left(N-1\right)^2 \left(\gamma_\sigma^2 +4
	\Delta_{\sigma}^2 
	\right) + 8 N^2 \Omega_{\sigma}^2}{8^{N}\,
	\Omega_{\sigma}^{2N} \left(\gamma_\sigma^2 +4
	\Delta_{\sigma}^2 \right)^{1-N}}\,.
\end{equation}


\begin{figure}[t!]
  \centering
  \includegraphics[width=\linewidth]{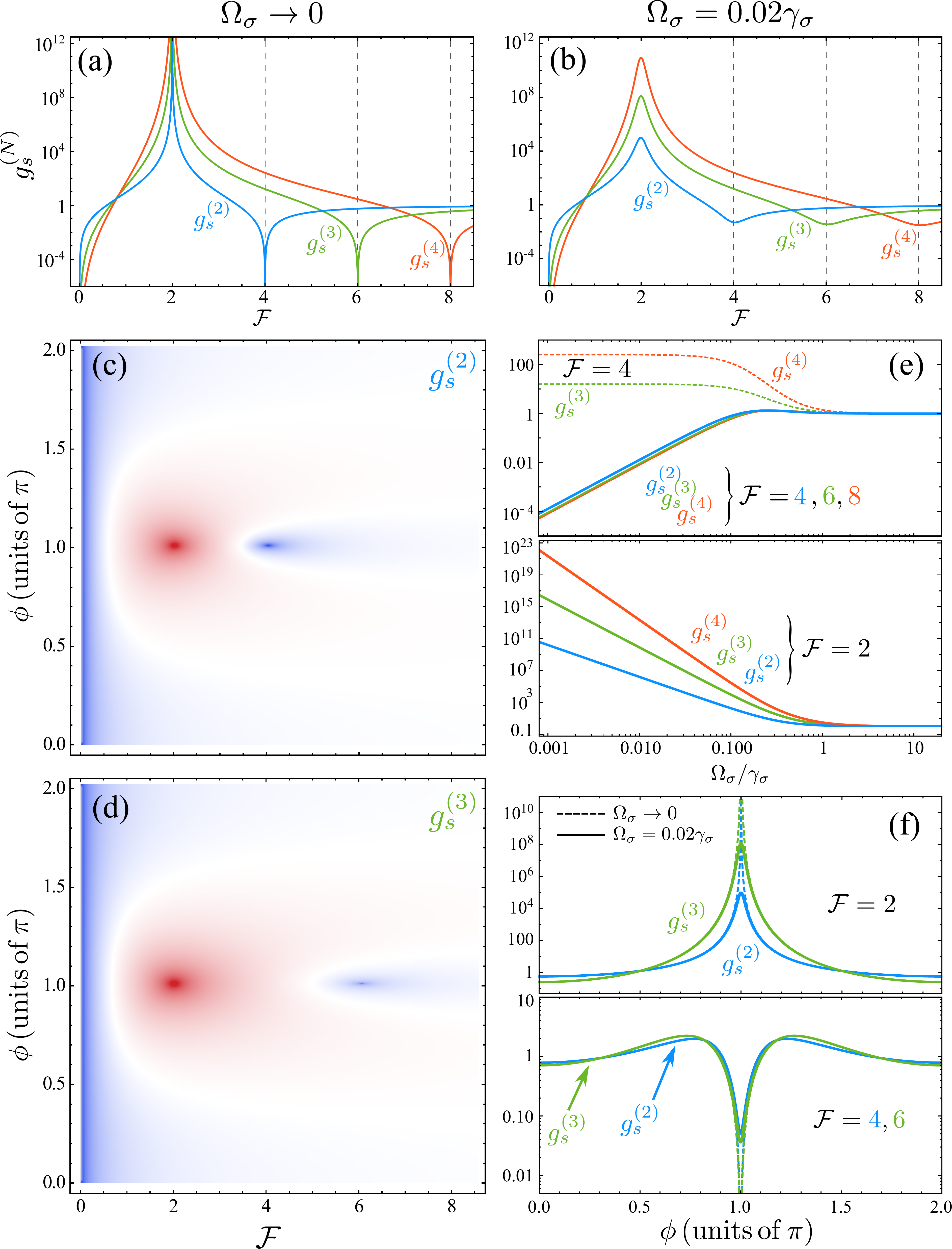}
  \caption{(Color online) $N$-photon statistics of a two-level system
    driven coherently at rate~$\Omega_\sigma$ an homodyned
    (laser corrected) with a field of phase~$\phi$ and amplitude a
    fraction~$\mathcal{F}$ of $\Omega_\sigma$, in the limit of
    vanishing driving [(a),
    Eq.~(\ref{eq:Fri28Feb055659GMT2020})] and for small but finite
    driving [(b), Eq.~(\ref{eq:Fri28Feb062937GMT2020})]. (c), (d) Maps of
    two- and three-photon statistics as a function of the homodyning field. 
    (e)~Shows how correlations are weakened with increasing driving (solid
    lines) and how antibunching is realized at a given photon-number
    at a time, as opposed to bunching. Limits are given in
    Table~\ref{tb:Mon20Apr111143CEST2020}.  (f) shows the dependence
    on the phase of the homodyning field, for both vanishing and
    finite drivings for the case of bunching ($\mathcal{F}=2$) and
    two and three photon antibunching ($\mathcal{F}=4, 6$). The only parameter has
    been taken as~$\gamma_\sigma=1$.}
  \label{fig:Fri28Feb062459GMT2020}
\end{figure}

\begin{table}[t]
\begin{tabular}{lcccc}
\cline{1-5}
        & $\mathcal{F}=2$ & 4     & 6 & 8 \\ \cline{1-5}
$\g{2}_s$ & $1/(64\Omega^4)$           & $128\Omega^2$ &   &   \\
$\g{3}_s$ & $1/(128\Omega^6)$          & 16  & $(729/8)\Omega^2$ &   \\
$\g{4}_s$ & $9/(4096\Omega^8)$             & 256    &   & $(524\,288/6561)\Omega^2$ \\ \cline{1-5}
\end{tabular}
\caption{$N$-photon statistics with homodyning for the cases shown in
  Fig.~\ref{fig:Fri28Feb062459GMT2020}(e), to leading-order
  in~$\Omega$, in units of~$\gamma_\sigma=1$.}
\label{tb:Mon20Apr111143CEST2020}
\end{table}

Interestingly, suppressing the coherent contribution of the emission is not
the only possibility. One can also tune the coherent contribution by
choosing $\beta' = e^{i \phi} |\beta'|$, where the amplitude is
parametrized as $|\beta'|= \frac{\mathrm{R}}{\mathrm{T}} |\beta|$.
The amplitude $|\beta'|$ can be expressed more suitably in terms of
the driving intensity of the laser:
$|\beta'| = \frac{\Omega_{\sigma}}{\gamma_{\sigma}} \mathcal{F}$.
Thus, we are broadening the range of possible output
configurations~\cite{lopezcarreno18b}, with $N$-particle correlators
for the resonance-fluorescence plus an external laser having the
following form, from which the population and two-photon statistics
follow as special cases~($N=1, 2$, respectively):
\begin{multline}
  \label{eq:Fri28Feb055659GMT2020}
  \g{N}_{s} = \frac{\mathrm{T}^{2N}}{\av{n_s}^N} \frac{\mathcal{F}^{2(N-1)}\Omega_\sigma^{2 N}}{\gamma_\sigma^{2 N} \left(\gamma_\sigma^2 + 4 \Delta_\sigma^2 + 8 \Omega_\sigma^2 \right)}\Big[4 N^2 \gamma_\sigma^2+{}\\
\mathcal{F}^2 \left(\gamma_\sigma^2 + 4 \Delta_\sigma^2 + 8 \Omega_\sigma^2 \right)
+ 4 N \mathcal{F} \gamma_\sigma \left(\gamma_\sigma \cos \phi - 2 \Delta_\sigma \sin \phi\right)
 \Big]
\end{multline}
where $\langle n_s \rangle \equiv \pop{s}$, cf. Eq.~\eqref{eq:FriFeb23210546CET2018b}.
These expressions are correct for any driving strength. However, in
the Heitler regime, the results become ideal in the sense that
antibunching becomes exactly zero and superbunching becomes
infinite. In this case, Eq.~(\ref{eq:Fri28Feb055659GMT2020}) takes the
simpler form:
\begin{equation}
\label{eq:Fri28Feb062937GMT2020}
\g{N}_s = 
\frac{|\beta'|^{2(N-1)} \left( |\beta'|^2 + N^2 \av{n_\sigma}
    +2 N \mathrm{Im} \lbrace \alpha |\beta'| e^{-i \phi} \rbrace	\right)}
{\left(|\beta'|^2 +  \av{n_\sigma}+2 
    \mathrm{Im} \lbrace \alpha |\beta'| e^{-i \phi} \rbrace	\right)^N}\,.
\end{equation}

Both Eqs.~(\ref{eq:Fri28Feb055659GMT2020})
and~(\ref{eq:Fri28Feb062937GMT2020}) are shown in
Fig.~\ref{fig:Fri28Feb062459GMT2020}, where one can see how homodyning
produces sharp resonances for various photon-numbers in antibunching
and a common superbunching when the population vanishes. The case with
no homodyning~($\mathcal{F}=0$) produces the best antibunching, both
in magnitude (closest to zero), and in quality, namely, all~$g^{(n)}$
go to zero simultaneously. This is antibuching in the sense tacitly
understood for a single-photon source, separating photons the ones
from the others, so it is apt to call it ``conventional
antibunching'', to differentiate it from the other type of two-photon
antibunching, realized with an homodyning of $\mathcal{F} \neq 0$
and~$\phi=0$. In the latter case, only~$g^{(2)}$ is small and
higher-order correlators are typically~$\gg1$. One can also, with
different choice of~$\mathcal{F}$, realize $n$-photon
antibunching~($n=3$ and~$4$ are shown in
Fig.~\ref{fig:Fri28Feb062459GMT2020} in green and orange,
respectively), but then again for a given photon number only. This
follows from admixing a squeezed and coherent states as discussed in
Section~\ref{sec:Tue14Apr124542CEST2020} and indeed the phenomenology
of the homodyning of resonance fluorescence as shown in
Fig.~\ref{fig:Fri28Feb062459GMT2020}(b) is similar to that shown in
Fig.~\ref{fig:Tue14Apr124339CEST2020}(a) (with the coherent and
squeezing fractions being tuned, respectively). In particular, and for
the same reason, one also finds in resonance fluorescence the
tunability of~$g^{(n)}_s$ between zero to infinity, at small enough
driving, simply by changing the coherent fraction. We call these
features ``unconventional.''  Besides, this terminology fits (and has
been chosen accordingly~\cite{zubizarretacasalengua20a}) with the
literature~\cite{ferretti10a,flayac13a,majumdar13a, gerace14a,
  flayac15a, zhou15a, tang15a, cheng17a, flayac17b, yu17a, wang17a,
  snijders18a,vaneph18a, shen18a, trivedi19a} which calls
``unconventional'' the supposed ``blockade'' that takes place when
interfering fields~\cite{lemonde14a}.  They do not arise from a
blockade from states of the system, as in the conventional
scenario~\cite{birnbaum05a, verger06a, faraon08a, faraon10a,
  hoffman11a, rabl11a, lang11a, muller15a, radulaski17a, deng17a,
  sarma17a, ghosh19a, delteil19a}, but from an interference. Note that
unconventional superbunching, unlike unconventional antibunching, is
simultaneously bunched at all photon-orders. Actually, this also
follows from the cancellation of a correlator, namely, the first-order
one (population) with the result of having all higher-order
correlators diverging.  For non-vanishing driving, the features are
qualitatively similar but strongly damped, as shown on the right
column of Fig.~\ref{fig:Fri28Feb062459GMT2020}. Panel~(e) gives a
quantitative account of how correlations weaken with driving. The
trend of antibunching/superbunching with~$\Omega$ is easily obtained
from a series expansion of Eq.~(\ref{eq:Fri28Feb055659GMT2020}) and is
given in Table~\ref{tb:Mon20Apr111143CEST2020} to leading order in
$\Omega$ (to next order, for instance,
$\g{2}_s(\mathcal{F}=2)=1/(64\Omega^4)+1/(4\Omega^2)$ and
$\g{3}_s(\mathcal{F}=4)=16-768\Omega^2$). This also shows for the case
of two-photon antibunching how only~$g^{(2)}_s$ improves while
higher-orders (dashed) saturate.  As shown in panel~(c) and~(d), in
this case, the phase of the homodyning is the same (for the anharmonic
oscillator, for instance, it
is~$N$-dependent~\cite{zubizarretacasalengua20a})). The sensitivity of
the effect to the phase and driving strength is shown quantitatively
in panel~(f).

\begin{figure}[t]
  \centering 
  \includegraphics[width=.75\linewidth]{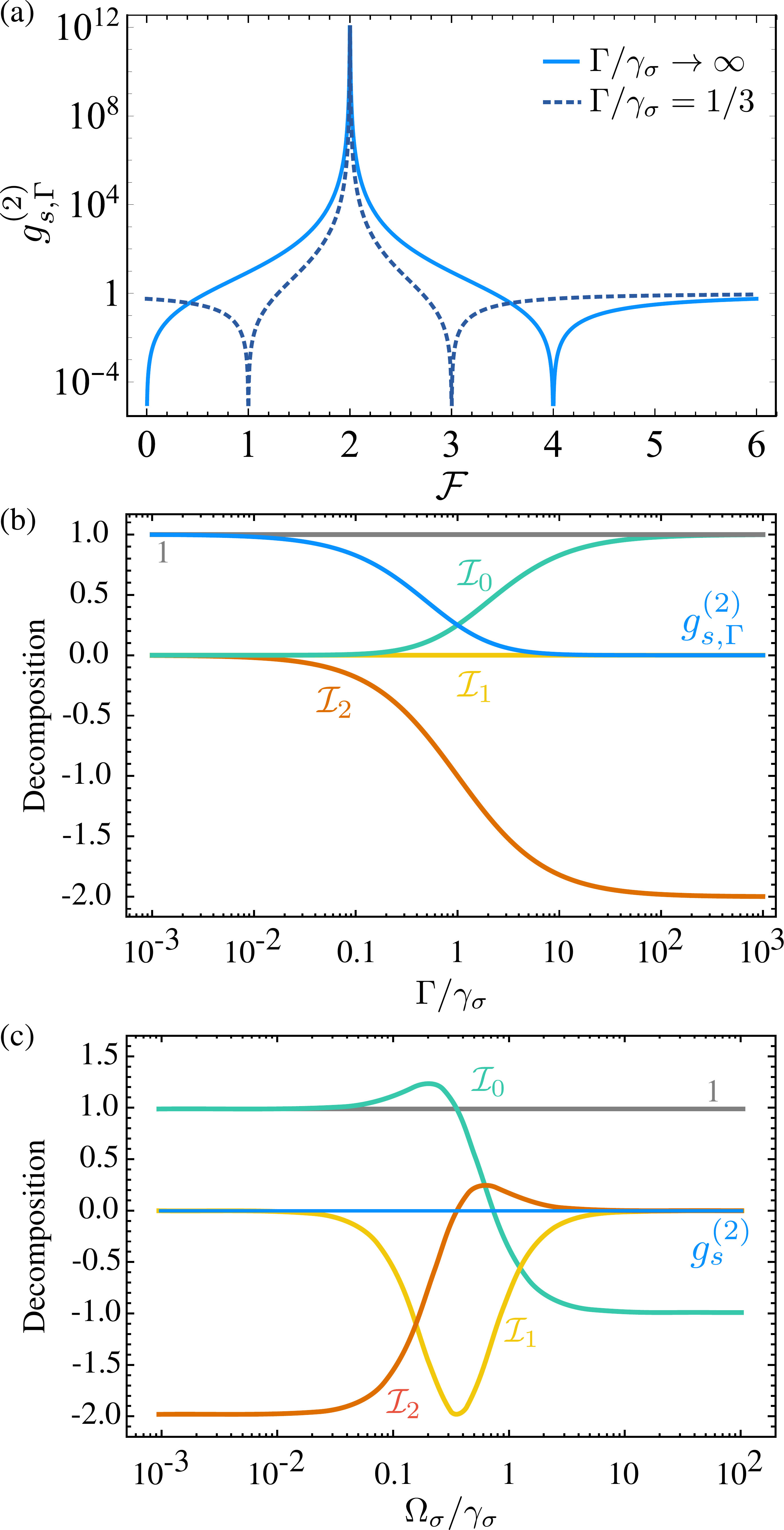}
  \caption{(Color online) (a) Two-photon statistics of a two-level
    system in the Heitler limit with (dashed-dark) and without
    (solid-light blue) filtering as a function of the homodyning
    amplitude~$\mathcal{F}$. Conventional antibunching gets spoiled by
    filtering but perfect unconventional antibunching is obtained with
    homodyning. (b) Evolution of the $\mathcal{I}$ parameters as a
    function of the filter's width~$\Gamma$ showing how filtering
    disrupts the perfect cancellation
    of~$g^{(2)}_{s,\Gamma}$. Homodyning restores the
    condition~$\mathcal{I}_2=-2\mathcal{I}_0$. (c) Evolution of the
    $\mathcal{I}$ parameters as a function of pumping
    strength~$\Omega_\sigma$. Without filtering, the cancellation is
    still perfect at all pumpings but involves~$\mathcal{I}_1$ and
    thus cannot be fully restored with homodyning.}
  \label{fig:Wed15Apr104949CEST2020}
\end{figure}

One could naturally question why going to the extent of homodyning
since the case~$\mathcal{F}=0$ may appear superior in several
respects, in particular, at non-vanishing pumping. Beyond the fact that
conventional and unconventional statistics stem from two different
types of light, which may have some interest per se, one obvious
application is to restore antibunching which has been lost as a result
of filtering. It is now amply demonstrated that frequency filtering
spoils antibunching of resonance
fluorescence~\cite{arXiv_phillips20a,arXiv_hanschke20a}. The reason
for this in the Heitler regime is shown in
Fig.~\ref{fig:Wed15Apr104949CEST2020}, that displays in Panel~(b) the
$\mathcal{I}$ coefficients in the Heitler regime in presence of
filtering~$\Gamma$ (this can also be obtained from the wavefunction
approximation method~\cite{visser95a} as detailed in the Supplementary
Material for this and other systems studied in this text):
\begin{subequations}
  \label{eq:Thu16Apr164453CEST2020}
  \begin{align}
    \mathcal{I}_0&={\Gamma^2\over(\Gamma+\gamma_\sigma)^2}\,,\\
    \mathcal{I}_1&=0\,,\\
    \mathcal{I}_2&=-{2\Gamma\over\Gamma+\gamma_\sigma}
  \end{align}
\end{subequations}
which shows how the filter perturbs the balance of the~$\mathcal{I}$
coefficients, with~$\mathcal{I}_2$ going to zero faster
than~$\mathcal{I}_0$, resulting in the sum taking off as
\begin{equation}
  g_{s,\Gamma}^{(2)}=\left({\gamma_\sigma\over\Gamma+\gamma_\sigma}\right)^2\,.
\end{equation}
While the conventional antibunching is irretrievably lost,
unconventional two-photon antibunching now gets two conditions to be
restored,
at~$\mathcal{F}=2(1\pm\sqrt{\Gamma/(\Gamma+\gamma_\sigma})$~\cite{lopezcarreno18b}.
The case~$\Gamma/\gamma_\sigma=1/3$ is shown in
Fig.~\ref{fig:Wed15Apr104949CEST2020}(a). At this sub-natural
linewidth filtering, conventional antibunching has reduced from
exactly zero in the Heitler limit to $9/16\approx0.56$, but perfect
antibunching can be restored with~$\mathcal{F}=2\pm1$. Both resonances
are unconventional in the sense that successive higher-order
correlators have non-matching resonances~\cite{lopezcarreno19a}, but
in both cases, perfect antibunching can be restored, regardless of the
filter's width, by bringing back~$\mathcal{I}_0$ and~$\mathcal{I}_2$
to~2 and~1, respectively, with the appropriate correction of the
coherent fraction.

Restoring such a perfect cancellation of the~$\mathcal{I}$
coefficients works exactly in the Heitler limit and at vanishing
driving. At non-vanishing driving, the~$\mathcal{I}$
parameters~(\ref{eq:decompositiontermswhole}) read
\begin{subequations}
	\label{eq:FriFeb23150914CET2018}
	\begin{align}
	\mathcal{I}_0 &= \frac{|\alpha|^2 \left( 6 \mean{n_\sigma}- 4
		|\alpha|^2 \right)}{n_\sigma^2} - 1\,,\\
	\mathcal{I}_1 &= -8 \frac{ |\alpha|^2 \left(
		\mean{n_\sigma}-   |\alpha|^2
		\right)}{n_\sigma^2}\,, \\ 
	\label{eq:2LSg2decompositiontermsI2}
	\mathcal{I}_2 &= 2 \frac{ |\alpha|^2 \left(
		\mean{n_\sigma}-  2 |\alpha|^2 \right)}{n_\sigma^2}\,.
	\end{align}
\end{subequations}
In this case, as seen in Fig.~\ref{fig:Wed15Apr104949CEST2020}(c),
perfect antibunching does not follow simply from
$\mathcal{I}_2=-2\mathcal{I}_0$ but also involves $\mathcal{I}_1$
which is related to anomalous quadrature moments, that a
displaced-thermal squeezed state does not possess, therefore causing a
breakdown of the Gaussian-states approximation. Since it also depends
on~$\alpha$, it becomes impossible to realize another perfect
cancellation, or restore it if spoiled, since the sum depends in
multiple ways on the one free parameter: the homodyning signal. As a
result, the minimum antibunching is now finite, as shown in
Fig.~\ref{fig:Fri28Feb062459GMT2020}(b). At still
  higher pumping, $\Omega_\sigma\gg\gamma_\sigma$, in the so-called
  Mollow regime, antibunching comes exclusively from~$\mathcal{I}_0$,
  which is the antibunching of an incoherently-pumped two-level system
  with non-Gaussian fluctuations. In this limit,
  $\av{d^\dagger d}=\av{n_s}$, which absorbs the term~1 in
  Eqs.~(\ref{eq:Mon20Apr190800CEST2020}). There is no way to correct
  for the coherent component in this case since there is no coherence
  involved.

We conclude this section by using the self-homodyning approach to
further characterize the nature of the admixing, which we now show,
corresponds to lowest order in the driving to a coherent state for
$\langle\sigma\rangle$, which is a tautology, and to a squeezed
thermal state for the fluctuations~$\mathcal{\epsilon}$. Two
quantities allow to identify squeezing in a quantum field~$\epsilon$,
namely, the mean $\av{X_{\epsilon,\chi}}$ of the quadratures
$X_{\epsilon,\chi}\equiv\frac{1}{2} (e^{i\chi} \ud{\epsilon} +
\mathrm{c.c.})$ for the operator~$\epsilon$ with phase~$\chi$, as well
as the variance (dispersion)
$\av{\Delta X_{\epsilon,\chi}^2} = \av{X_{\epsilon,\chi}^2 -
  \av{X_{\epsilon,\chi}}^2}$. Note that this could also be done
directly for the full field~$\sigma$ since this only adds the
comparatively trivial contribution of the coherent state
$\av{\sigma}=\alpha$.  The maximum and minimum of the normal-ordered
quadrature variance for a single-mode can be computed independently of
the specific nature of the field:
\begin{multline}
\label{eq:sQuadDisp}
\av{:\Delta X_{\epsilon}^2:}_{\mathrm{max}/\mathrm{min}} =
\av{\Delta X_{\epsilon}^2}_{\mathrm{max}/\mathrm{min}} - \frac{1}{4} \\
{}=\frac{1}{2} \left[\pm |\av{\epsilon^2}-\av{\epsilon}^2| + \pop{\epsilon} - |\av{\epsilon}|^2\right]\,,
\end{multline}
where the sign corresponds to the maximum and minimum, respectively.
While the variance is positive, its normal-ordered counterpart does
not have to be.  The deviation of the variance from the vacuum value
(which is $\frac{1}{4}$) to negative values reveals some degree of
quadrature squeezing. Likewise, the angle of squeezing is generically
given by
$\theta = \mathrm{arg} \left[\av{\epsilon^2} -
  \av{\epsilon}^2\right]$. One gets, by substituting the correlators
\eqref{eq:2LSincohCorrs} in \eqref{eq:sQuadDisp},
\begin{subequations}
\begin{align}
  &\av{:\Delta X_{\epsilon}^2:}_\mathrm{min} = -\frac{2 \Omega_{\sigma}^2
    \left(\gamma_{\sigma}^2 + 4 \Delta_\sigma^2 - 8 \Omega_\sigma^2 \right)}{\left(\gamma_{\sigma}^2 + 4 \Delta_\sigma^2 + 8 \Omega_\sigma^2\right)^2}, \\
&\av{:\Delta X_{\epsilon}^2:}_\mathrm{max} = \frac{2\Omega_{\sigma}^2}{\gamma_{\sigma}^2 + 4 \Delta_\sigma^2 + 8 \Omega_\sigma^2}, 
\end{align}
\end{subequations}
with the angle of squeezing
$\theta = \mathrm{arg} [(\gamma_{\sigma}-2 i \Delta_\sigma)^2]$.
%
Analyzing the sign of these quantities would therefore allow us to
infer squeezing. This is made particularly clear in the low-driving
regime ($\Omega_{\sigma} \rightarrow 0$), where the previous
expressions at the lowest order in $\Omega_\sigma$ simplify to
\begin{equation}
\label{eq:2LSlowpumpDisp}
\av{:\Delta X^2_\epsilon:}_{\mathrm{max}/\mathrm{min}} \approx \pm 
\frac{2  \Omega_{\sigma}^2}{ \gamma_{\sigma}^2 + 4 \Delta^2_\sigma}\,.	
\end{equation}
In this limit, the two extrema of the normal-ordered variance are
simply symmetrical around zero, so $\av{:\Delta X^2_\epsilon:}$ is
always negative and fluctuations thus always bear some indication of
squeezing.  We can furthermore recognize these expressions as a limit
of low squeezing from a displaced squeezed thermal state. When
$r \rightarrow 0$, such states have the variance
\begin{equation}
\label{eq:DSTlowsqueezedDispersion}
\av{:\Delta X^2:}_{\mathrm{max}/\mathrm{min}}^{\mathrm{DST}} \approx
\frac{1}{4}
\left[\left(1 \pm 2 r\right) \left(1 + 2 \av{n_\mathrm{th}}\right)-1\right]
\approx \pm \, \frac{r}{2},
\end{equation}
where the superscript DST means that the observable corresponds to an
exact \textit{displaced squeezed thermal} state.  We have approximated
$1 + 2 \av{n_\mathrm{th}}$ to $1$ since the thermal population grows
like $\Omega_{\sigma}^4$ (which comes from the first-order of the
incoherent population). Comparing Eq.~\eqref{eq:2LSlowpumpDisp} with
Eq.~\eqref{eq:DSTlowsqueezedDispersion} shows that the incoherent
population in the Heitler regime behaves like a squeezed thermal state
with effective squeezing parameter~$r_\mathrm{eff}$
and effective thermal population $\mathrm{p}_\mathrm{th}$
\begin{equation}
\label{eq:2LSreff}
r_\mathrm{eff} = 
\frac{4 \Omega_{\sigma}^2}{ \gamma_{\sigma}^2 + 4 \Delta_{\sigma}^2}\quad{\text{and}}\quad
\mathrm{p}_\mathrm{th} \approx \frac{16	\Omega_\sigma^4}{\left( \gamma_{\sigma}^2 + 4 \Delta_{\sigma}^2\right)^2 }\,.
\end{equation}
%


%
From these two parameters, an effective $\g{2}$, namely
$\g{2}_\mathrm{eff}$, can be obtained for the fluctuations.  Supposing
that, in the low-excitation regime, the state of fluctuations is that
of a squeezed thermal state, then $\g{2}_\epsilon$ should have the
same form.  Fixing $|\alpha| = 0$ in Eq.~\eqref{eq:DST_g2} and taking
the limit $r^2 \rightarrow 0$ and
$\mathrm{p}_\mathrm{th} \rightarrow 0$ (both go to 0 with the same
power dependence), we get
\begin{equation}
\g{2}_\mathrm{eff} \approx \frac{r_\mathrm{eff}^2}{\left(r_\mathrm{eff}^2 +\mathrm{p}_\mathrm{th}\right)^2},
\end{equation}
which, after substituting Eqs.~\eqref{eq:2LSreff}, reads as
\begin{equation}
\label{eq:2LSg2eff}
\g{2}_\mathrm{eff} \approx \frac{\left(\gamma_{\sigma}^2 + 4 \Delta_{\sigma}^2\right)^2}
{64\Omega_{\sigma}^4}\,.
\end{equation}
This simple expression is a good approximation to the exact result
Eq.~\eqref{eq:FriFeb23174312CET2018} that gives the statistics of the
fluctuations. As was the case when admixing a pure coherent-state to a
coherent state in Section~\ref{sec:Tue14Apr124542CEST2020},
antibunching of the total signal~$g_s^{(2)}<1$ is obtained from a
coherent state and a \emph{superbunched}~$\g{2}_\epsilon\gg 1$ (even
diverging~$\g{2}_\mathrm{eff}\rightarrow\infty$) squeezed state.

\section{Jaynes--Cummings statistics}

The same analysis as above can be transported to a wealth of other
systems. For instance, in the case of an anharmonic oscillator, new
resonances appear and with richer phase conditions than those of a
two-level system~\cite{zubizarretacasalengua20a}. We go directly to a
fundamental and natural system where to apply the concepts above,
namely, the Jaynes--Cummings model~\cite{jaynes63a,shore93a}, since
this adds to the two-level system a quantized optical mode (a cavity)
with a total emission that, therefore, consists intrinsically of the
mixing of a quantum (two-level system) and a coherent (cavity)
signals. The photon statistics of this system has been for decades
observed in one way or another to exhibit resonances which are a
simple and direct manifestation of self-homodyning in the wake of the
previous sections, but which have often been merely taken as the brute
result of numerical simulations. We now provide what we believe is the
appropriate physical picture to unify, classify, and understand such
results. Given the role of the amplitude and phase of the fields in
phenomena that are ultimately interferences, we include in the
Hamiltonian two driving terms, one for the emitter, $\Omega_\sigma$,
the other for the cavity~$\Omega_a$. Their relative phase~$\phi$ and
the ratio of their amplitude $\chi\equiv\Omega_\sigma/\Omega_a$ will
play a role in tuning the statistics.  The Hamiltonian therefore
reads as
\begin{multline}
  \label{eq:Thu31May103357CEST2018}
  H_\mathrm{jc} = \Delta_\sigma \ud{\sigma} \sigma + \Delta_a \ud{a} a
  +
  g \left(\ud{a} \sigma + \ud{\sigma} a \right) +{}\\
  {}+\Omega_a \left(e^{i \phi} \ud{a} + e^{- i \phi} a\right) +
  \Omega_\sigma \left(\ud{\sigma} + \sigma\right).
\end{multline}
Solving for the steady state in the low-driving
regime,~i.e., when $\Omega_{a, \sigma}\ll \gamma_a\,,\gamma_\sigma$, yields for
the populations:
\begin{widetext}
  \begin{equation}
    \label{eq:Tue29May182623CEST2018}
    \av{n_{\substack{a\\\sigma}}} =  4 \, \frac{4 g^2 \Omega_{\substack{a\\\sigma}}^2
      + \Gamma_{\substack{\sigma\\a}}^2 \Omega_{\substack{a\\\sigma}}^2  - 4 g \Omega_a \Omega_\sigma
      \left(\pm 2 \Delta_{\substack{a\\\sigma}}\cos \phi + \gamma_{\substack{\sigma\\a}} \sin \phi \right)}
    {16 g^4 + 8 g^2 \left(\gamma_a \gamma_\sigma - 4 \Delta_a \Delta_\sigma\right) +
      \Gamma_a^2 \Gamma_\sigma^2	} \, ,
  \end{equation}
  with matching upper/lower indices (including~$\pm$) and with
  $\Gamma^2_i = \gamma_i^2 + 4 \Delta_i^2$ (for
  $i = a, \sigma$).
  Similarly, the two-photon coherence function from the cavity
  can be found as:
  \begin{equation}
    \label{eq:Tue29May184632CEST2018}
    \begin{split}
      \g{2}_a = & \Big\lbrace \Big[ 16 g^4 + 8 g^2 \left(\gamma_a \gamma_\sigma -4 \Delta_a
	\Delta_\sigma \right) + \Gamma_a^2 \Gamma^2_\sigma \Big]
      \Big[16 g^4 \left(1 + \chi^4 \right) + 8 g^2 
      \big(2 \chi^2 \Gamma_{11}^2  + 4 \Delta_\sigma \tilde{\Delta}_{11}- \gamma_\sigma
      \tilde{\gamma}_{11}\big) + \Gamma^2_{\sigma} \Gamma_{11}^2 -\\
      & 16 g \chi \big( \Delta_\sigma \Gamma_{11}^2 + 4 g^2 \tilde{\Delta}_{11}
      [1+\chi^2] \big) \cos \phi + 8 g^2 \chi^2
      \big(4 g^2 - \gamma_\sigma \tilde{\gamma}_{11} + 4 \Delta_\sigma 
      \tilde{\Delta}_{11}\big) \cos 2 \phi \, - \\
      & 8 g \chi \big(\gamma_\sigma \Gamma_{11}^2 + 4 g^2 \tilde{\gamma}_{11} [\chi^2-1] \big)
      \sin \phi + 16 g^2 \chi^2  \big(\gamma_a \Delta_\sigma + \gamma_\sigma \tilde{\Delta}_{12}\big)
      \sin 2 \phi  \Big] \Big\rbrace \Big/ \\
      & \Big\lbrace \Big[16 g^4 + 8 g^2 \Big(\gamma_a \tilde{\gamma}_{11}- 4
      \Delta_a \tilde{\Delta}_{11}\Big) + \Gamma_a^2 \Gamma_{11}^2\Big] \Big[4 g^2 \chi^2 + \Gamma_{\sigma}^2 - 4 g \chi \big(2 \Delta_\sigma \cos \phi +
      \gamma_\sigma \sin \phi\big)\Big]^2 \Big\rbrace ,
    \end{split}
  \end{equation}
  where $\tilde{\Delta}_{ij} \equiv i \Delta_a + j \Delta_\sigma$,
  $\tilde{\gamma}_{ij} = i \gamma_a + j \gamma_\sigma$ and
  $\Gamma_{ij}^2 \equiv \tilde{\gamma}_{ij}^2 + 4
  \tilde{\Delta}^2_{ij}$. The range of $\chi$ extends from 0 to
  $\infty$ so that it is convenient to use the derived
  quantity~$\tilde{\chi}=\frac{2}{\pi} \atan(\chi)$ which varies
  between 0 and 1. The expressions above are cumbersome but they are
  covering a considerable amount of phenomenology, each variation of
  which could give rise to an independent numerical study of its
  own. Let us start with the much simpler-looking particular case of
  one pumping only, namely, with cavity-pumping only, which is the
  case most discussed in the literature. Then
  Eq.~(\ref{eq:Tue29May184632CEST2018}) reduces to:
  \begin{equation}
    \label{eq:JCg2}
    \g{2}_a = [16 g^4 + 8 g^2 \left(\gamma_{\sigma } \gamma_a - 4 \Delta_a \Delta_{\sigma} \right) + \Gamma_a^2 \Gamma_\sigma^2] [ 16 g^4 -8 g^2 (\gamma_\sigma \tilde\gamma_{11} - 4 \Delta_\sigma \tilde\Delta_{11} ) + \Gamma_{\sigma}^2\tilde\Gamma_{11}^2 ]\big/\Gamma_\sigma ^4 [ 16 g^4 +8 g^2 (\gamma_a \tilde\gamma_{11} - 4 \Delta_a \tilde\Delta_{11}) + \Gamma_a^2 \tilde\Gamma_{11}^2]\,.
  \end{equation}
\end{widetext}

\begin{figure}
  \centering
  \includegraphics[width=.9\linewidth]{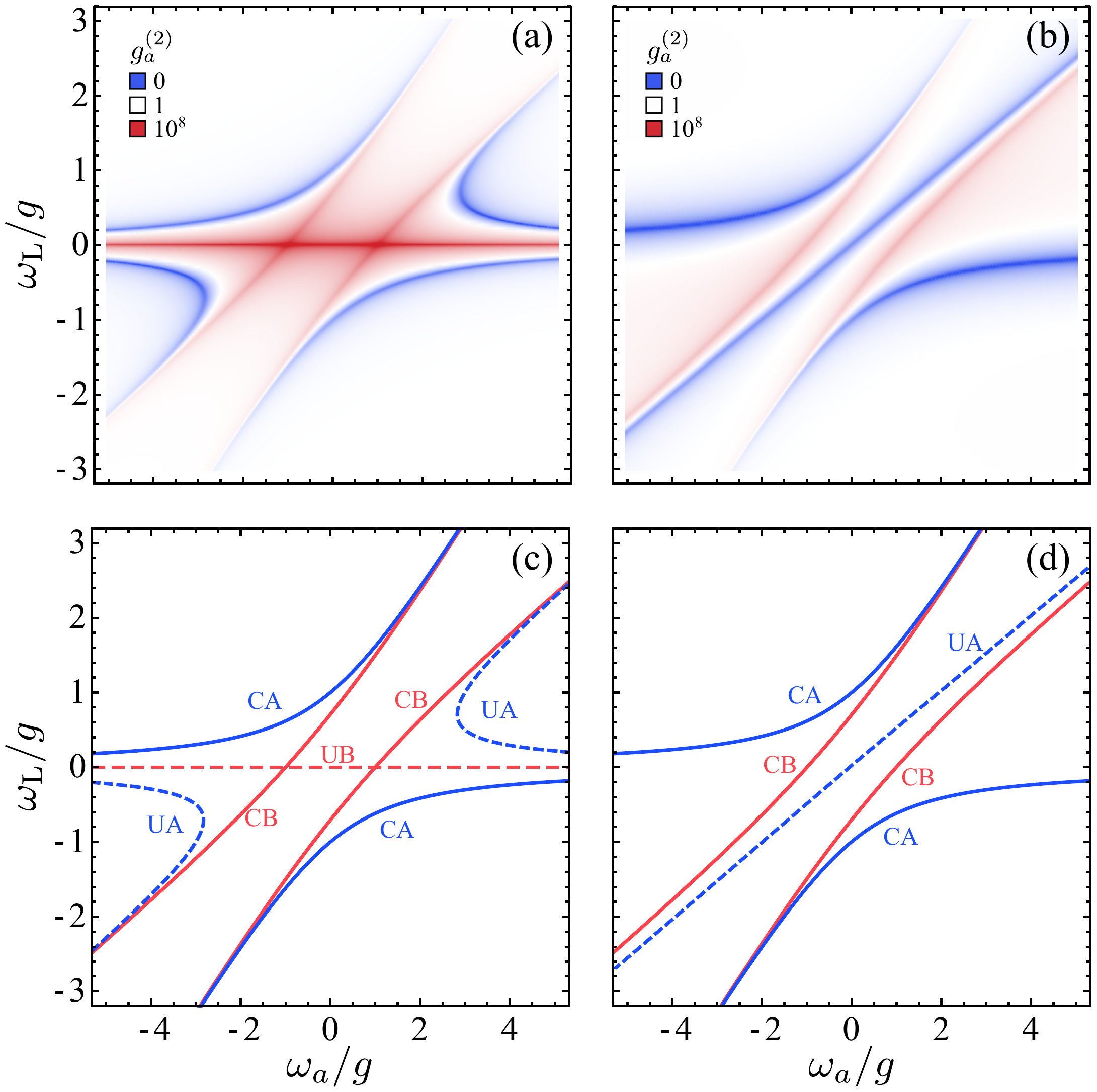}
  \caption{(Color online) Landscape of two-photon correlations in the
    Jaynes--Cummings system, for cavity driving only (left, with
    $\tilde\chi=0$) and mixed driving (right, with $\tilde\chi=0.5$
    and~$\phi=\pi/2$), as a function of where the system is driven
    ($\omega_\mathrm{L}$) and where it is emitting ($\omega_a$). The
    top row shows the exact results,
    Eq.~(\ref{eq:Tue29May184632CEST2018}) left and Eq.~(\ref{eq:JCg2})
    right, and the bottom row its classification in terms of
    Conventional (C) and Unconventional (U) features of Antibunching
    (A) and Bunching (B), namely, Eq.~(\ref{eq:eigenstates})
    with~$N=1$ for CA (solid blue) and~$N=2$ for CB (solid red), and
    Eq.~(\ref{eq:Tue24Mar173546CET2020}) for UA (dashed blue). For the
    left case with cavity-pumping only, the UA simplifies to
    Eq.~(\ref{eq:UAcondition}).  Parameters: $g=1$, $\gamma_a=0.1$
    and~$\gamma_\sigma=0.01$.}
  \label{fig:Fri28Feb070402GMT2020}
\end{figure}

A density plot of Eq.~(\ref{eq:JCg2}) is shown in
Fig.~\ref{fig:Fri28Feb070402GMT2020}(a) where one sees that the
formula produces simple features in the form of well-defined lines of
antibunching (blue in our color code) and bunching (red), as a
function of the relevant parameters (pumping, lifetimes, etc.)  The
general expression, Eq.~(\ref{eq:Tue29May184632CEST2018}) is shown in
the facing panel~\ref{fig:Fri28Feb070402GMT2020}(b) for the case of a
balanced driving~$\Omega_a=\Omega_\sigma$ ($\tilde\chi=0.5$) with also
a relative phase of~$\pi/2$ between the two drivings. Other cases can
be visualised interactively with an
applet~\cite{wolfram_casalengua18a}. Depending on the configuration,
one sees that some features appear while other disappear, e.g., the
horizontal superbunched line disappears and a diagonal antibunched
line appears, with also two antibunched hyperbolas now absent. We
remind that the change from one case to the other comes merely from
switching on a second and out-of-phase driving term from
$\Omega_\sigma=0$ (left) to~$\Omega_\sigma=\Omega_a$ (right). One can
see how, as a result, in the configuration of driving the system and
detecting the photons both at resonance
($\omega_\mathrm{L}=\omega_{A}=0$), there is a drastic change from
giant superbunching ($g^{(2)}=1.6\times10^9$) when driving the cavity,
to strong antibunching ($g^{(2)}=0.01$) when also driving the
emitter. This is an illustration of how greatly tunable is the photon
statistics, this times through the balance of the coherent fields
involved.

\begin{figure*}[th]
  \centering \includegraphics[width=\linewidth]{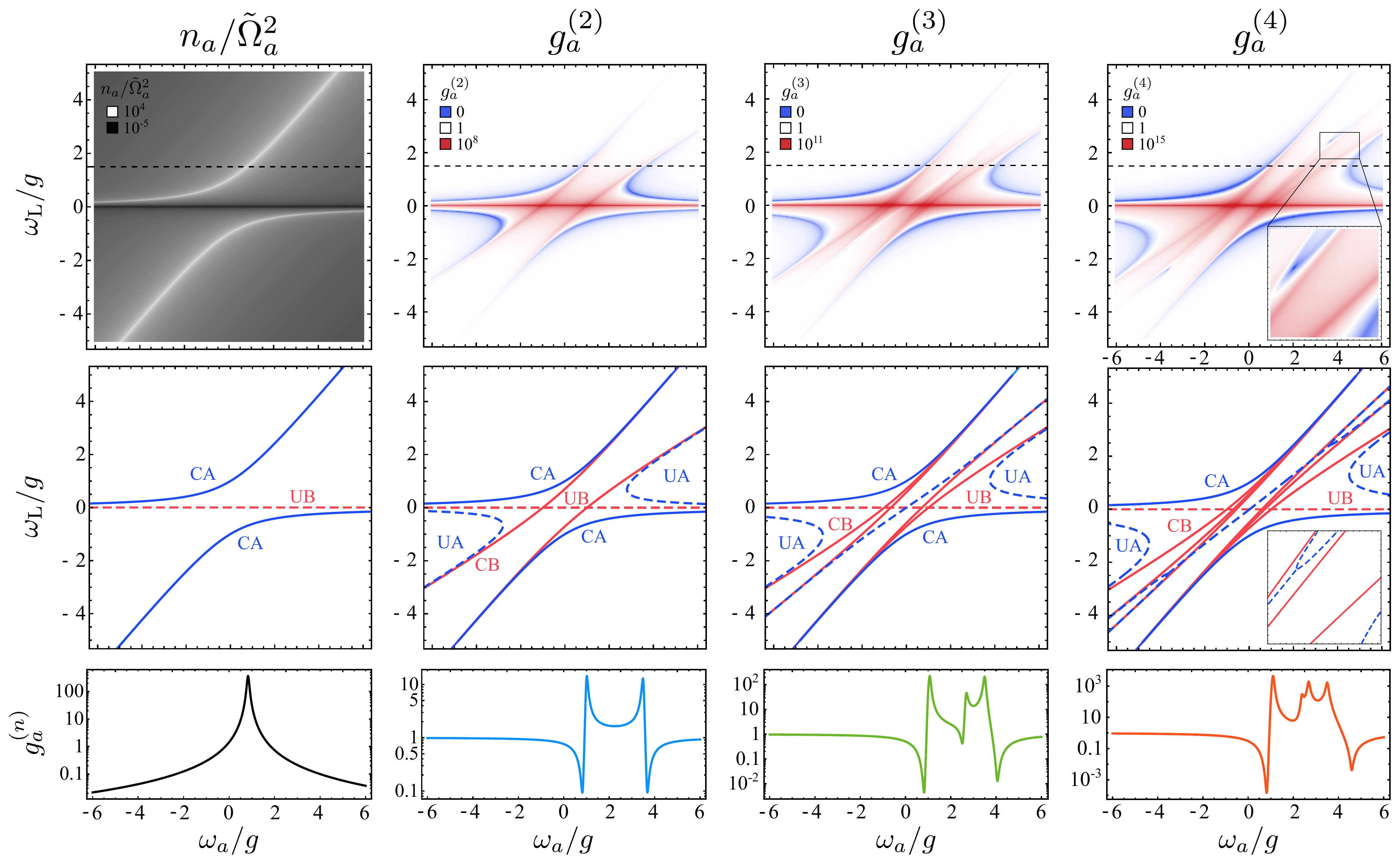}
  \caption{(Color online) Conventional and unconventional features at
    the $N=1, 2, 3$ and~$4$-photon level. Upper row shows the
    numerically exact landscapes of $N$-photon observables, from the
    population normalized to the relative laser intensity
    $\tilde{\Omega}_a^2 = \Omega_a^2 /\gamma_a^2$ ($N=1$, left) till
    four-photon correlations ($N=4$ right). Middle-row shows the
    theoretical lines that reproduce these structures and their
    classifications as conventional (C, solid) and unconventional (U,
    dashed) bunching (red) and antibunching (blue),
    respectively. Bottom row shows the cuts along the horizontal
    dashed line in the top row.  The inset in $\g{4}_a$ magnifies a
    forking of antibunching. Parameters are the same as in
    Fig.~\ref{fig:Fri28Feb070402GMT2020}(a).}
  \label{fig:Fri28Feb133015GMT2020}
\end{figure*}

We now address the qualitative meaning of each line. Some of the
features, shown in Fig.~\ref{fig:Fri28Feb070402GMT2020}, are easily
recognized, namely, the lower and upper polaritons, with their
characteristic anticrossing, and even more simply, the bare
states of the cavity (horizontal line) and two-level system
(diagonal). Their expressions are consequently easily found, as
$\omega_{a}$, $\omega_\sigma$ for the bare states
and~\cite{delvalle09a,laussy12e}
\begin{multline}
  \label{eq:eigenstates}
  E^{(N)}_{\pm}=N\omega_{a}+\frac{\omega_\sigma-\omega_a}{2}\\
  \pm\mathrm{Re}\sqrt{(\sqrt{N}g)^2+\left(
      \frac{\omega_a-\omega_\sigma}{2}-i\frac{\gamma_a-\gamma_\sigma}{4}\right)^2
  }\,,
\end{multline}
with~$N=1$ for the single polaritons and~$N=2$ for the two-excitation
polaritons. Equation~(\ref{eq:eigenstates}) with~$N=1$ yields the blue
solid lines labelled CA, for ``conventional antibunching'', in
Fig.~\ref{fig:Fri28Feb070402GMT2020}(c--d). At the~$\g{2}$ level, only
features up to~$N=2$ show up, but if one considers higher-order photon
correlations, then higher rungs of the Jaynes--Cummings ladder are
probed and the traces formed by Eq.~(\ref{eq:eigenstates})
for~$N\le k$ are seen in~$\g{k}$, as is shown in
Fig.~\ref{fig:Fri28Feb133015GMT2020} for~$N$ up to~$4$. Although these
features can appear only at a given photon-number $N$, their position
is otherwise fixed. 
We note also that although we consider throughout strong-coupling
configurations, and its underlying dressed-states structure, there is
not such a clear-cut distinction between strong and weak-coupling, as
discussed in more details in Ref.~\cite{zubizarretacasalengua20a},
where a critical coupling strength~$g_P$ between the cavity and the
2LS that results in Poissonian statistics ($\g{2}_a =1$), is compared
to bunching/antibunching in the system. We do not discuss
this further but give its closed-form expression:
\begin{multline}
  \label{eq:FriMar2104433CET2018}
  g_{P} = \frac{1}{2} \Big \lbrace \big [ 16 \Delta_\sigma^4
  + 32 \Delta_a \Delta_\sigma^3 - 8(\gamma_a^2 +3\gamma_a
  \gamma_\sigma + \gamma_\sigma^2 - 4\Delta_a^2)\Delta_\sigma^2 \\ 
  {}-8 \gamma_\sigma (4\gamma_a+3\gamma_\sigma)
  \Delta_a \Delta_\sigma + \gamma_\sigma^2 (2\gamma_a^2 +
  2\gamma_a \gamma_\sigma + \gamma_\sigma^2 - 8\Delta_a^2 )
  \big]^{1/2} + {}\\\gamma_\sigma^2 - 4\Delta_\sigma^2 \Big
  \rbrace^{1/2}.
\end{multline}
A smaller coupling~$g<g_\mathrm{P}$ produces antibunched light while a
larger coupling~$g>g_\mathrm{P}$ produces bunched light.

Less immediate to identify are the other features, not accounted for
by Eq.~(\ref{eq:eigenstates}), but that can be extracted from
Eq.~(\ref{eq:Tue29May184632CEST2018}). This can be conveniently done
since the features already identified do not provide the best
antibunching, which is produced by the unconventional mechanism
instead and we have seen that this reaches exactly zero in the
vanishing driving limit.  One can thus hope to find the condition for
the other lines simply by solving $\g{2}_a = 0$
with~$\Omega_{a,\sigma}\rightarrow0$, which yields the following
condition (see also the Supplemental Material):
  \begin{multline}
    \label{eq:Sun3May130904CEST2020}
    \Delta_a = \Big[i \big(\gamma_\sigma + 2 i \Delta_\sigma \big) \big(\tilde{\gamma}_{11} + 2 i \Delta_\sigma \big) + {}\\
     {{}+ 4 e^{-i \phi} g \, \chi \big(\tilde{\gamma}_{11} + 2 i \Delta_\sigma \big)
      - 4 i g^2 \big(1 + e^{-2 i \phi} \chi^2\big)}\Big]\bigg/\\
    ({2 \gamma_\sigma + 4 i \Delta_\sigma - 8 i e^{-i \phi} g \, \chi })\,.
  \end{multline}
  The expression is, in general, complex, which means that one fails
  to get exactly $\g{2}_a=0$. Given that we are now dealing with
  self-homodyning, there is no guarantee indeed that the system would
  interfere its coherent and quantum component so as to cancel exactly
  a given photon-number statistics. Instead, there is the need for a
  fine tuning, which, if not enforced externally as was the case in
  the previous Sections, can only be realized fortuitously. This
  account for the sharpness of the resonances as the exact conditions
  to produce a perfect cancellation requires a careful balancing which
  is realized at an isolated point of the configuration space. Taking
  the real part, however, happens to provide the condition for
  Unconventional Antibunching (UA) that accounts for the features not
  produced by Eq.~(\ref{eq:eigenstates}). The expression then reads as
\begin{widetext}
  \begin{equation}
    \label{eq:Tue24Mar173546CET2020}
    \Delta_a = \frac{4 g \chi \big\lbrace 2 \cos \phi \big[2 \Delta_\sigma^2 + g^2 \big(1+ \chi^2 \big)\big]
      - g \chi \Delta_\sigma \cos 2 \phi - \gamma_\sigma \sin  \phi \big(g \chi \cos \phi - 2 \Delta_\sigma \big)\big\rbrace - \Delta_{\sigma} \big[\Gamma^2_\sigma + 4 g^2 \big(1+ 4 \chi^2 \big)\big]}
    {\gamma_\sigma^2 + 4 \big(\Delta_\sigma^2 + 4 g^2 \chi^2 \big) - 8 g \chi \big( 2 \Delta_\sigma \cos \phi
      + \gamma_\sigma \sin \phi \big)}\,.
  \end{equation}
  Equation~(\ref{eq:Tue24Mar173546CET2020}) yields the blue dashed
  lines labelled UA in Fig.~\ref{fig:Fri28Feb070402GMT2020}(c--d).
  All CA, UA, CB and UB lines are easily recognized in the numerically
  exact plots in Panels~(a--b), which fit perfectly with the
  theoretical lines for small enough~$\Omega_{a,\sigma}$.  Further
  imposing the imaginary part to be also zero finds the isolated
  points where $\g{2}_a=0$ exactly, which provides a second
  expression:
  \begin{equation}
    \label{eq:Fri28Feb113846GMT2020}
    \Delta_\sigma = \frac{4 g \chi \cos \phi \, (\tilde{\gamma}_{11} - g \chi \sin \phi)
      \pm \sqrt{-\tilde{\gamma}_{11} (\gamma_\sigma - 4 g \chi \sin \phi) [-4 g^2 +
        \gamma_\sigma \tilde{\gamma}_{11}- 4 g \chi (g \chi \cos 2 \phi + \tilde{\gamma}_{11} \sin \phi)	] + 4 g^2\chi^2 \sin^2 2 \phi
      } }{2 \tilde{\gamma}_{11}}\,.
  \end{equation}
\end{widetext}
Since Eq.~(\ref{eq:Fri28Feb113846GMT2020}) has to be real, the
radicand has to be positive.  For the particular case $\chi = 0$,
which is that of main interest, these expressions simplify
considerably, namely, Eq.~(\ref{eq:Tue24Mar173546CET2020}) becomes
\begin{equation}
\label{eq:UAcondition}
\Delta_a  = -\Delta_\sigma \left(1 + \frac{4 g^2}{\gamma_\sigma^2 + 4 \Delta_\sigma^2}\right) \,, 
\end{equation}
and Eq.~(\ref{eq:Fri28Feb113846GMT2020}) becomes
\begin{equation}
\label{eq:Sun3May131441CEST2020}
\Delta_\sigma  = \pm \sqrt{\frac{\gamma_\sigma g^2}{\tilde{\gamma}_{11}} - \frac{\gamma_\sigma^2}{4}} \,. 
\end{equation}
These conditions give, first, the UA dashed lines shown, in
Fig.~\ref{fig:Fri28Feb070402GMT2020}(c) and, second, the optimum
points along these curves. Thus, with mixed driving
(Eqs.~(\ref{eq:Tue24Mar173546CET2020}--\ref{eq:Fri28Feb113846GMT2020}))
or cavity-only driving
(Eqs.~(\ref{eq:UAcondition}--\ref{eq:Sun3May131441CEST2020})), both
conditions taken together provide where to drive and detect the Jaynes--Cummings
system to reach perfect $\g{2}_a$ cancellation.

The complexity of photon correlations when including all orders can
hardly be exaggerated. Figure~\ref{fig:Fri28Feb133015GMT2020} shows
how the configuration of Fig.~\ref{fig:Fri28Feb070402GMT2020}(a)
appears when resolved to different photon numbers ($1\le N\le 4$). At
the single-photon level, left column, which is simply luminescence, or
any measurement of the population~$n_a$ of the system, one only
resolves the familiar anticrossing of the two dressed states, or
polaritons. A cut as shown in the bottom-left panel is simply a
Lorentzian function whose width is given by the effective lifetime of
the system
$\gamma_U= \big(\gamma_a + \frac{4 g^2
  \gamma_\sigma}{\Gamma_{\sigma}^2}\big)/2$. There is actually one
feature which is not typically considered given its intrinsically
impractical measurement, namely, the horizontal black line
at~$\omega_\mathrm{L}=0$ which exhibits a suppression of the
population. There is much less light emitted there than at any random
point. This anomalously faint light, however, comes with very strong
correlations, to all-orders, as is revealed in the other panels where
this line shows up as unconventional bunching. This results precisely
from the self-homodyning of the system cancelling largely the coherent
contribution of the emission, leaving mainly quantum fluctutations,
or, here one could say, quantum noise, that has superbunched
statistics. 

At the two-photon level, on the second column of
Fig.~\ref{fig:Fri28Feb133015GMT2020}, one finds again the polariton
lines, which are antibunched, according to the conventional blockade
scenario, whence the label CA. They are supplemented by two
unconventional, self-homodyning antibunching lines UA, in dashes, as
well as the~$N=2$ polariton dressed states of
Eq.~(\ref{eq:eigenstates}) which, by two-photon absorption, now
exhibit conventional bunching CB. There is a beautiful symmetry and
even proximity of these resonances, although they can be of a
different character (antibunching and bunching) and origin
(conventional and unconventional). Note, in particular, in the bottom
panel how the CA and UA exhibit an essentially identical value. At the
three-photon level, on the third column, and even more so at the
four-photon level, fourth column, one now finds a proliferation of CB
features, due to involving the higher rungs of the Jaynes-Cummings
ladder, although these simply add to the lines already existing. In
contrast, as already commented, the UA lines shift positions. There
also appears more UA lines, as can be seen in $\g{3}_a$ with the
appearance of a diagonal UA line, that further exhibits a fork at the
four-photon level. One can check that this complicated structure,
predicted by the theory, is reproduced and easily identified in the
numerically exact landscape of correlation, as shown in the inset of
$\g{4}_a$ where the branch crossing of antibunching is clearly
resolvable, despite being surrounded by conventional bunching
lines. The cuts in the bottom row consequently exhibit extremely
complex resonances alternating between giant bunching and
antibunching, whose relative interplay account for the relative values
found in each case.

\begin{figure}
  \centering \includegraphics[width=.66\linewidth]{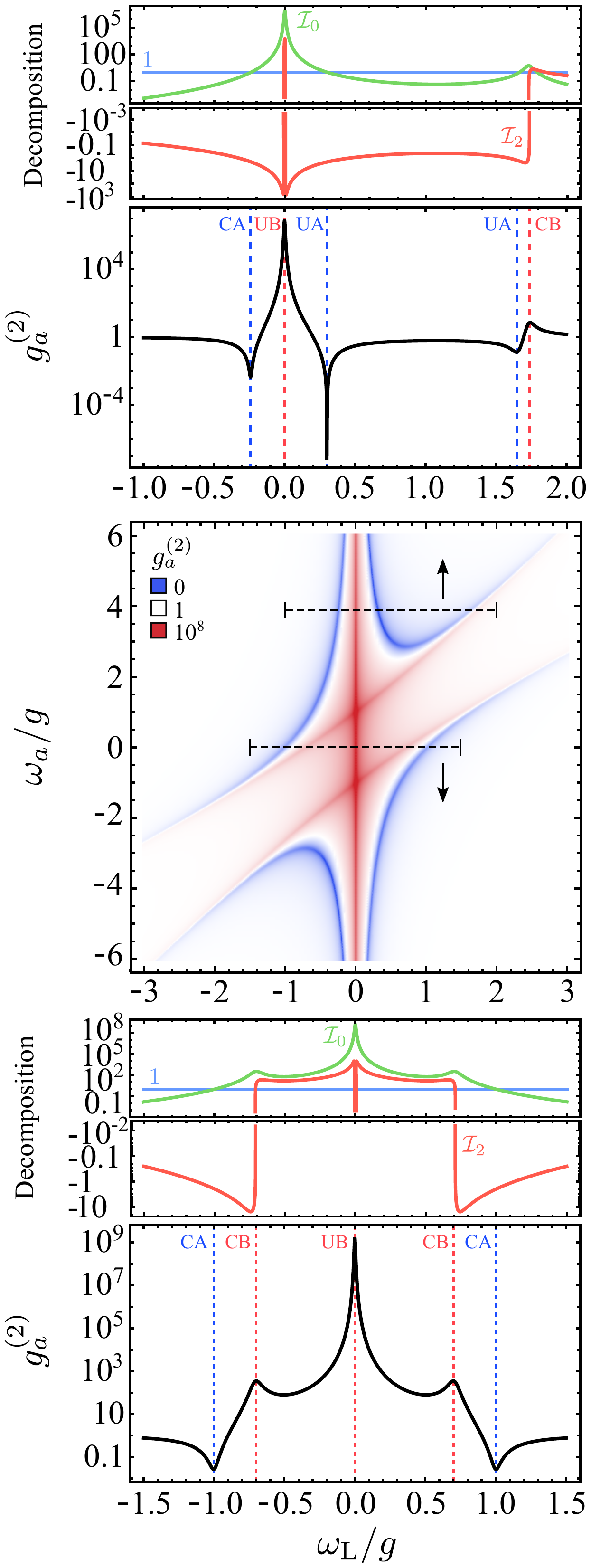}
  \caption{(Color online) Decomposition of the Jaynes--Cummings
    statistics into its~$\mathcal{I}$ coefficients (in log-scales,
    separating the positive and negative components) for the two cuts
    shown in the middle panel. The upper case captures one of the
    exactly-zero antibunching, on the UA line. Parameters are the same
    as in Fig.~\ref{fig:Fri28Feb070402GMT2020}(a).}
  \label{fig:Fri28Feb162129GMT2020}
\end{figure}

This complex phenomenology fits with the simple classification above
and can also be simply understood as multiphoton interferences, as can
be illustrated by their decomposition in terms of the~$\mathcal{I}$
parameters of Eq.~(\ref{eq:Mon20Apr190800CEST2020}) (the same could be
done for the $\mathcal{J}$, $\mathcal{K}$ parameters, etc.)
%
In this dynamical case, the
decomposition~\eqref{eq:decompositiontermswhole} yields the following
expressions, when the cavity alone is driven ($\chi = 0$) at vanishing
pumping (although the general case could also be provided, there is no
need to for the present discussion):
\begin{subequations}
  \label{eq:Mon4May142336CEST2020}
  \begin{align}
    \mathcal{I}_0 = & \ 256 g^8  \bigm/f_1\left(g,\Delta_{a,\sigma}, \gamma_{a,\sigma} \right)\,,\\
    \mathcal{I}_1 = & \ 0\,,\\
    \begin{split}
      \mathcal{I}_2 = & \ 32 g^4 \bigg[-\gamma _{\sigma }^2 \Big(4 g^2 +
      \gamma_a \left(\gamma_a +\gamma_{\sigma }\right) - 4 \Delta_{a}^2\Big)\\
      &{}+4 \gamma_{\sigma } \left(4 \gamma_a + 3 \gamma_{\sigma }\right) \Delta_a \Delta_{\sigma}- 16 \Delta_a \Delta_{\sigma}^3 +{}\\
      4 \Delta_{\sigma}^2 &\Big(4 g^2 + \gamma_a 
      \left( \gamma_a + \gamma_{\sigma }\right) - 4 \Delta_{a}^2\Big) \bigg] \biggm/ 
      f_1\left(g, \Delta_{a,\sigma}, \gamma_{a,\sigma}\right)\,,
    \end{split}
\end{align}
\end{subequations}
where the function
$f_1\left(g,\Delta_{a,\sigma},\gamma_{a,\sigma}\right)$ is defined as:
\begin{multline}
  f_1\left(g, \Delta_{a,\sigma}, \gamma_{a,\sigma}\right) = \Big(\gamma_{\sigma }^2 + 4
  \Delta_{\sigma}^2\Big)^2 \Big(16 g^4 \\+8 g^2 \big[\gamma_a
  \left(\gamma_a + \gamma_{\sigma }\right) - 4 \Delta_a
  \left(\Delta_{a}+\Delta_{\sigma}\right)\big] +\\{} \big[\gamma_a^2
  +4 \Delta_{a}^2\big] \big[\left(\gamma_a + \gamma_\sigma\right)^2 +
  4 \left(\Delta_a + \Delta_{\sigma}\right)^2\big]\Big)\,.
\end{multline}

As previously, at vanishing driving, $\mathcal{I}_1=0$ and we are
therefore back to the paradigm of the above sections of admixing a
squeezed and coherent state.  Simply, the admixing varies
self-consistently with detunings depending on the system
parameters. Figure~\ref{fig:Fri28Feb162129GMT2020} shows how
the~$\mathcal{I}$ parameters balance each other to produce the
various features. As is also the case from a squeezing-coherent state
admixture, the sub-Poissonian $\mathcal{I}_0$ parameter is always
positive, even for CA lines, and it is for self-homodyning to bring an
overall negative~$\g{2}_a$. This decomposition makes particularly
obvious something which will be confirmed even more in the next
section with the more complex case of polaritons, and that one can
check was the case in the simpler previous cases, namely, once one
recognizes the common denominator, $\mathcal{I}_2$ encodes most of the
complexity of the problem.  The similarities in how various features
get decomposed can be deceptive. Note how the UA resonance in the top
panel of Fig.~\ref{fig:Fri28Feb162129GMT2020} is much sharper than
the CA one, namely, $\g{2}_a(\mathrm{CA})=0.0045$ as compared to
$\g{2}_a(\mathrm{UA})=0$ exactly, to leading order (the cut
at~$\omega_\mathrm{L}\approx 3.92g$ has been chosen to intercept this
global minimum). Although the $\mathcal{I}$ lines seem symmetric
around~0, $\mathcal{I}_2$ is steeper for the UA line than for the CA
line and conversely $\mathcal{I}_0$ is steeper for the CA line than
for the UA line, causing the sharper resonance for the unconventional
line. Adding the higher-order correlations, one would also see how the
antibunching is pinned at the same position for CA and takes place in
different places depending on the photon-order for UA. Note also how
the characteristic dispersive-like shape of antibunching and bunching
as seen on the right of the upper panel, that one can understand as
the meeting of two lines (here UA and CB), arises due to a change of
sign of~$\mathcal{I}_2$. Likewise, these changes of sign are notable
when bunching is produced, but instead of discussing them further, we
turn to what is possibly the most interesting consequence of all these
considerations, which, in the Jaynes--Cummings system, occurs in the
case of mixed driving~\cite{xu14a}. 
\begin{figure}
  \centering \includegraphics[width=\linewidth]{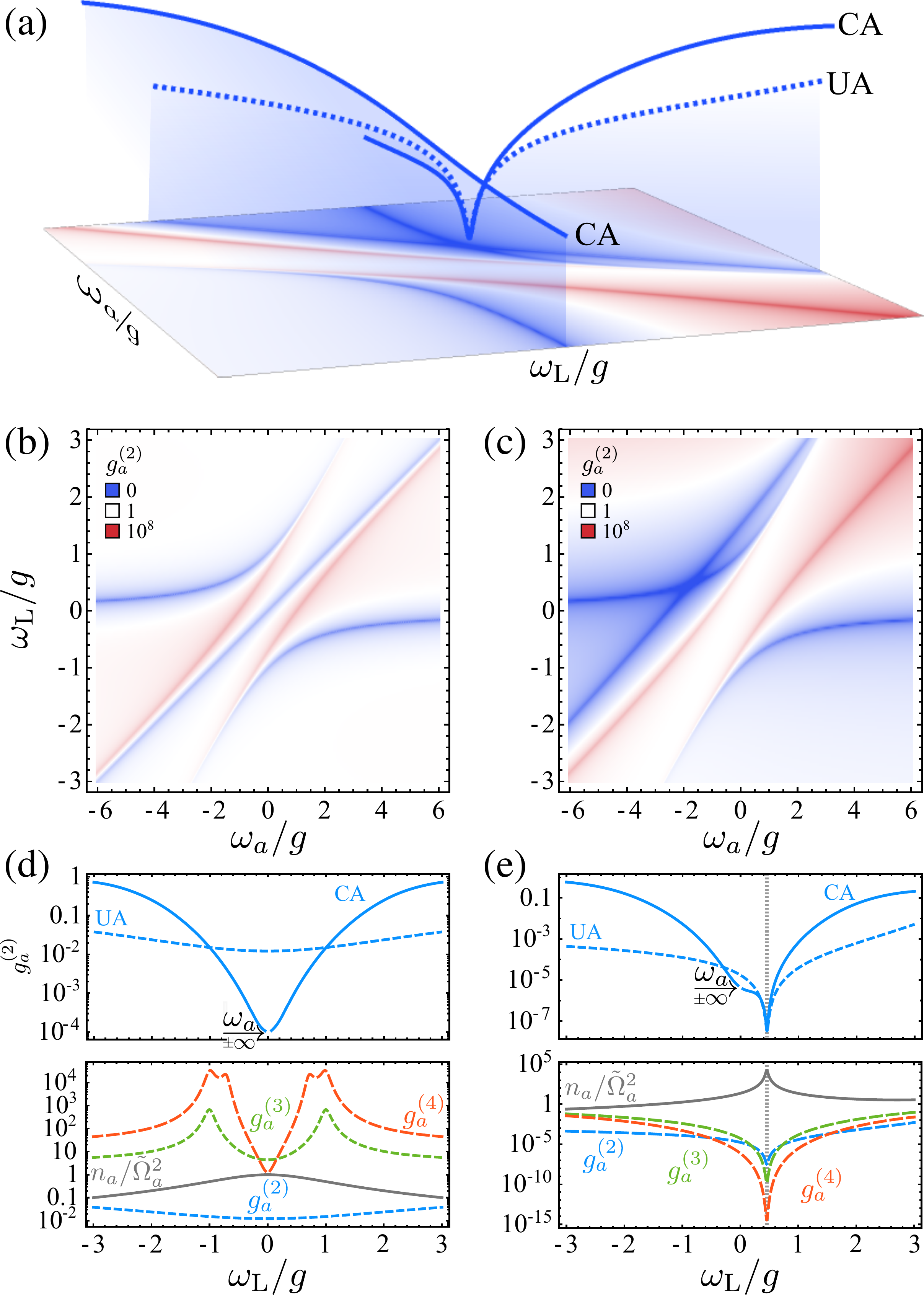}
  \caption{(Color online) Intersection of the UA and CA lines in the
    Jaynes--Cummings model. (a) Representation to show simultaneously
    the lines' shape in the correlation landscape and their magnitude,
    with two of them intersecting with the effect of the UA line
    dragging down the CA line. (b--c) Correlation landscapes with (b)
    no intersection, when~$\chi=0.87$ and~$\phi=\pi$ and~(c) [case
    also shown in~(a)] with intersection, when~$\chi=0.5$
    and~$\phi=\pi/2$. (d--e) Magnitude of the UA and CA along their
    respective lines, as a function of~$\omega_\mathrm{L}$ as the
    scanning parameter. Note that, as a consequence, the
    case~$\omega_\mathrm{L}=0$ corresponds to infinite cavity
    detuning, indicated by a gap opening which is otherwise
    continuous. In the intersecting case, note the highly populated,
    strong and all-order antibunching. Parameters are the same as
    in Fig.~\ref{fig:Fri28Feb070402GMT2020}(a).}
  \label{fig:Fri28Feb172456GMT2020}
\end{figure}
In this case, one can find the peculiar situation where conventional
and unconventional features intersect. This can happen for the
superbunching as seen in Fig.~\ref{fig:Fri28Feb070402GMT2020}(a), with
the effect of maximizing it drastically when UB and CB meet, but more
notably, it can also happen involving a polariton line, meaning, with
a lot of signal. Namely, the CA from the upper polariton branch can
meet the UA line, as shown in Fig.~\ref{fig:Fri28Feb172456GMT2020}
that compares the case of balanced driving, namely,
$\Omega_a=\Omega_\sigma$ ($\chi=1$) with both~$\rightarrow 0$ (left
column), with unbalanced driving, $\Omega_a\approx\Omega_\sigma/5$
($\chi\approx5$), and going to zero in this ratio.  In the
correlation-landscapes, one can recognize the CA polariton lines,
displaying the characteristic anticrossing shapes, and the straight
line of UA (all in blue, being antibunched). In the $\chi=0$ case, the
UA line fits between the two CA line and all remain distinct. In
the~$\chi=0.87$ case, the UA line is shifted to negative $\omega_a$
and intersects the CA line (at~$\omega_a\approx-1.73g$ and
$\omega_\mathrm{L}\approx 0.46g$). At this intersection, one finds the
advantageous configuration combining the best of two worlds, namely, a
large population since the emission comes from a real state of the
system (CA), the antibunching is very strong (UA) and occurs to
all-orders (CA). This is shown in Panels~(d--e) of
Fig.~\ref{fig:Fri28Feb172456GMT2020}. In the non-intersecting case
(left column), antibunching is better (smaller) in the CA case
when~$|\omega_\mathrm{L}|<g$, because self-homodyning, with
$\g{2}_a(\mathrm{UA})\approx0.012$, happens in this case to be far
from its optimum cancellation, and by symmetry, also have its minima
degenerate, thus bearing resemblances, although superficially only,
with conventional antibunching. One features that remains is the small
population, namely, $n_a/\tilde\Omega_a^2\approx0.99$. Although CA is
better in this case, it reaches its minimum of
$\g{2}_a(\mathrm{CA})\approx10^{-4}$
when~$\omega_a\rightarrow\pm\infty$, so this is an asymptotic optimum
(which we indicate on the figure by opening a gap in the curve pointed
out by
$\mathrel{\substack{\omega_a\\\longrightarrow\\\pm\infty}}$). Now,
comparing with the intersecting case (right column), the UA exhibits
one of its typical sharp and strong resonances, here
with~$\g{2}_a\approx 3.3\times10^{-8}$, and this drags the CA line in
its wake, as seen in Fig.~8(a) in the full-space of correlation and
more quantitatively in Fig.~8(e). Also, all the $n$-photon antibunching
are now degenerate, as is typical of a CA resonance. And as the final
asset, the population is, this time,
$n_a/\tilde\Omega_a^2\approx1.67\times10^4$, that is, almost 20\,000
times higher than in the non-intersecting case. We have, therefore, a
considerably enhanced situation due to the intersection of UA and CA
lines: a much stronger antibunching as compared to CA alone, with a
much stronger signal as compared to UA alone. We further discuss this
peculiarity in the next section, where it becomes even more
attractive. We conclude this section by noting that the
$\mathrel{\substack{\omega_a\\\longrightarrow\\\pm\infty}}$ gap opened
in the CA line does not produce a discontinuity, suggesting underlying
symmetries and a peculiar parameter-space topology supporting the
landscape of correlations, which we are simply laying down on a plane
for simplicity and by force of habit, but such a continuity, which
becomes compelling in absence of a symmetry as in
Fig.~\ref{fig:Fri28Feb172456GMT2020}(e), points at a
better parametrization.\break

\section{Microcavity polaritons statistics}
We now contrast the previous results with a platform---namely,
microcavity polaritons---where the effects just discussed become even
more relevant given the intense activity in realizing the two types of
antibunching: conventional and unconventional. We will show in
particular how the rich interplay of self-homodyning of the anharmonic
oscillator~\cite{zubizarretacasalengua20a} with a cavity leads to
surprising results going against the community's expectations
regarding the role of interactions and the experimental configurations
to consider. The Hamiltonian is similar to the Jaynes--Cumming's
Eq.~(\ref{eq:Thu31May103357CEST2018}), except that the emitter
(excitons) has annihilation operator~$b$ also following Bose algebra
(like~$a$) but with a nonlinear quartic nonlinearity of strength~$U$
(describing exciton-exciton self-interactions):
\begin{align}
  \label{eq:Mon5Jun145532BST2017}
  H={}&\hbar\omega_a\ud{a}a+\hbar\omega_b\ud{b}b+\frac{U}{2}\ud{b}\ud{b}bb+\hbar g(\ud{a}b+a\ud{b})+{}\nonumber\\
      &+\Omega_a e^{i \omega_a t}a+\Omega_b e^{i\omega_b t}b + \mathrm{h.c.}
\end{align}
In the limit~$U\rightarrow\infty$, the results
recover those of the Jayne--Cummings model. Polariton systems have
weak nonlinearities~$U\ll\gamma_a$ and we will consider both limits.
The antibunching for both the cavity and exciton emission (since this
is one is not trivially zero anymore) can be obtained in
closed-form. Again, despite the result not being of immediate interest
in such a bulky form, given that it covers in a unified
single-formula a plethora of results that one otherwise finds
scattered in the literature, we nevertheless give it in its entirety
as follows:
\begin{figure}
  \centering
  \includegraphics[width=.66\linewidth]{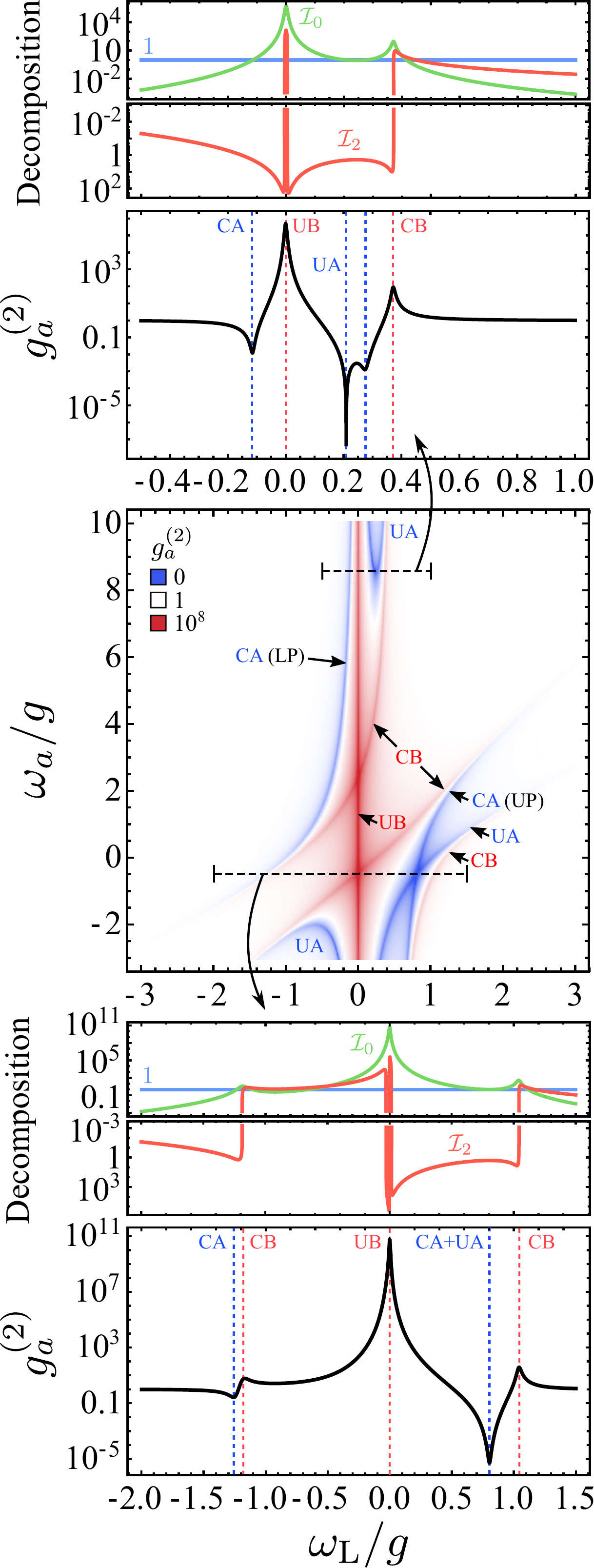}
  \caption{(Color online) Decomposition of~$\g{2}_a$ in terms of the
    $\mathcal{I}$ coefficients (Supplemental Material), for
    microcavity polaritons, along the two cuts shown in the central
    panel, which intersect the points where $\g{2}_a$ becomes exactly
    zero to leading order for the upper cut and corresponds to the
    CA--UA intersection point for the lower cut. Parameters: $g=1$,
    $\gamma_a=0.1$, $\gamma_b=0.01$, $U=1$, $\chi=0$ ($\phi=0$). The
    cuts are at~$\omega_a/g\approx 8.63$ where UA is exactly zero and
    $\omega_a/g\approx -0.44$ where UA and CA intersect.}
  \label{fig:Wed22Apr182034CEST2020}
\end{figure}

\begin{widetext}
  \begin{subequations}
    \label{eq:Thu31May102314CEST2018}  	
    \begin{equation}
      \label{eq:Mon2Mar143716GMT2020}
      \begin{split}
        \g{2}_a = & \Big\lbrace \Big[16 g^4 + 8 g^2 \big(\gamma_a \gamma_b - 4 \Delta_a
        \Delta_b \big) + \Gamma_a^2 \Gamma_b^2 \Big]
        \Big[\Gamma_b^2 \Gamma_{11}^2 \big(\gamma_b^2 + \tilde{U}_{12}^2\big)
        + 8 g^2 \Big(U^2 [4 \Delta_b \tilde{\Delta}_{11}- \gamma_b \tilde{\gamma}_{11}]+ 2 \Gamma_{11}^2 [\gamma_b^2 + \tilde{U}_{12}^2] \chi^2 +\\
        & 8 U \Delta_b^2 \tilde{\Delta}_{11} -2 U \gamma_b^2 \tilde{\Delta}_{13} - 4 U \gamma_a \gamma_b
        \Delta_b \Big) + 16 g^4 \Big(U^2 + [\tilde{\gamma}_{11}^2 + (U + 2 \tilde{\Delta}_{11})^2] \chi^4\Big) - 16 g \chi \cos \phi \, \Big( \Delta_b \Gamma_{11}^2 [\gamma_b^2 + 
        \tilde{U}_{12}^2] \\ 
        & + 2 g^2 [U(2 \tilde{\Delta}_{11} \tilde{U}_{12}- \gamma_b \tilde{\gamma}_{11})
        + (2 U^2 \tilde{\Delta}_{11}+ 2 \Delta_b \Gamma_{11}^2+ U \lbrace \gamma_a
        \tilde{\gamma}_{11} + 4 \tilde{\Delta}_{11} \tilde{\Delta}_{12} \rbrace ) \chi^2 ]\Big) +
        8 g^2 \chi^2 \cos 2 \phi \, 
        \Big(4 g^2 U [U + 2 \tilde{\Delta}_{11}] -  \\
        & U^2 [\gamma_b \tilde{\gamma}_{11}- 4 \Delta_b \tilde{\Delta}_{11}] - [\gamma_b^2 - 4 
        \Delta_b^2] \Gamma_{11}^2 + 2 U [\gamma_a^2 \Delta_b + \tilde{\Delta}_{12}
        (4 \Delta_b \tilde{\Delta}_{11} - \gamma_b^2)] \Big) -8 g \chi \sin \phi \, 
        \Big( \gamma_b \Gamma_{11}^2 [\gamma_b^2 + \tilde{U}_{12}^2] + \\
        & 4 g^2 [\gamma_b \Gamma_{11}^2 \chi^2 + U (\chi^2-1) (U \tilde{\gamma}_{11}  + 2 
        \gamma_b \Delta_a +2 \tilde{\gamma}_{12} \Delta_b)] \Big) + 8 g^2 \chi^2 \sin 2 \phi \,
        \Big( -4 g^2 U \tilde{\gamma}_{11} + 4 \gamma_b \Delta_b \Gamma_{11}^2 + 2 U^2
        [\gamma_a \Delta_b + \gamma_b \tilde{\Delta}_{12}] + \\
        & U [\gamma_a^2 \gamma_b + 4 \gamma_b \tilde{\Delta}_{12}^2 + \gamma_a \Gamma_b^2] \Big) \Big]
        \Big\rbrace \Big/ 
         \Big\lbrace \Big( \Gamma_a^2 \Gamma_{11}^2 \big[\gamma_b^2 + \tilde{U}_{12}^2 \big] + 16 g^4 \big[\tilde{\gamma}_{11}^2 + \big(U + 2 
        \tilde{\Delta}_{11}\big)^2\big] + 8 g^2 \big[U^2 \big(\gamma_a \tilde{\gamma}_{11} -
        4 \Delta_a \tilde{\Delta}_{11}\big) + {} \\
        & \Gamma_{11}^2 \big(\gamma_a \gamma_b - 4 \Delta_a \Delta_b\big) - 2U \big(\gamma_a^2 \tilde{\Delta}_{1 \bar{1}} - 2 \gamma_a \gamma_b \Delta_b
        + 4 \Delta_a \tilde{\Delta}_{11} \tilde{\Delta}_{12}\big) \big] \Big)
        \Big(4 g^2 \chi^2 + \Gamma_b^2  - 4 g \chi
        \big[2 \Delta_b \cos \phi + \gamma_b \sin \phi \big]\Big)^2 \Big\rbrace
        \, , 
      \end{split}
    \end{equation}
    %
    \begin{equation}
      \begin{split}
        \g{2}_b = & \big\lbrace \Gamma_{11}^2  \big[16 g^4 + 8 g^2 \big( \gamma_a 
        \gamma_b - 4 \Delta_a \Delta_b \big) + \Gamma_a^2 \Gamma_b^2\big] \big\rbrace  
        \Big/ 
        \big\lbrace \Gamma_a^2 \Gamma_{11}^2 \big[\gamma_b^2 +\tilde{U}_{12}^2 \big] + 16 g^4 \big[\tilde{\gamma}_{11}^2 + \big(U + 2 
        \tilde{\Delta}_{11}\big)^2\big] + {}\\
        &8 g^2 \big[U^2 \big(\gamma_a \tilde{\gamma}_{11} -
        4 \Delta_a \tilde{\Delta}_{11}\big) + \Gamma_{11}^2 \big(\gamma_a \gamma_b -
        4 \Delta_a \Delta_b\big) - 2U \big(\gamma_a^2 \tilde{\Delta}_{1 \bar{1}} - 2 \gamma_a \gamma_b \Delta_b
        + 4 \Delta_a \tilde{\Delta}_{11} \tilde{\Delta}_{12}\big) \big] \big\rbrace
        \, ,
      \end{split}
    \end{equation}
  \end{subequations}
  where we have used the short-hand
  notation~$\Gamma_{c}^2 = \gamma_c^2 + 4\Delta_c^2$ for $c=a,b$ as
  well as $\tilde{\Delta}_{ij} \equiv i \Delta_a + j \Delta_b$,
  $\tilde{\gamma}_{ij} \equiv i \gamma_a + j \gamma_\sigma$,
  $\Gamma_{ij}^2 \equiv \tilde{\gamma}_{ij}^2 + 4
  \tilde{\Delta}^2_{ij}$, $\tilde{U}_{ij} \equiv i U + j \Delta_b$ and
  $\bar{j}$ denotes negative integer values~($\bar{j} \equiv -j$).  As
  before, considerable simplifications are obtained when focusing on
  particular cases, e.g., by considering cavity pumping only, i.e.,
  $\chi=0$ (the case usually assumed in the literature), in which
  case, Eq.~(\ref{eq:Mon2Mar143716GMT2020}) reduces to
  \begin{multline}
    \label{eq:PBg2}
    \g{2}_a = [16g^4 + 8g^2( \gamma_a \gamma_b - 4\Delta_a
    \Delta_b)+\Gamma_a^2 \Gamma_b^2 ]\times{}\\ {}\times \lbrace
    16g^4U^2 + \Gamma_b^2 \Gamma_{11}^2 [\gamma_b^2 + (U+2\Delta_b)^2]
    -8g^2U [4\gamma_a\gamma_b \Delta_b -8 \Delta_b^2 \Delta_{11} +
    2\gamma_b^2 (\Delta_{11}+2\Delta_b) + U(\gamma_b \gamma_{11}-
    4\Delta_b\Delta_{11})]\rbrace \bigm/ \\\big[ 8g^2 \Gamma_b^4\{ U^2
    (\gamma_a\gamma_{11}- 4 \Delta_a\Delta_{11}) + \Gamma_{11}^2
    (\gamma_a\gamma_b -4 \Delta_a \Delta_b) -2 U [ \gamma_a^2
    \Delta_{1\bar 1} - 2 \gamma_a \gamma_b \Delta_b + 4 \Delta_a
    \Delta_{11} \Delta_{12}]\} + {}\\ \Gamma_b^4
    \{\Gamma_a^2 \Gamma_{11}^2 [\gamma_b^2 +(U+2\Delta_b)^2] + 16g^4
    [\gamma_{11}^2+(U+2\Delta_{11})^2] \} \big]\,.
\end{multline}
\end{widetext}
The same decomposition of this two-photon correlation function can be
made in terms of the~$\mathcal{I}$
parameters~(\ref{eq:decompositiontermswhole}). These results are also
exact but become extremely heavy even for the particular
case~$\chi = 0$, so we give them in the Supplementary Material. They
are plotted for two cuts of interest in the landscape of correlations
of polaritons in Fig.~\ref{fig:Wed22Apr182034CEST2020}.  We need not
describe in much details the structure of this landscape for polaritons
since it is so closely
 related to the Jaynes--Cummings case, with
Conventional~C and Unconventional~U, Bunching~B and Antibunching~A
combinations giving rise to CB, CA, UB and UA lines, with the same
origins and consequently identical properties, such as C lines being
attributable to dressed states of the systems and U lines to
interferences at a given photon number.
 These are labelled directly on
the figure and one can compare with
Figs.~\ref{fig:Fri28Feb070402GMT2020}--\ref{fig:Fri28Feb162129GMT2020}
from the Jaynes--Cummings limit to see both similarities and
departures. Some are quantitative only, such as the two CB lines that
were two parallel lines in the Jaynes--Cummings case now become curved
and drifting away in the UP region, where the UA line also gets
squeezed and sent away to large cavity detunings. Other are
qualitative, like the apparition of a new CB line, due to a dressed
state from the second manifold (purely upper-polaritonic) whose energy
grows like the interaction strength, being sent away to infinity in
the Jaynes--Cummings limit~$U\rightarrow\infty$ where the line has
thus completely disappeared.  The decomposition in terms of the
$\mathcal{I}$ parameters also presents mainly quantitative departures
from its Jaynes--Cummings counterpart. The strong oscillations of
$\mathcal{I}_2$ are concomitant with the intersection of the CB and UB
lines here, which produces, like in the Jaynes--Cummings case, a boost
of superbunching by several orders of magnitudes, peaking
at~$\g{2}_a\approx 6.9\times10^{10}$ for the lower intersection. More
importantly, we find again the intersection between the UA and CA
lines, already discussed in the previous section. Here, an important
deviation is that this can happen without the need of mixed-driving,
which is understandably a complication to implement
experimentally. Beside in the polariton platform even more so than in
the Jaynes--Cummings system, one is interested in finding the optimum
antibunching (i.e., smallest $g^{(2)}_a$)~\cite{sanvitto19a}. The value of this
intersection point in this regard will be stressed in the
following. As previously, the perfect-antibunching conditions can be
derived from the equation $\g{2}_a=0$, which we will do here for the
$\chi=0$ case given by the expression Eq.~\eqref{eq:PBg2}. Then,
clearing $\Delta_a$ from the previous equation, leads to:
\begin{multline}
\label{eq:UApbcond}
\Delta_a = \big\lbrace
e^{i \phi} \big[4 g^2U - \big(\gamma_b + 2 i \Delta_b\big)
	\big(\tilde{\gamma}_{11} + 2 i \Delta_b\big) \big(U + 2 \Delta_b - i \gamma_b\big)
	\big] \\
	 + 4 i  g \chi \big(U + 2 \Delta_b - i \gamma_b\big) \big(\tilde{\gamma}_{11} + 2 i \Delta_b\big)\\
	+ 4 e^{-i \phi} g^2 \chi^2 \big(U + 2 \Delta_b - i \tilde{\gamma}_{11}\big)
\big\rbrace
/ \mathcal{N}
\end{multline}
where $\mathcal{N}$ is defined as
\begin{multline}
\mathcal{N} =
2 \big[ e^{i \phi} \big(\gamma_b + 2 i \Delta_b\big) \big(\gamma_b + i U + 2 i \Delta_b\big)\\
+4 g \chi \big(U + 2 \Delta_b - i \gamma_b \big) - 4 g^2 \chi^2 e^{-i \phi}\big].
\end{multline}

Again, although by definition $\Delta_a$ must be real, we arrive to a
complex-valued condition, but the real part of Eq.~(\ref{eq:UApbcond})
gives the equation for the UA lines. Moreover, the cancellation of its
imaginary part also provides a second condition that allows to
identify $\g{2}_a = 0$ exactly, to lowest-order in the driving. Like in
the Jaynes--Cummings case, since there are two conditions, it is not
possible to fulfil both simultaneously except at isolated points in
the parameter space.  As an illustration, we present here the case of
cavity excitation ($\chi = 0$).  Splitting both real and imaginary
parts from Eq.~\eqref{eq:UApbcond}, we find:
\begin{subequations}
\begin{align}
&\Delta_a  = - \Delta_b - \frac{4 g^2 \Delta_b}{\gamma_b^2 + 4 \Delta_b^2}
           + \frac{2 g^2 (U + 2 \Delta_b)}{\gamma_b^2 + (U + 2 \Delta_b)^2}\,,\label{eq:Thu23Apr164141CEST2020} \\
& 0 = \gamma_a + \gamma_b + 4 g^2 \gamma_b \left( - \frac{1}{\gamma_b^2 + 4 \Delta_b^2}
+ \frac{1}{\gamma_b^2 + (U + 2 \Delta_b)^2} \right) .
\end{align}
\end{subequations}

\begin{figure*}
  \centering
  \includegraphics[width=\linewidth]{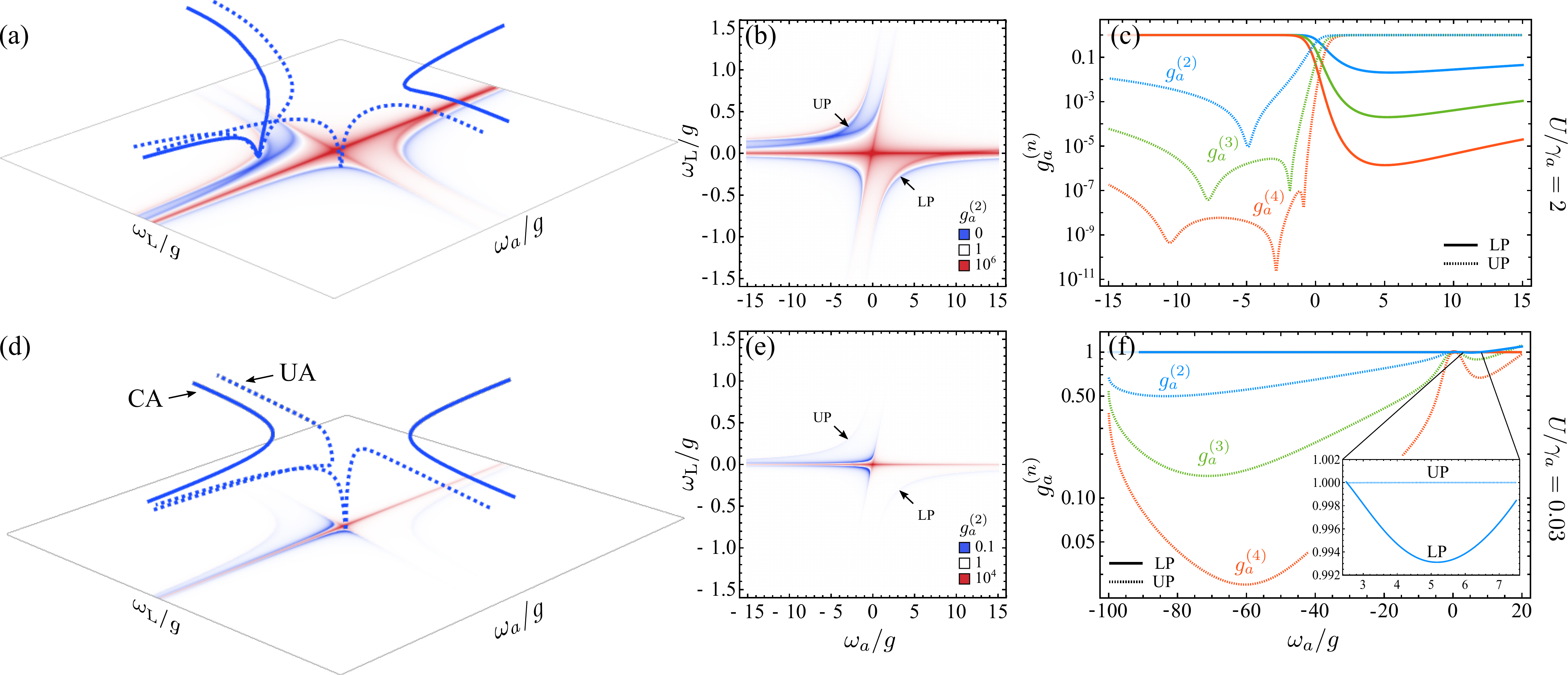}
  \caption{(Color online) Conventional and unconventional statistics
    in a microcavity polariton system, with strong (top row,
    $U/\gamma_a=0.2$) and weak (bottom row, $U/\gamma_a=0.03$)
    interactions. (a, d) Conventional (solid) and unconventional
    (dashed) antibunching as a 3D representation of the landscapes of
    correlations shown in panels (b--e). Note how the CA line gets
    pulled down by the UA one if they intersect. (c--f) $\g{n}_a$
    for~$n=2,3,4$ along the upper (UP) and lower (LP) polariton lines
    (i.e., CA). The upper polariton line gets a much better
    antibunching thanks to its proximity, or even intersection, with
    the unconventional antibunching. The inset in panel~(f) shows for
    comparison the much smaller antibunching of the upper line (the
    one so far reported experimentally).}
  \label{fig:Mon2Mar184713GMT2020}
\end{figure*}

The first expression, Eq.~(\ref{eq:Thu23Apr164141CEST2020}), provides
an implicit equation for the three distinct curves of UA shown in
Figs.~\ref{fig:Wed22Apr182034CEST2020}
and~\ref{fig:Mon2Mar184713GMT2020}, whereas the latter gives the exact
location where $\g{2}_a$ becomes exactly zero, to lowest-order in the
driving.

To appreciate how these several aspects compete with each others, we
show in Fig.~\ref{fig:Mon2Mar184713GMT2020}(a,d) a 3D representation
of the joint magnitude and shapes of the antibunching lines in the
polaritonic landscapes of correlations, for strong (upper row) and
weak (lower row) polariton interactions. The correlation landscapes
are also shown in Panels~(b) and~(e) with the two polariton branches
identified, and the magnitudes of antibunching on the polariton
branches are given up to the fourth-order in Panels~(c) and~(f). There
is a crossing of the CA and UA lines in the upper case, as indeed the
conditions for the intersection to take place require that the
interactions~$U$ are neither too large nor too weak but be in the
range
\begin{equation}
  \label{eq:Thu23Apr181531CEST2020}
  2.57\gamma_b+9.30\gamma_b^3/g^2\lesssim U\lesssim 2(g^2/\gamma_b)-5\gamma_b
\end{equation}
which is obtained by studying the solutions of
$\mathrm{UA}=\mathrm{CA}$ in the limit of small~$\gamma_b$, yielding
exact but surprisingly awkward solutions for the lower bound: for
instance $2.57$ is really $\sqrt{1+6\cos(\pi/9)}$. The exact solution
for 9.30 is a similar but more complex expression in terms of
trigonometric functions of multiples of~$\pi/9$. When the intersection
exists, as is clear in Panel~(b) of
Fig.~\ref{fig:Mon2Mar184713GMT2020}, one sees (Panels~(a) and~(c)) how
the CA line, typically of fairly modest antibunching, gets sucked-in
by the UA line and exhibits, as a result, record-value of antibunching
as compared to typical CA values, here with
$\g{2}_a\approx9.2\times10^{-6}$, as compared to $\g{2}_a\approx0.02$
only on the other (LP) polariton branch.  Note, interestingly, that
this effect takes place on the UP polariton branch, which is typically
discarded by experimentalists for practical reasons (it is typically
less bright and not as defined as its lower sister). In both cases,
being on the branches, the population is very large, namely,
$n_a/\tilde\Omega_a^2\approx31.05$ on the UP as compared to
$n_a/\tilde\Omega_a^2\approx0.02$ for the optimum UA off-branch, even
though in this case the antibunching is perfect, being exactly zero.
Also, although the minima for higher-order correlators are not
degenerate with this crossing point, they are at least also very
small, unlike UA features alone where two-photon antibunching comes
with higher-order photon bunching
(cf.~Fig.~\ref{fig:Wed15Apr104949CEST2020}). 

Although the intersection is not always guaranteed, it is interesting
that the proximity alone of the UA and CA lines tends to produce a
similar result of a considerable improvement of the CA. 
This is shown in the bottom row of
Fig.~\ref{fig:Mon2Mar184713GMT2020}, where there is no intersection,
but as the two lines converge asymptotically towards each other with
increasing cavity (negative) detuning, the CA on the UP line becomes
much smaller than would be expected from conventional polariton
blockade. With~$U/\gamma_a=0.03$, which is about the experimental
value assumed in several systems, this proximity leads to a value of
$\g{2}_a\approx 0.499$ which, to be appreciated, has to be compared
with its pure CA counterpart (on the LP branch), which is
$\g{2}_a\approx 0.993$ only, that corresponds to the results recently
experimentally reported~\cite{munozmatutano19a,delteil19a}. Our
picture shows the considerable antibunching improvement that is in
principle within reach merely by changing branches. The large detuning
required to reach the minimum, namely for our parameters,
$\omega_a\approx-83.4g$, means a smaller population, which can become
as detrimental to the signal as being off-branch, and indeed,
$n_a/\tilde\Omega_a^2\approx 5 \times 10^{-4}$ on the highly-detuned
UP branch is much less than $n_a/\tilde\Omega_a^2\approx 0.02$ for the
optimum UA of $\g{2}_a=0$, off-branch at $\omega_a\approx-0.48g$.  But
since the resonance is very broad, one can still obtain sizable
all-order antibunching on the UP branch at smaller detunings, as seen
in Fig.~10(f).  For instance, at $\omega_a=-10 g $, a routine-detuning,
$\g{2}_a\approx 0.86$, which would be a clear-cut, compelling
measurement still over 20 times larger than the CA on the
LP branch, and with, this time, the considerable population
$n_a/\tilde\Omega_a^2\approx 3.2$, that is about 160 higher than the
off-branch UA. At $\omega_a=-5 g$, with a still neatly-resolvable
$\g{2}_a\approx0.92$ (similar to the values actually reported in the
literature on the lower branch\cite{munozmatutano19a,delteil19a}), one
gains another order of magnitude in signal.

\begin{figure*}
  \centering \includegraphics[width=.66\linewidth]{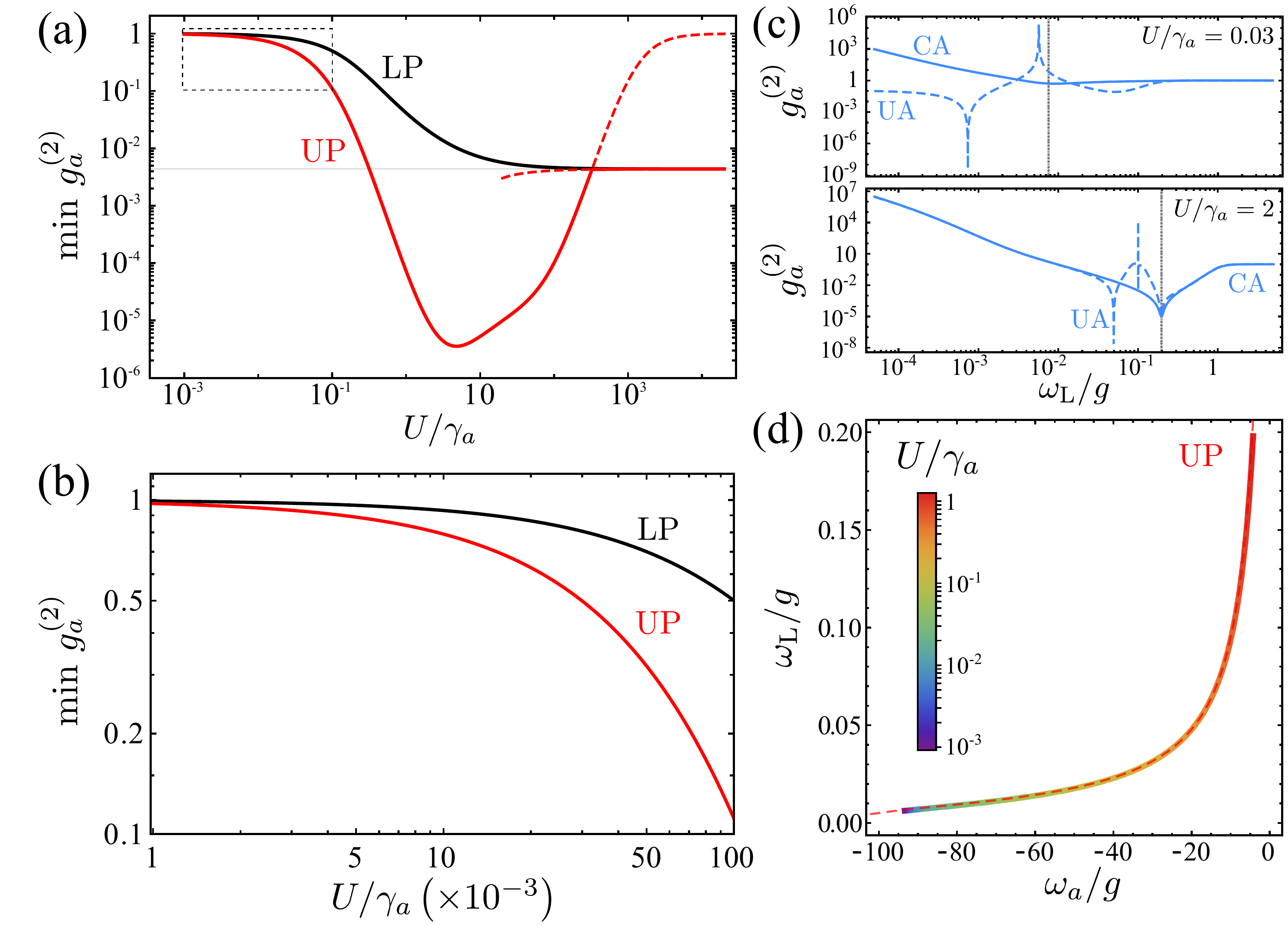}
  \caption{(Color online) Polariton antibunching as a function of
    polariton interactions. (a) Antibunching on the two polariton
    branches, showing the much greater UP antibunching due to the
    proximity to the UA line, even at very small interactions (zoom in
    Panel~(b)). An optimum is obtained for a finite value of
    interactions, in contrast to antibunching on the LP line which is
    optimum in the Jaynes--Cummings limit~$U\rightarrow\infty$. The
    solid lines in~(a) show the optimum value of~$\g{2}_a$ on the LP
    (black) and the UP (red). On the UP, where it is significantly
    stronger, this is due to the intersection, or proximity, between
    UA and CA, until interactions get very large, $U/\gamma\gtrsim10$,
    where another local minimum overtakes and the UA+CA
    carries on as the dashed line. (c) Details of the proximity
    (top) and intersection (bottom) effect for the small and high
    interactions, respectively. The vertical line shows the minimum
    antibunching on the UP line, which is CA. (d) Location on the UP
    line where to drive the system depending on its
    interaction~$U/\gamma_a$ to optimize its two-photon antibunching.}
  \label{fig:Tue21Apr155051CEST2020}
\end{figure*}

We leave these interesting prospects of this intersection for an even
more surprising result, that concerns the strength of the interactions
required to optimize antibunching. It is widely assumed that as strong
as possible interactions are required to maximise antibunching (i.e.,
to minimise the value of~$\g{2}_a$). This is true for CA, but not for
UA. Consequently, this limitation also gets lifted at the UA--CA
intersection, with an antibunching on the polariton branch, with a
large population that is maximum for a finite interaction strength,
close to~$U\approx g$. This coincidental value appears to be favouring
the level degeneracy and thus to optimize the unconventional type of
the joint antibunching. This is shown in
Fig.~\ref{fig:Tue21Apr155051CEST2020}(a), that provides the optimum
two-photon antibunching~$\g{2}_a$ along the LP line (black), where it
is purely CA, and along the UP line (red), where it intersects with
the UA line. The LP antibunching behaves as expected, steadily
decreasing until it reaches its optimum value which is that of the
Jaynes--Cummings system when~$U\rightarrow\infty$, with
$\g{2}_a\approx4.4\times10^{-3}$. In contrast, the UP antibunching
achieves several orders of magnitude smaller values for finite
interactions, namely, $\g{2}_a\approx 3.5\times10^{-6}$ at
$U/\gamma_a \approx 4.8$. While this optimum antibunching is obtained
for a very strong interaction strength by today's polaritonic
standards, one still get a considerable improvement for the typical
values of current experiments, as shown in Panel~(b). We already
discussed the case~$U/\gamma_a=0.03$, that is shown again with the
strong-interaction case~$U/\gamma_a=2$ in Panel~(c), where one sees
how the LP optimum antibunching (shown with the dotted vertical grey
line) arises from the UA--CA intersection in the strongly-interacting
case and the pull-down effect from their proximity in the
low-interaction one. The detuning that minimizes $\g{2}_a$ on the UP
line, that is, where to drive the system to optimize its bright
antibunched emission, is shown in in Panel~(d). Here again, one can
compromise antibunching for a stronger signal by reducing the
detuning. More than the antibunching per se, these results could be
particularly attractive to accurately estimate the strength of
polariton interactions.

\begin{figure}[t!]
  \centering
  \includegraphics[width=.75\linewidth]{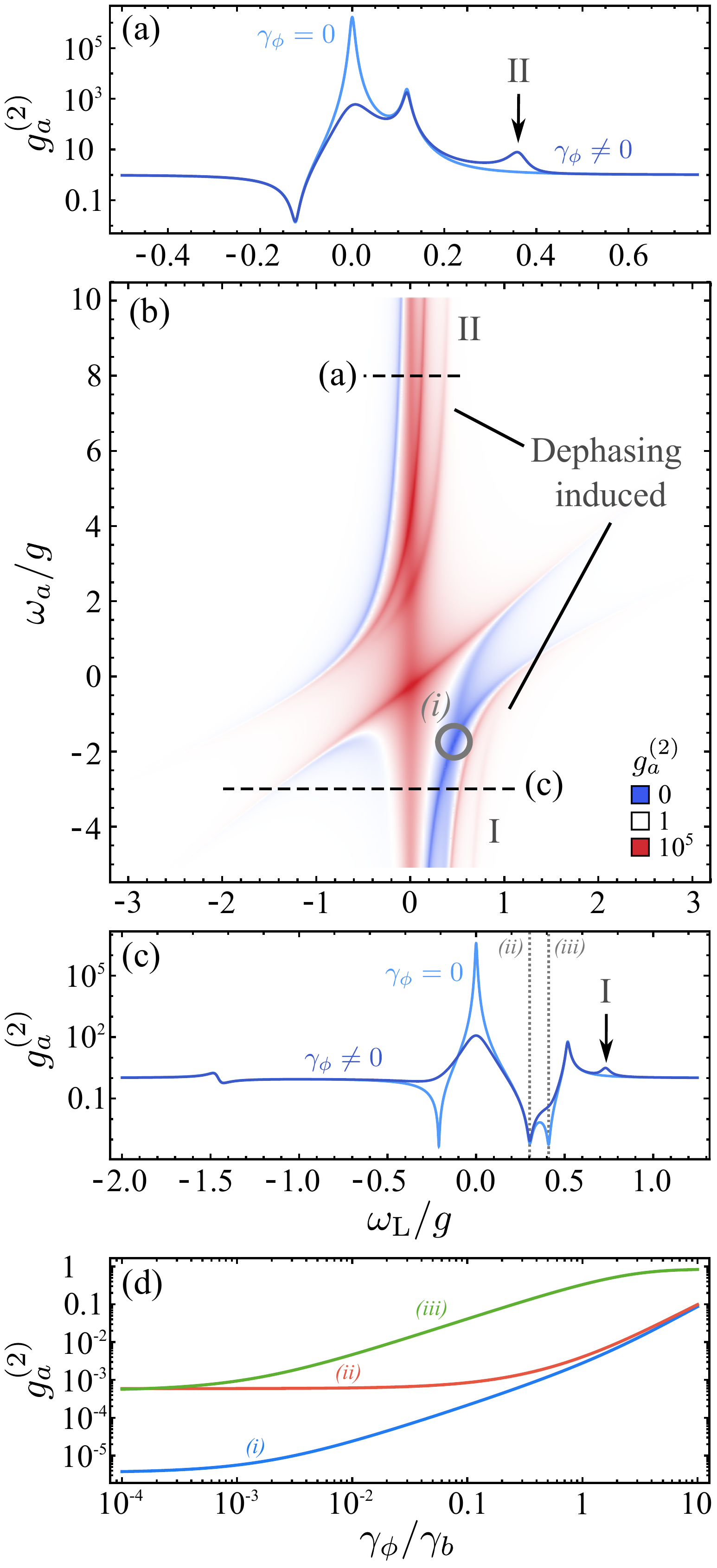}
  \caption{Effect of dephasing on photon correlations $\g{2}_a$.  In
    Panel~(b), the correlation landscape scanning in frequencies, with
    two cuts shown in Panels~(a) and~(c) at $\omega_a/g = -3$ and~$8$,
    respectively, along with their dephasing-free counterparts (light
    blue). Some of the antibunching/bunching peaks, related to
    unconventional features, are strongly suppressed whereas
    conventional lines are left almost untouched. Additionally, new
    dephasing-induced lines emerge, labelled~I and~II in (a--c).
    Panel~(d) shows the robustness of the antibunching against
    dephasing: (i) for the $\mathrm{UA}+\mathrm{CA}$ intersection
    (marked with a gray circle in Panel~(b)), (ii) for CA and (iii)
    for UA at the position shown in Panel~(c).  Parameters: $g = 1$,
    $\gamma_a = 0.1$, $\gamma_b = 0.01$, $U = 0.5$, $\chi = 0$ (cavity
    driven) and $\gamma_\phi = 0.1\gamma_b$.}
  \label{fig:Fri24Apr084553CEST2020}
\end{figure}

We conclude with a brief consideration on the role of dephasing on the
effects we have discussed in this text. Although they also are of a
general character, we discuss them in the context of polaritons only.
Pure dephasing can be included to a system's dynamics by adding to the
master equation the Lindblad term
$(\gamma_\phi/2)\mathcal{L}_{\ud{b} b} \rho $ with~$\gamma_\phi$ the
dephasing rate. This describes loss of quantum coherence, and has the
overall effect of spoiling correlations: damping superbunching and
antibunching, both getting closer to~1 from their respective sides
of~$\g{2}=1$. This affects as well our homodyning configuration, where
we can restore or impose infinite superbunching and exactly zero
antibunching to leading-order in the driving, in absence of
dephasing. This becomes impossible when~$\gamma_\phi\neq0$, even in
the vanishing driving regime~\cite{lopezcarreno19a}. On the other
hand, the response of conventional and unconventional features to
dephasing is very different, confirming their distinct nature and
character. Namely, conventional features are much less sensitive to
dephasing and remain mostly undisturbed for small values of dephasing
as compared to the linewidths of the bare states, i.e., when
$\gamma_\phi / \gamma_b \lesssim 1$~\cite{zubizarretacasalengua20a}.
Above that point, corresponding to considerable dephasing rates,
conventional features start to fade away although slowly and still
exhibiting a remarkable resilience. In comparison, unconventional
features, which are due to interferences, are extremely fragile and
their very-good values without dephasing are strongly affected by its
presence. The behaviour of the intersection of CA and UA is
interesting in this regard. Panel~(d) of
Fig.~\ref{fig:Fri24Apr084553CEST2020} shows the evolution of~$\g{2}_a$
for three types of antibunching: $(i)$ the intersection of UA and CA,
identified by a circle in Fig.~\ref{fig:Fri24Apr084553CEST2020}(b),
and $(ii)$ the CA, and $(iii)$ the UA on the
cut~(c). One can see, again, how CA is more robust to dephasing,
remaining essentially unaffected till $\gamma_\phi$ becomes a
significant fraction of~$\gamma_b$. In contrast, UA is quickly spoiled
by $\gamma_\phi$. Interestingly, the intersection behaves like both
types, being strongly affected when it is very small and exhibiting
the same type of resilience to dephasing as CA when its antibunching
becomes of the same magnitude.

Another interesting role of dephasing goes beyond affecting the
existing features, and brings new ones, namely, additional bunching
resonances appear with~$\gamma_\phi\neq0$, which could also be
considered for spectroscopic applications so as to estimate the amount
of dephasing present in a system. This is shown in
Fig.~\ref{fig:Fri24Apr084553CEST2020}, where the correlation landscape
is shown for a small dephasing rate
($\gamma_\phi/g=0.1\gamma_b$). Arrows point at the two new lines that
are dephasing-induced. They are also shown in two cuts at
$\omega_a/g=-3$ and $\omega_a/g=8$, in Panels~(a) and~(c), that
compare the case with (dark-blue) and without (light-blue)
dephasing. The additional bunching peaks correspond to transitions
between polariton energy levels.  Consequently, the line~I (at
negative detunings) fulfills
\begin{equation}
  \label{eq:Fri24Apr090542CEST2020}
  \omega_\mathrm{L}=\tilde E_0^{(2)}-\tilde E_-^{(1)}
\end{equation}
while line~II (at positive detunings) is given by
\begin{equation}
  \label{eq:Fri24Apr090549CEST2020}
  \omega_\mathrm{L}=\tilde E_-^{(2)}-\tilde E_-^{(1)}\,,
\end{equation}
where~$\tilde E^{(k)}_l$ are the energy levels of the few-particle
(up to~$k=2$) polariton states, which can be
  obtained in closed-form but yielding awkward expressions.  We give
  here the first-order term in the interaction~$U$ under
  strong-coupling conditions~($g\gg \gamma_a$, $\gamma_b$):
\begin{subequations}
  \label{eq:pb2ndman} 
  \begin{align}
    \tilde E_0^{(2)}&=\omega_a + \omega_b +\frac{g^2}{2 R^2}U\,,\\
    \tilde E^{(2)}_{\pm}&=\omega_a + \omega_b \pm 2R {}\nonumber\\
    &\quad{}+\frac{2g^2+(\omega_a - \omega_b)[(\omega_a - \omega_b)\mp 2R]}{8 R^2}U\,,
  \end{align}
\end{subequations}
with~$R=\sqrt{g^2+(\omega_a - \omega_b)^2/4}$, and
$\tilde E_-^{(1)}=E_-^{(1)}$ of Eq.~(\ref{eq:eigenstates}) since in
absence of interactions and transitions, the lower-polariton energies
coincide with the one-polariton excitations. One polariton, from the
upper or lower branch, is emitted, leaving some room for a second
polariton to be radiated. The incoherent contribution to the field is
no longer depleted to second-order, whence the origin of these
transitions.
  
\section{Summary and Conclusions}

If one mixes a coherent state with a quantum state, such as a squeezed
state, depending on their relative amplitudes and phases, one can tune 
the statistics of the resulting state
from superbunching to antibunching, although a coherent state has no
correlations and a squeezed state is always super-Poissonian. This is
the manifestation at the multiphoton level of the well-known ``not
just the sum'' principle of interfering fields, whereby their
combination can result in a total that has opposite
characteristics. Indeed, bringing the two-photon nonzero amplitudes
out of phase will result in two-photon antibunching in the total
field, if they otherwise differ in their one and three-photon
components, so the effect is, again, neatly produced with a squeezed
and coherent states. As amplitudes go to zero, the correlations can
actually diverge or vanish completely. This simple
phenomenology turns out to be at the foundation of a large body of
results in coherently-driven systems, where they have been branded as
an ``unconventional photon blockade'', in reference to the
conventional blockade scenario due to nonlinearities that act as a
photon turnstile. This unconventional mechanism produces much stronger
correlations and is particularly noteworthy in not relying on strong
nonlinearities. For these reasons, it has triggered an intense
research activity with a thriving literature~\cite{xu13a, tang15a,
  shen15a, shen15b, li15c, xu16a, wang16b, zhou16a, zhou16b, liu16a,
  kryuchkyan16a, wang17a, cheng17a, deng17a, yu17a, zhou18a} that
studied considerable variations of the effect at the two-photon level,
due to an exaggerated appreciation of~$\g{2}$ to
quantify single-photon emission~\cite{grunwald19a}. We have shown how
this can be fruitfully understood in several platforms (including
resonance fluorescence, the Jaynes--Cummings model and microcavity
polaritons) as the simple admixing of a coherent state with a quantum
state which, to lowest-order in the driving, is a squeezed state,
thereby realizing self-consistently and in a dynamical setting the
paradigm of feeding a beam-splitter with a coherent state and a
squeezed state, producing a displaced squeezed thermal state on one
output port. We have shown in particular how one can control
externally the coherent field to tune the statistics of the overall
emission.  In coherently-driven quantum optical systems, the coherent
state does not have to be provided externally but can be seen to come
directly from the mean field, while the squeezed state, or other types
of quantum states, come from the fluctuations. For given system
parameters, a particular configuration of detunings will realize the
admixture that yields optimum antibunching and another the optimum
superbunching. Scanning over the full range of these parameters, one
can thus reveal features in a landscape of correlations, that are
clearly explained in terms of the coherent and squeezed states
admixture. At low driving, we have shown through a decomposition of
the Glauber coherence functions~$\g{n}$ in terms of
coefficients~$\mathcal{I}$, $\mathcal{J}$, etc., that embed various
types of quantum correlations, how these features are further tightly
related to so-called conventional blockade correlations, which pertain
to the dressed-states of the system. Since both mechanisms rely on
low-driving, even when one is dealing with a two-level system, the
driving is too weak to bring the nonlinearity to imprint a
non-Gaussian character to the field. Still, both the conventional and
unconventional scenarios, that we prefer to tag on the statistics
itself rather than on an alleged ``blockade,'' indeed provide two
fairly distinct families of features, with their own characteristics
that one can recognise throughout platforms. The two mechanisms
combined with the two types of correlations bring us to a
classification of UA, UB, CA and CB for Unconventional Antibunching,
Unconventional Bunching, Conventional Antibunching and Conventional
Bunching, respectively. Their main characteristics are as follows: unconventional
features are photon-number dependent, i.e., are realized for a
given~$n$ that can be chosen, but one at a time, with a typical
scenario being a small~$\g{2}$ but large~$\g{n}$ for~$n\ge2$, and this
takes place in different locations of the parameter space.  In
contrast, CA occurs to all orders simultaneously and is pinned to the
same underlying structure. Unconventional features are typically
stronger than conventional ones, also not requiring, as already
observed, strong nonlinearities. In the limit of vanishing driving,
they become exactly zero (antibunching) and infinite (superbunching)
to leading-order in the driving. This manifests itself in much sharper
resonances, in particular when contrasted to the conventional ones
sitting nearby.  But unconventional features are also more fragile, to
dissipation, dephasing and driving. Because of the latter, they
also suffer from a weak signal. Nonetheless, in the full landscape of
correlations, we have shown the existence of intersection points of U
and C lines where, remarkably, both qualities of the two types are
produced, namely, for the case of antibunching, one can get bright,
robust and all-order antibunching, these being features of CA, that is
also very strong even in weakly interacting systems, this being the
main feature of UA. This is particularly appealing for a platform such
as microcavity polaritons, where antibunching is highly sought but
remains so-far extremely weak. In the configuration we propose, which
involves the upper polariton branch instead of the usually favoured
lower one, we predict the existence of a highly-populated,
strongly-antibunched emission. While the interest of such a point for
applications is unclear given the Gaussian character of the emission,
it would certainly be valuable for spectroscopic purposes to help in
measuring the polariton-polariton interaction and/or dephasing rate.
Overall, at the theoretical level, our picture unifies a considerable
amount of phenomenology related to photon statistics in coherently
driven systems, which, whether of interest or not for applications,
should help be valuable to synthesise the gigantic number of
minor variations of a fairly simple theme.

\bibliography{sci,Books,pb,arXiv} 

\begin{thebibliography}{83}%
\makeatletter
\providecommand \@ifxundefined [1]{%
 \@ifx{#1\undefined}
}%
\providecommand \@ifnum [1]{%
 \ifnum #1\expandafter \@firstoftwo
 \else \expandafter \@secondoftwo
 \fi
}%
\providecommand \@ifx [1]{%
 \ifx #1\expandafter \@firstoftwo
 \else \expandafter \@secondoftwo
 \fi
}%
\providecommand \natexlab [1]{#1}%
\providecommand \enquote  [1]{``#1''}%
\providecommand \bibnamefont  [1]{#1}%
\providecommand \bibfnamefont [1]{#1}%
\providecommand \citenamefont [1]{#1}%
\providecommand \href@noop [0]{\@secondoftwo}%
\providecommand \href [0]{\begingroup \@sanitize@url \@href}%
\providecommand \@href[1]{\@@startlink{#1}\@@href}%
\providecommand \@@href[1]{\endgroup#1\@@endlink}%
\providecommand \@sanitize@url [0]{\catcode `\\12\catcode `\$12\catcode
  `\&12\catcode `\#12\catcode `\^12\catcode `\_12\catcode `\%12\relax}%
\providecommand \@@startlink[1]{}%
\providecommand \@@endlink[0]{}%
\providecommand \url  [0]{\begingroup\@sanitize@url \@url }%
\providecommand \@url [1]{\endgroup\@href {#1}{\urlprefix }}%
\providecommand \urlprefix  [0]{URL }%
\providecommand \Eprint [0]{\href }%
\providecommand \doibase [0]{http://dx.doi.org/}%
\providecommand \selectlanguage [0]{\@gobble}%
\providecommand \bibinfo  [0]{\@secondoftwo}%
\providecommand \bibfield  [0]{\@secondoftwo}%
\providecommand \translation [1]{[#1]}%
\providecommand \BibitemOpen [0]{}%
\providecommand \bibitemStop [0]{}%
\providecommand \bibitemNoStop [0]{.\EOS\space}%
\providecommand \EOS [0]{\spacefactor3000\relax}%
\providecommand \BibitemShut  [1]{\csname bibitem#1\endcsname}%
\let\auto@bib@innerbib\@empty
\bibitem [{\citenamefont {Young}(1807)}]{young_book1807a}%
  \BibitemOpen
  \bibfield  {author} {\bibinfo {author} {\bibfnamefont {T.}~\bibnamefont
  {Young}},\ }\href@noop {} {\emph {\bibinfo {title} {A Course of Lectures on
  Natural Philosophy and the Mechanical Arts}}}\ (\bibinfo  {publisher} {J.
  Johnson, London},\ \bibinfo {year} {1807})\BibitemShut {NoStop}%
\bibitem [{\citenamefont {Ficek}\ and\ \citenamefont
  {Swain}(2004)}]{ficek_book04a}%
  \BibitemOpen
  \bibfield  {author} {\bibinfo {author} {\bibfnamefont {Z.}~\bibnamefont
  {Ficek}}\ and\ \bibinfo {author} {\bibfnamefont {S.}~\bibnamefont {Swain}},\
  }\href@noop {} {\emph {\bibinfo {title} {Quantum Interference and Coherence:
  Theory and Experiments}}}\ (\bibinfo  {publisher} {Springer, New York},\
  \bibinfo {year} {2004})\BibitemShut {NoStop}%
\bibitem [{\citenamefont {Ritze}\ and\ \citenamefont
  {Bandilla}(1979)}]{ritze79a}%
  \BibitemOpen
  \bibfield  {author} {\bibinfo {author} {\bibfnamefont {H.-H.}\ \bibnamefont
  {Ritze}}\ and\ \bibinfo {author} {\bibfnamefont {A.}~\bibnamefont
  {Bandilla}},\ }\href {\doibase 10.1016/0030-4018(79)90152-4} {\bibfield
  {journal} {\bibinfo  {journal} {Opt. Commun.}\ }\textbf {\bibinfo {volume}
  {29}},\ \bibinfo {pages} {126} (\bibinfo {year} {1979})}\BibitemShut
  {NoStop}%
\bibitem [{\citenamefont {Paul}(1982)}]{paul82a}%
  \BibitemOpen
  \bibfield  {author} {\bibinfo {author} {\bibfnamefont {H.}~\bibnamefont
  {Paul}},\ }\href {doi:10.1103/RevModPhys.54.1061} {\bibfield  {journal}
  {\bibinfo  {journal} {Rev. Mod. Phys.}\ }\textbf {\bibinfo {volume} {54}},\
  \bibinfo {pages} {1061} (\bibinfo {year} {1982})}\BibitemShut {NoStop}%
\bibitem [{\citenamefont {Bu{\v z}ek}\ and\ \citenamefont
  {Knight}(1995)}]{buzek95a}%
  \BibitemOpen
  \bibfield  {author} {\bibinfo {author} {\bibfnamefont {V.}~\bibnamefont
  {Bu{\v z}ek}}\ and\ \bibinfo {author} {\bibfnamefont {P.~L.}\ \bibnamefont
  {Knight}},\ }\href {doi:10.1016/S0079-6638(08)70324-X} {\bibfield  {journal}
  {\bibinfo  {journal} {Progress in Optics}\ }\textbf {\bibinfo {volume}
  {34}},\ \bibinfo {pages} {1} (\bibinfo {year} {1995})}\BibitemShut {NoStop}%
\bibitem [{\citenamefont {Prakash}\ and\ \citenamefont
  {Mishra}(2005)}]{prakash05a}%
  \BibitemOpen
  \bibfield  {author} {\bibinfo {author} {\bibfnamefont {H.}~\bibnamefont
  {Prakash}}\ and\ \bibinfo {author} {\bibfnamefont {D.~K.}\ \bibnamefont
  {Mishra}},\ }\href {doi:10.1088/0953-4075/38/6/005} {\bibfield  {journal}
  {\bibinfo  {journal} {J. Phys. B.: At. Mol. Opt. Phys.}\ }\textbf {\bibinfo
  {volume} {38}},\ \bibinfo {pages} {665} (\bibinfo {year} {2005})}\BibitemShut
  {NoStop}%
\bibitem [{\citenamefont {Majumdar}\ \emph {et~al.}(2012)\citenamefont
  {Majumdar}, \citenamefont {Englund}, \citenamefont {Bajcsy},\ and\
  \citenamefont {\Vuckovic}}]{majumdar12a}%
  \BibitemOpen
  \bibfield  {author} {\bibinfo {author} {\bibfnamefont {A.}~\bibnamefont
  {Majumdar}}, \bibinfo {author} {\bibfnamefont {D.}~\bibnamefont {Englund}},
  \bibinfo {author} {\bibfnamefont {M.}~\bibnamefont {Bajcsy}}, \ and\ \bibinfo
  {author} {\bibfnamefont {J.}~\bibnamefont {\Vuckovic}},\ }\href
  {doi:10.1103/PhysRevA.85.033802} {\bibfield  {journal} {\bibinfo  {journal}
  {Phys. Rev. A}\ }\textbf {\bibinfo {volume} {85}},\ \bibinfo {pages} {033802}
  (\bibinfo {year} {2012})}\BibitemShut {NoStop}%
\bibitem [{\citenamefont {Boddeda}\ \emph {et~al.}(2019)\citenamefont
  {Boddeda}, \citenamefont {Glorieux}, \citenamefont {Bramati},\ and\
  \citenamefont {Pigeon}}]{boddeda19a}%
  \BibitemOpen
  \bibfield  {author} {\bibinfo {author} {\bibfnamefont {R.}~\bibnamefont
  {Boddeda}}, \bibinfo {author} {\bibfnamefont {Q.}~\bibnamefont {Glorieux}},
  \bibinfo {author} {\bibfnamefont {A.}~\bibnamefont {Bramati}}, \ and\
  \bibinfo {author} {\bibfnamefont {S.}~\bibnamefont {Pigeon}},\ }\href
  {doi:10.1088/1361-6455/ab3e98} {\bibfield  {journal} {\bibinfo  {journal} {J.
  Phys. B.: At. Mol. Opt. Phys.}\ }\textbf {\bibinfo {volume} {52}},\ \bibinfo
  {pages} {215401} (\bibinfo {year} {2019})}\BibitemShut {NoStop}%
\bibitem [{\citenamefont {Vogel}(1991)}]{vogel91a}%
  \BibitemOpen
  \bibfield  {author} {\bibinfo {author} {\bibfnamefont {W.}~\bibnamefont
  {Vogel}},\ }\href {doi:10.1103/PhysRevLett.67.2450} {\bibfield  {journal}
  {\bibinfo  {journal} {Phys. Rev. Lett.}\ }\textbf {\bibinfo {volume} {67}},\
  \bibinfo {pages} {2450} (\bibinfo {year} {1991})}\BibitemShut {NoStop}%
\bibitem [{\citenamefont {Vogel}(1995)}]{vogel95a}%
  \BibitemOpen
  \bibfield  {author} {\bibinfo {author} {\bibfnamefont {W.}~\bibnamefont
  {Vogel}},\ }\href {doi:10.1103/PhysRevA.51.4160} {\bibfield  {journal}
  {\bibinfo  {journal} {Phys. Rev. A}\ }\textbf {\bibinfo {volume} {51}},\
  \bibinfo {pages} {4160} (\bibinfo {year} {1995})}\BibitemShut {NoStop}%
\bibitem [{\citenamefont {Carmichael}(1985)}]{carmichael85a}%
  \BibitemOpen
  \bibfield  {author} {\bibinfo {author} {\bibfnamefont {H.~J.}\ \bibnamefont
  {Carmichael}},\ }\href {doi:10.1103/PhysRevLett.55.2790} {\bibfield
  {journal} {\bibinfo  {journal} {Phys. Rev. Lett.}\ }\textbf {\bibinfo
  {volume} {55}},\ \bibinfo {pages} {2790} (\bibinfo {year}
  {1985})}\BibitemShut {NoStop}%
\bibitem [{\citenamefont {Fischer}\ \emph {et~al.}(2016)\citenamefont
  {Fischer}, \citenamefont {M\"uller}, \citenamefont {Rundquist}, \citenamefont
  {Sarmiento}, \citenamefont {Piggott}, \citenamefont {Kelaita}, \citenamefont
  {Dory}, \citenamefont {Lagoudakis},\ and\ \citenamefont
  {\Vuckovic}}]{fischer16a}%
  \BibitemOpen
  \bibfield  {author} {\bibinfo {author} {\bibfnamefont {K.~A.}\ \bibnamefont
  {Fischer}}, \bibinfo {author} {\bibfnamefont {K.}~\bibnamefont {M\"uller}},
  \bibinfo {author} {\bibfnamefont {A.}~\bibnamefont {Rundquist}}, \bibinfo
  {author} {\bibfnamefont {T.}~\bibnamefont {Sarmiento}}, \bibinfo {author}
  {\bibfnamefont {A.~Y.}\ \bibnamefont {Piggott}}, \bibinfo {author}
  {\bibfnamefont {Y.}~\bibnamefont {Kelaita}}, \bibinfo {author} {\bibfnamefont
  {C.}~\bibnamefont {Dory}}, \bibinfo {author} {\bibfnamefont {K.~G.}\
  \bibnamefont {Lagoudakis}}, \ and\ \bibinfo {author} {\bibfnamefont
  {J.}~\bibnamefont {\Vuckovic}},\ }\href {doi:10.1038/nphoton.2015.276}
  {\bibfield  {journal} {\bibinfo  {journal} {Nature Photon.}\ }\textbf
  {\bibinfo {volume} {10}},\ \bibinfo {pages} {163} (\bibinfo {year}
  {2016})}\BibitemShut {NoStop}%
\bibitem [{\citenamefont {M\"{u}ller}\ \emph {et~al.}(2016)\citenamefont
  {M\"{u}ller}, \citenamefont {Fischer}, \citenamefont {Dory}, \citenamefont
  {Sarmiento}, \citenamefont {Lagoudakis}, \citenamefont {Rundquist},
  \citenamefont {Kelaita},\ and\ \citenamefont {Vu\v{c}kovi\'{c}}}]{muller16a}%
  \BibitemOpen
  \bibfield  {author} {\bibinfo {author} {\bibfnamefont {K.}~\bibnamefont
  {M\"{u}ller}}, \bibinfo {author} {\bibfnamefont {K.~A.}\ \bibnamefont
  {Fischer}}, \bibinfo {author} {\bibfnamefont {C.}~\bibnamefont {Dory}},
  \bibinfo {author} {\bibfnamefont {T.}~\bibnamefont {Sarmiento}}, \bibinfo
  {author} {\bibfnamefont {K.~G.}\ \bibnamefont {Lagoudakis}}, \bibinfo
  {author} {\bibfnamefont {A.}~\bibnamefont {Rundquist}}, \bibinfo {author}
  {\bibfnamefont {Y.~A.}\ \bibnamefont {Kelaita}}, \ and\ \bibinfo {author}
  {\bibfnamefont {J.}~\bibnamefont {Vu\v{c}kovi\'{c}}},\ }\href
  {doi:10.1364/OPTICA.3.000931} {\bibfield  {journal} {\bibinfo  {journal}
  {Optica}\ }\textbf {\bibinfo {volume} {3}},\ \bibinfo {pages} {931} (\bibinfo
  {year} {2016})}\BibitemShut {NoStop}%
\bibitem [{\citenamefont {Dory}\ \emph {et~al.}(2017)\citenamefont {Dory},
  \citenamefont {Fischer}, \citenamefont {M\"uller}, \citenamefont
  {Lagoudakis}, \citenamefont {Sarmiento}, \citenamefont {Rundquist},
  \citenamefont {Zhang}, \citenamefont {Kelaita}, \citenamefont {Sapra},\ and\
  \citenamefont {\Vuckovic}}]{dory17a}%
  \BibitemOpen
  \bibfield  {author} {\bibinfo {author} {\bibfnamefont {C.}~\bibnamefont
  {Dory}}, \bibinfo {author} {\bibfnamefont {K.~A.}\ \bibnamefont {Fischer}},
  \bibinfo {author} {\bibfnamefont {K.}~\bibnamefont {M\"uller}}, \bibinfo
  {author} {\bibfnamefont {K.~G.}\ \bibnamefont {Lagoudakis}}, \bibinfo
  {author} {\bibfnamefont {T.}~\bibnamefont {Sarmiento}}, \bibinfo {author}
  {\bibfnamefont {A.}~\bibnamefont {Rundquist}}, \bibinfo {author}
  {\bibfnamefont {J.~L.}\ \bibnamefont {Zhang}}, \bibinfo {author}
  {\bibfnamefont {Y.}~\bibnamefont {Kelaita}}, \bibinfo {author} {\bibfnamefont
  {N.~V.}\ \bibnamefont {Sapra}}, \ and\ \bibinfo {author} {\bibfnamefont
  {J.}~\bibnamefont {\Vuckovic}},\ }\href {doi:10.1103/PhysRevA.95.023804}
  {\bibfield  {journal} {\bibinfo  {journal} {Phys. Rev. A}\ }\textbf {\bibinfo
  {volume} {95}},\ \bibinfo {pages} {023804} (\bibinfo {year}
  {2017})}\BibitemShut {NoStop}%
\bibitem [{\citenamefont {Fischer}\ \emph {et~al.}(2017)\citenamefont
  {Fischer}, \citenamefont {Kelaita}, \citenamefont {Sapra}, \citenamefont
  {Dory}, \citenamefont {Lagoudakis}, \citenamefont {M\"uller},\ and\
  \citenamefont {\Vuckovic}}]{fischer17a}%
  \BibitemOpen
  \bibfield  {author} {\bibinfo {author} {\bibfnamefont {K.~A.}\ \bibnamefont
  {Fischer}}, \bibinfo {author} {\bibfnamefont {Y.~A.}\ \bibnamefont
  {Kelaita}}, \bibinfo {author} {\bibfnamefont {N.~V.}\ \bibnamefont {Sapra}},
  \bibinfo {author} {\bibfnamefont {C.}~\bibnamefont {Dory}}, \bibinfo {author}
  {\bibfnamefont {K.~G.}\ \bibnamefont {Lagoudakis}}, \bibinfo {author}
  {\bibfnamefont {K.}~\bibnamefont {M\"uller}}, \ and\ \bibinfo {author}
  {\bibfnamefont {J.}~\bibnamefont {\Vuckovic}},\ }\href
  {doi:10.1103/PhysRevApplied.7.044002} {\bibfield  {journal} {\bibinfo
  {journal} {Phys. Rev. Appl.}\ }\textbf {\bibinfo {volume} {7}},\ \bibinfo
  {pages} {044002} (\bibinfo {year} {2017})}\BibitemShut {NoStop}%
\bibitem [{\citenamefont {Fischer}\ \emph {et~al.}(2018)\citenamefont
  {Fischer}, \citenamefont {Sun}, \citenamefont {Lukin}, \citenamefont
  {Kelaita}, \citenamefont {Trivedi},\ and\ \citenamefont
  {\Vuckovic}}]{fischer18a}%
  \BibitemOpen
  \bibfield  {author} {\bibinfo {author} {\bibfnamefont {K.}~\bibnamefont
  {Fischer}}, \bibinfo {author} {\bibfnamefont {S.}~\bibnamefont {Sun}},
  \bibinfo {author} {\bibfnamefont {D.}~\bibnamefont {Lukin}}, \bibinfo
  {author} {\bibfnamefont {Y.}~\bibnamefont {Kelaita}}, \bibinfo {author}
  {\bibfnamefont {R.}~\bibnamefont {Trivedi}}, \ and\ \bibinfo {author}
  {\bibfnamefont {J.}~\bibnamefont {\Vuckovic}},\ }\href
  {doi:10.1103/PhysRevA.98.021802} {\bibfield  {journal} {\bibinfo  {journal}
  {Phys. Rev. A}\ }\textbf {\bibinfo {volume} {98}},\ \bibinfo {pages}
  {021802(R)} (\bibinfo {year} {2018})}\BibitemShut {NoStop}%
\bibitem [{\citenamefont {Foster}\ \emph {et~al.}(2019)\citenamefont {Foster},
  \citenamefont {Hallett}, \citenamefont {Iorsh}, \citenamefont {Sheldon},
  \citenamefont {Godsland}, \citenamefont {Royall}, \citenamefont {Clarke},
  \citenamefont {Shelykh}, \citenamefont {Fox}, \citenamefont {Skolnick},
  \citenamefont {Itskevich},\ and\ \citenamefont {Wilson}}]{foster19a}%
  \BibitemOpen
  \bibfield  {author} {\bibinfo {author} {\bibfnamefont {A.}~\bibnamefont
  {Foster}}, \bibinfo {author} {\bibfnamefont {D.}~\bibnamefont {Hallett}},
  \bibinfo {author} {\bibfnamefont {I.}~\bibnamefont {Iorsh}}, \bibinfo
  {author} {\bibfnamefont {S.}~\bibnamefont {Sheldon}}, \bibinfo {author}
  {\bibfnamefont {M.}~\bibnamefont {Godsland}}, \bibinfo {author}
  {\bibfnamefont {B.}~\bibnamefont {Royall}}, \bibinfo {author} {\bibfnamefont
  {E.}~\bibnamefont {Clarke}}, \bibinfo {author} {\bibfnamefont
  {I.}~\bibnamefont {Shelykh}}, \bibinfo {author} {\bibfnamefont
  {A.}~\bibnamefont {Fox}}, \bibinfo {author} {\bibfnamefont {M.}~\bibnamefont
  {Skolnick}}, \bibinfo {author} {\bibfnamefont {I.}~\bibnamefont {Itskevich}},
  \ and\ \bibinfo {author} {\bibfnamefont {L.}~\bibnamefont {Wilson}},\ }\href
  {doi:10.1103/PhysRevLett.122.173603} {\bibfield  {journal} {\bibinfo
  {journal} {Phys. Rev. Lett.}\ }\textbf {\bibinfo {volume} {122}},\ \bibinfo
  {pages} {173603} (\bibinfo {year} {2019})}\BibitemShut {NoStop}%
\bibitem [{\citenamefont {{Zubizarreta Casalengua}}\ \emph
  {et~al.}(2020)\citenamefont {{Zubizarreta Casalengua}}, \citenamefont
  {{L\'opez Carre\~no}}, \citenamefont {Laussy},\ and\ \citenamefont {del
  Valle}}]{zubizarretacasalengua20a}%
  \BibitemOpen
  \bibfield  {author} {\bibinfo {author} {\bibfnamefont {E.}~\bibnamefont
  {{Zubizarreta Casalengua}}}, \bibinfo {author} {\bibfnamefont {J.~C.}\
  \bibnamefont {{L\'opez Carre\~no}}}, \bibinfo {author} {\bibfnamefont
  {F.~P.}\ \bibnamefont {Laussy}}, \ and\ \bibinfo {author} {\bibfnamefont
  {E.}~\bibnamefont {del Valle}},\ }\href {doi: 10.1002/lpor.201900279}
  {\bibfield  {journal} {\bibinfo  {journal} {Laser Photon. Rev.}\ } (\bibinfo
  {year} {2020})}\BibitemShut {NoStop}%
\bibitem [{\citenamefont {Lvovsky}\ and\ \citenamefont
  {Babichev}(2002)}]{lvovsky02a}%
  \BibitemOpen
  \bibfield  {author} {\bibinfo {author} {\bibfnamefont {A.~I.}\ \bibnamefont
  {Lvovsky}}\ and\ \bibinfo {author} {\bibfnamefont {S.~A.}\ \bibnamefont
  {Babichev}},\ }\href {doi:10.1103/PhysRevA.66.011801} {\bibfield  {journal}
  {\bibinfo  {journal} {Phys. Rev. A}\ }\textbf {\bibinfo {volume} {66}},\
  \bibinfo {pages} {011801(R)} (\bibinfo {year} {2002})}\BibitemShut {NoStop}%
\bibitem [{\citenamefont {Windhager}\ \emph {et~al.}(2011)\citenamefont
  {Windhager}, \citenamefont {Suda}, \citenamefont {Pacher}, \citenamefont
  {Peev},\ and\ \citenamefont {Poppe}}]{windhager11a}%
  \BibitemOpen
  \bibfield  {author} {\bibinfo {author} {\bibfnamefont {A.}~\bibnamefont
  {Windhager}}, \bibinfo {author} {\bibfnamefont {M.}~\bibnamefont {Suda}},
  \bibinfo {author} {\bibfnamefont {C.}~\bibnamefont {Pacher}}, \bibinfo
  {author} {\bibfnamefont {M.}~\bibnamefont {Peev}}, \ and\ \bibinfo {author}
  {\bibfnamefont {A.}~\bibnamefont {Poppe}},\ }\href
  {doi:10.1016/j.optcom.2010.12.019} {\bibfield  {journal} {\bibinfo  {journal}
  {Opt. Commun.}\ }\textbf {\bibinfo {volume} {284}},\ \bibinfo {pages} {1907}
  (\bibinfo {year} {2011})}\BibitemShut {NoStop}%
\bibitem [{\citenamefont {Xu}\ \emph {et~al.}(2012)\citenamefont {Xu},
  \citenamefont {Jia}, \citenamefont {Hu}, \citenamefont {Duan}, \citenamefont
  {Guo},\ and\ \citenamefont {Ma}}]{xu12a}%
  \BibitemOpen
  \bibfield  {author} {\bibinfo {author} {\bibfnamefont {X.}~\bibnamefont
  {Xu}}, \bibinfo {author} {\bibfnamefont {F.}~\bibnamefont {Jia}}, \bibinfo
  {author} {\bibfnamefont {L.}~\bibnamefont {Hu}}, \bibinfo {author}
  {\bibfnamefont {Z.}~\bibnamefont {Duan}}, \bibinfo {author} {\bibfnamefont
  {Q.}~\bibnamefont {Guo}}, \ and\ \bibinfo {author} {\bibfnamefont
  {S.}~\bibnamefont {Ma}},\ }\href {doi:10.1080/09500340.2012.733435}
  {\bibfield  {journal} {\bibinfo  {journal} {J. Mod. Opt.}\ }\textbf {\bibinfo
  {volume} {59}},\ \bibinfo {pages} {1624} (\bibinfo {year}
  {2012})}\BibitemShut {NoStop}%
\bibitem [{\citenamefont {Mehringer}\ \emph {et~al.}(2018)\citenamefont
  {Mehringer}, \citenamefont {M\"ahrlein}, \citenamefont {von Zanthier},\ and\
  \citenamefont {Agarwal}}]{mehringer18a}%
  \BibitemOpen
  \bibfield  {author} {\bibinfo {author} {\bibfnamefont {T.}~\bibnamefont
  {Mehringer}}, \bibinfo {author} {\bibfnamefont {S.}~\bibnamefont
  {M\"ahrlein}}, \bibinfo {author} {\bibfnamefont {J.}~\bibnamefont {von
  Zanthier}}, \ and\ \bibinfo {author} {\bibfnamefont {G.~S.}\ \bibnamefont
  {Agarwal}},\ }\href {doi:10.1364/OL.43.002304} {\bibfield  {journal}
  {\bibinfo  {journal} {Opt. Lett.}\ }\textbf {\bibinfo {volume} {43}},\
  \bibinfo {pages} {2304} (\bibinfo {year} {2018})}\BibitemShut {NoStop}%
\bibitem [{\citenamefont {Blaizot}\ and\ \citenamefont
  {Ripka}(1985)}]{blaizot_book85a}%
  \BibitemOpen
  \bibfield  {author} {\bibinfo {author} {\bibfnamefont {J.-P.}\ \bibnamefont
  {Blaizot}}\ and\ \bibinfo {author} {\bibfnamefont {G.}~\bibnamefont
  {Ripka}},\ }\href@noop {} {\emph {\bibinfo {title} {Quantum Theory of Finite
  Systems}}}\ (\bibinfo  {publisher} {MIT Press, Cambridge, MA},\ \bibinfo
  {year} {1985})\BibitemShut {NoStop}%
\bibitem [{\citenamefont {Gerry}\ and\ \citenamefont
  {Knight}(2005)}]{gerry_book05a}%
  \BibitemOpen
  \bibfield  {author} {\bibinfo {author} {\bibfnamefont {C.~C.}\ \bibnamefont
  {Gerry}}\ and\ \bibinfo {author} {\bibfnamefont {P.~L.}\ \bibnamefont
  {Knight}},\ }\href@noop {} {\emph {\bibinfo {title} {Introductory Quantum
  Optics}}}\ (\bibinfo  {publisher} {Cambridge University Press, Cambridge},\
  \bibinfo {year} {2005})\BibitemShut {NoStop}%
\bibitem [{\citenamefont {{L\'opez Carre{\~n}o}}\ \emph
  {et~al.}(2018)\citenamefont {{L\'opez Carre{\~n}o}}, \citenamefont
  {Casalengua}, \citenamefont {del Valle},\ and\ \citenamefont
  {Laussy}}]{lopezcarreno18b}%
  \BibitemOpen
  \bibfield  {author} {\bibinfo {author} {\bibfnamefont {J.~C.}\ \bibnamefont
  {{L\'opez Carre{\~n}o}}}, \bibinfo {author} {\bibfnamefont {E.~Z.}\
  \bibnamefont {Casalengua}}, \bibinfo {author} {\bibfnamefont
  {E.}~\bibnamefont {del Valle}}, \ and\ \bibinfo {author} {\bibfnamefont
  {F.~P.}\ \bibnamefont {Laussy}},\ }\href {doi:10.1088/2058-9565/aacfbe}
  {\bibfield  {journal} {\bibinfo  {journal} {Quantum Science and Technology}\
  }\textbf {\bibinfo {volume} {3}},\ \bibinfo {pages} {045001} (\bibinfo {year}
  {2018})}\BibitemShut {NoStop}%
\bibitem [{\citenamefont {Glauber}(1963)}]{glauber63a}%
  \BibitemOpen
  \bibfield  {author} {\bibinfo {author} {\bibfnamefont {R.~J.}\ \bibnamefont
  {Glauber}},\ }\href {doi:10.1103/PhysRevLett.10.84} {\bibfield  {journal}
  {\bibinfo  {journal} {Phys. Rev. Lett.}\ }\textbf {\bibinfo {volume} {10}},\
  \bibinfo {pages} {84} (\bibinfo {year} {1963})}\BibitemShut {NoStop}%
\bibitem [{\citenamefont {Mandel}(1982)}]{mandel82a}%
  \BibitemOpen
  \bibfield  {author} {\bibinfo {author} {\bibfnamefont {L.}~\bibnamefont
  {Mandel}},\ }\href {doi:10.1103/PhysRevLett.49.136} {\bibfield  {journal}
  {\bibinfo  {journal} {Phys. Rev. Lett.}\ }\textbf {\bibinfo {volume} {49}},\
  \bibinfo {pages} {136} (\bibinfo {year} {1982})}\BibitemShut {NoStop}%
\bibitem [{\citenamefont {Mandel}(1979)}]{mandel79a}%
  \BibitemOpen
  \bibfield  {author} {\bibinfo {author} {\bibfnamefont {L.}~\bibnamefont
  {Mandel}},\ }\href {doi:10.1364/OL.4.000205} {\bibfield  {journal} {\bibinfo
  {journal} {Opt. Lett.}\ }\textbf {\bibinfo {volume} {4}},\ \bibinfo {pages}
  {205} (\bibinfo {year} {1979})}\BibitemShut {NoStop}%
\bibitem [{\citenamefont {Lemonde}\ \emph {et~al.}(2014)\citenamefont
  {Lemonde}, \citenamefont {Didier},\ and\ \citenamefont {Clerk}}]{lemonde14a}%
  \BibitemOpen
  \bibfield  {author} {\bibinfo {author} {\bibfnamefont {M.-A.}\ \bibnamefont
  {Lemonde}}, \bibinfo {author} {\bibfnamefont {N.}~\bibnamefont {Didier}}, \
  and\ \bibinfo {author} {\bibfnamefont {A.~A.}\ \bibnamefont {Clerk}},\ }\href
  {doi:10.1103/PhysRevA.90.063824} {\bibfield  {journal} {\bibinfo  {journal}
  {Phys. Rev. A}\ }\textbf {\bibinfo {volume} {90}},\ \bibinfo {pages} {063824}
  (\bibinfo {year} {2014})}\BibitemShut {NoStop}%
\bibitem [{\citenamefont {Liew}\ and\ \citenamefont {Savona}(2010)}]{liew10a}%
  \BibitemOpen
  \bibfield  {author} {\bibinfo {author} {\bibfnamefont {T.~C.~H.}\
  \bibnamefont {Liew}}\ and\ \bibinfo {author} {\bibfnamefont {V.}~\bibnamefont
  {Savona}},\ }\href {doi:10.1103/PhysRevLett.104.183601} {\bibfield  {journal}
  {\bibinfo  {journal} {Phys. Rev. Lett.}\ }\textbf {\bibinfo {volume} {104}},\
  \bibinfo {pages} {183601} (\bibinfo {year} {2010})}\BibitemShut {NoStop}%
\bibitem [{\citenamefont {Bamba}\ \emph {et~al.}(2011)\citenamefont {Bamba},
  \citenamefont {\Imamoglu}, \citenamefont {Carusotto},\ and\ \citenamefont
  {Ciuti}}]{bamba11a}%
  \BibitemOpen
  \bibfield  {author} {\bibinfo {author} {\bibfnamefont {M.}~\bibnamefont
  {Bamba}}, \bibinfo {author} {\bibfnamefont {A.}~\bibnamefont {\Imamoglu}},
  \bibinfo {author} {\bibfnamefont {I.}~\bibnamefont {Carusotto}}, \ and\
  \bibinfo {author} {\bibfnamefont {C.}~\bibnamefont {Ciuti}},\ }\href
  {doi:10.1103/PhysRevA.83.021802} {\bibfield  {journal} {\bibinfo  {journal}
  {Phys. Rev. A}\ }\textbf {\bibinfo {volume} {83}},\ \bibinfo {pages}
  {021802(R)} (\bibinfo {year} {2011})}\BibitemShut {NoStop}%
\bibitem [{\citenamefont {Flayac}\ and\ \citenamefont
  {Savona}(2017)}]{flayac17b}%
  \BibitemOpen
  \bibfield  {author} {\bibinfo {author} {\bibfnamefont {H.}~\bibnamefont
  {Flayac}}\ and\ \bibinfo {author} {\bibfnamefont {V.}~\bibnamefont
  {Savona}},\ }\href {doi:10.1103/PhysRevA.96.053810} {\bibfield  {journal}
  {\bibinfo  {journal} {Phys. Rev. A}\ }\textbf {\bibinfo {volume} {96}},\
  \bibinfo {pages} {053810} (\bibinfo {year} {2017})}\BibitemShut {NoStop}%
\bibitem [{\citenamefont {Ferretti}\ \emph {et~al.}(2010)\citenamefont
  {Ferretti}, \citenamefont {Andreani}, \citenamefont {T\"ureci},\ and\
  \citenamefont {Gerace}}]{ferretti10a}%
  \BibitemOpen
  \bibfield  {author} {\bibinfo {author} {\bibfnamefont {S.}~\bibnamefont
  {Ferretti}}, \bibinfo {author} {\bibfnamefont {L.~C.}\ \bibnamefont
  {Andreani}}, \bibinfo {author} {\bibfnamefont {H.~E.}\ \bibnamefont
  {T\"ureci}}, \ and\ \bibinfo {author} {\bibfnamefont {D.}~\bibnamefont
  {Gerace}},\ }\href {doi:10.1103/PhysRevA.82.013841} {\bibfield  {journal}
  {\bibinfo  {journal} {Phys. Rev. A}\ }\textbf {\bibinfo {volume} {82}},\
  \bibinfo {pages} {013841} (\bibinfo {year} {2010})}\BibitemShut {NoStop}%
\bibitem [{\citenamefont {Flayac}\ and\ \citenamefont
  {Savona}(2013)}]{flayac13a}%
  \BibitemOpen
  \bibfield  {author} {\bibinfo {author} {\bibfnamefont {H.}~\bibnamefont
  {Flayac}}\ and\ \bibinfo {author} {\bibfnamefont {V.}~\bibnamefont
  {Savona}},\ }\href {doi:10.1103/PhysRevA.88.033836} {\bibfield  {journal}
  {\bibinfo  {journal} {Phys. Rev. A}\ }\textbf {\bibinfo {volume} {88}},\
  \bibinfo {pages} {033836} (\bibinfo {year} {2013})}\BibitemShut {NoStop}%
\bibitem [{\citenamefont {Majumdar}\ and\ \citenamefont
  {Gerace}(2013)}]{majumdar13a}%
  \BibitemOpen
  \bibfield  {author} {\bibinfo {author} {\bibfnamefont {A.}~\bibnamefont
  {Majumdar}}\ and\ \bibinfo {author} {\bibfnamefont {D.}~\bibnamefont
  {Gerace}},\ }\href {doi:10.1103/PhysRevB.87.235319} {\bibfield  {journal}
  {\bibinfo  {journal} {Phys. Rev. B}\ }\textbf {\bibinfo {volume} {87}},\
  \bibinfo {pages} {235319} (\bibinfo {year} {2013})}\BibitemShut {NoStop}%
\bibitem [{\citenamefont {Gerace}\ and\ \citenamefont
  {Savona}(2014)}]{gerace14a}%
  \BibitemOpen
  \bibfield  {author} {\bibinfo {author} {\bibfnamefont {D.}~\bibnamefont
  {Gerace}}\ and\ \bibinfo {author} {\bibfnamefont {V.}~\bibnamefont
  {Savona}},\ }\href {doi:10.1103/PhysRevA.89.031803} {\bibfield  {journal}
  {\bibinfo  {journal} {Phys. Rev. A}\ }\textbf {\bibinfo {volume} {89}},\
  \bibinfo {pages} {031803(R)} (\bibinfo {year} {2014})}\BibitemShut {NoStop}%
\bibitem [{\citenamefont {Flayac}\ \emph {et~al.}(2015)\citenamefont {Flayac},
  \citenamefont {Gerace},\ and\ \citenamefont {Savona}}]{flayac15a}%
  \BibitemOpen
  \bibfield  {author} {\bibinfo {author} {\bibfnamefont {H.}~\bibnamefont
  {Flayac}}, \bibinfo {author} {\bibfnamefont {D.}~\bibnamefont {Gerace}}, \
  and\ \bibinfo {author} {\bibfnamefont {V.}~\bibnamefont {Savona}},\ }\href
  {doi:10.1038/srep11223} {\bibfield  {journal} {\bibinfo  {journal} {Sci.
  Rep.}\ }\textbf {\bibinfo {volume} {5}},\ \bibinfo {pages} {11223} (\bibinfo
  {year} {2015})}\BibitemShut {NoStop}%
\bibitem [{\citenamefont {Zhou}\ \emph {et~al.}(2015)\citenamefont {Zhou},
  \citenamefont {Shen},\ and\ \citenamefont {Yi}}]{zhou15a}%
  \BibitemOpen
  \bibfield  {author} {\bibinfo {author} {\bibfnamefont {Y.~H.}\ \bibnamefont
  {Zhou}}, \bibinfo {author} {\bibfnamefont {H.~Z.}\ \bibnamefont {Shen}}, \
  and\ \bibinfo {author} {\bibfnamefont {X.~X.}\ \bibnamefont {Yi}},\ }\href
  {doi:10.1103/PhysRevA.92.023838} {\bibfield  {journal} {\bibinfo  {journal}
  {Phys. Rev. A}\ }\textbf {\bibinfo {volume} {92}},\ \bibinfo {pages} {023838}
  (\bibinfo {year} {2015})}\BibitemShut {NoStop}%
\bibitem [{\citenamefont {Tang}\ \emph {et~al.}(2015)\citenamefont {Tang},
  \citenamefont {Geng},\ and\ \citenamefont {Xu}}]{tang15a}%
  \BibitemOpen
  \bibfield  {author} {\bibinfo {author} {\bibfnamefont {J.}~\bibnamefont
  {Tang}}, \bibinfo {author} {\bibfnamefont {W.}~\bibnamefont {Geng}}, \ and\
  \bibinfo {author} {\bibfnamefont {X.}~\bibnamefont {Xu}},\ }\href
  {doi:10.1038/srep09252} {\bibfield  {journal} {\bibinfo  {journal} {Sci.
  Rep.}\ }\textbf {\bibinfo {volume} {5}},\ \bibinfo {pages} {9252} (\bibinfo
  {year} {2015})}\BibitemShut {NoStop}%
\bibitem [{\citenamefont {Cheng}\ \emph {et~al.}(2017)\citenamefont {Cheng},
  \citenamefont {Ye},\ and\ \citenamefont {Yu}}]{cheng17a}%
  \BibitemOpen
  \bibfield  {author} {\bibinfo {author} {\bibfnamefont {X.}~\bibnamefont
  {Cheng}}, \bibinfo {author} {\bibfnamefont {H.}~\bibnamefont {Ye}}, \ and\
  \bibinfo {author} {\bibfnamefont {Z.}~\bibnamefont {Yu}},\ }\href
  {doi:10.1016/j.spmi.2017.03.002} {\bibfield  {journal} {\bibinfo  {journal}
  {Superlatt. Microstruct.}\ }\textbf {\bibinfo {volume} {105}},\ \bibinfo
  {pages} {81} (\bibinfo {year} {2017})}\BibitemShut {NoStop}%
\bibitem [{\citenamefont {Yu}\ and\ \citenamefont {Liu}(2017)}]{yu17a}%
  \BibitemOpen
  \bibfield  {author} {\bibinfo {author} {\bibfnamefont {Y.}~\bibnamefont
  {Yu}}\ and\ \bibinfo {author} {\bibfnamefont {H.-Y.}\ \bibnamefont {Liu}},\
  }\href {doi:10.1080/09500340.2017.1285067} {\bibfield  {journal} {\bibinfo
  {journal} {J. Mod. Opt.}\ }\textbf {\bibinfo {volume} {64}},\ \bibinfo
  {pages} {1342} (\bibinfo {year} {2017})}\BibitemShut {NoStop}%
\bibitem [{\citenamefont {Wang}\ \emph {et~al.}(2017)\citenamefont {Wang},
  \citenamefont {Shen}, \citenamefont {Sun}, \citenamefont {Wu}, \citenamefont
  {Chen},\ and\ \citenamefont {Xue}}]{wang17a}%
  \BibitemOpen
  \bibfield  {author} {\bibinfo {author} {\bibfnamefont {G.}~\bibnamefont
  {Wang}}, \bibinfo {author} {\bibfnamefont {H.~Z.}\ \bibnamefont {Shen}},
  \bibinfo {author} {\bibfnamefont {C.}~\bibnamefont {Sun}}, \bibinfo {author}
  {\bibfnamefont {C.}~\bibnamefont {Wu}}, \bibinfo {author} {\bibfnamefont
  {J.-L.}\ \bibnamefont {Chen}}, \ and\ \bibinfo {author} {\bibfnamefont
  {K.}~\bibnamefont {Xue}},\ }\href {doi:10.1080/09500340.2016.1249978}
  {\bibfield  {journal} {\bibinfo  {journal} {J. Mod. Opt.}\ }\textbf {\bibinfo
  {volume} {64}},\ \bibinfo {pages} {583} (\bibinfo {year} {2017})}\BibitemShut
  {NoStop}%
\bibitem [{\citenamefont {Snijders}\ \emph {et~al.}(2018)\citenamefont
  {Snijders}, \citenamefont {Frey}, \citenamefont {Norman}, \citenamefont
  {Flayac}, \citenamefont {Savona}, \citenamefont {Gossard}, \citenamefont
  {Bowers}, \citenamefont {van Exter}, \citenamefont {Bouwmeester},\ and\
  \citenamefont {Löffler}}]{snijders18a}%
  \BibitemOpen
  \bibfield  {author} {\bibinfo {author} {\bibfnamefont {H.}~\bibnamefont
  {Snijders}}, \bibinfo {author} {\bibfnamefont {J.}~\bibnamefont {Frey}},
  \bibinfo {author} {\bibfnamefont {J.}~\bibnamefont {Norman}}, \bibinfo
  {author} {\bibfnamefont {H.}~\bibnamefont {Flayac}}, \bibinfo {author}
  {\bibfnamefont {V.}~\bibnamefont {Savona}}, \bibinfo {author} {\bibfnamefont
  {A.}~\bibnamefont {Gossard}}, \bibinfo {author} {\bibfnamefont
  {J.}~\bibnamefont {Bowers}}, \bibinfo {author} {\bibfnamefont
  {M.}~\bibnamefont {van Exter}}, \bibinfo {author} {\bibfnamefont
  {D.}~\bibnamefont {Bouwmeester}}, \ and\ \bibinfo {author} {\bibfnamefont
  {W.}~\bibnamefont {Löffler}},\ }\href {doi:10.1103/PhysRevLett.121.043601}
  {\bibfield  {journal} {\bibinfo  {journal} {Phys. Rev. Lett.}\ }\textbf
  {\bibinfo {volume} {121}},\ \bibinfo {pages} {043601} (\bibinfo {year}
  {2018})}\BibitemShut {NoStop}%
\bibitem [{\citenamefont {Vaneph}\ \emph {et~al.}(2018)\citenamefont {Vaneph},
  \citenamefont {Morvan}, \citenamefont {Aiello}, \citenamefont {F\'echant},
  \citenamefont {Aprili}, \citenamefont {Gabelli},\ and\ \citenamefont
  {Est\`eve}}]{vaneph18a}%
  \BibitemOpen
  \bibfield  {author} {\bibinfo {author} {\bibfnamefont {C.}~\bibnamefont
  {Vaneph}}, \bibinfo {author} {\bibfnamefont {A.}~\bibnamefont {Morvan}},
  \bibinfo {author} {\bibfnamefont {G.}~\bibnamefont {Aiello}}, \bibinfo
  {author} {\bibfnamefont {M.}~\bibnamefont {F\'echant}}, \bibinfo {author}
  {\bibfnamefont {M.}~\bibnamefont {Aprili}}, \bibinfo {author} {\bibfnamefont
  {J.}~\bibnamefont {Gabelli}}, \ and\ \bibinfo {author} {\bibfnamefont
  {J.}~\bibnamefont {Est\`eve}},\ }\href {doi:10.1103/PhysRevLett.121.043602}
  {\bibfield  {journal} {\bibinfo  {journal} {Phys. Rev. Lett.}\ }\textbf
  {\bibinfo {volume} {121}},\ \bibinfo {pages} {043602} (\bibinfo {year}
  {2018})}\BibitemShut {NoStop}%
\bibitem [{\citenamefont {Shen}\ \emph {et~al.}(2018)\citenamefont {Shen},
  \citenamefont {Shang}, \citenamefont {Zhou},\ and\ \citenamefont
  {Yi}}]{shen18a}%
  \BibitemOpen
  \bibfield  {author} {\bibinfo {author} {\bibfnamefont {H.~Z.}\ \bibnamefont
  {Shen}}, \bibinfo {author} {\bibfnamefont {C.}~\bibnamefont {Shang}},
  \bibinfo {author} {\bibfnamefont {Y.~H.}\ \bibnamefont {Zhou}}, \ and\
  \bibinfo {author} {\bibfnamefont {X.~X.}\ \bibnamefont {Yi}},\ }\href
  {doi:10.1103/PhysRevA.98.023856} {\bibfield  {journal} {\bibinfo  {journal}
  {Phys. Rev. Lett.}\ }\textbf {\bibinfo {volume} {98}},\ \bibinfo {pages}
  {023856} (\bibinfo {year} {2018})}\BibitemShut {NoStop}%
\bibitem [{\citenamefont {Trivedi}\ \emph {et~al.}(2019)\citenamefont
  {Trivedi}, \citenamefont {Radulaski}, \citenamefont {Fischer}, \citenamefont
  {Fan},\ and\ \citenamefont {\Vuckovic}}]{trivedi19a}%
  \BibitemOpen
  \bibfield  {author} {\bibinfo {author} {\bibfnamefont {R.}~\bibnamefont
  {Trivedi}}, \bibinfo {author} {\bibfnamefont {M.}~\bibnamefont {Radulaski}},
  \bibinfo {author} {\bibfnamefont {K.~A.}\ \bibnamefont {Fischer}}, \bibinfo
  {author} {\bibfnamefont {S.}~\bibnamefont {Fan}}, \ and\ \bibinfo {author}
  {\bibfnamefont {J.}~\bibnamefont {\Vuckovic}},\ }\href
  {doi:10.1103/PhysRevLett.122.243602} {\bibfield  {journal} {\bibinfo
  {journal} {Phys. Rev. Lett.}\ }\textbf {\bibinfo {volume} {122}},\ \bibinfo
  {pages} {243602} (\bibinfo {year} {2019})}\BibitemShut {NoStop}%
\bibitem [{\citenamefont {Birnbaum}\ \emph {et~al.}(2005)\citenamefont
  {Birnbaum}, \citenamefont {Boca}, \citenamefont {Miller}, \citenamefont
  {Boozer}, \citenamefont {Northup},\ and\ \citenamefont
  {Kimble}}]{birnbaum05a}%
  \BibitemOpen
  \bibfield  {author} {\bibinfo {author} {\bibfnamefont {K.}~\bibnamefont
  {Birnbaum}}, \bibinfo {author} {\bibfnamefont {A.}~\bibnamefont {Boca}},
  \bibinfo {author} {\bibfnamefont {R.}~\bibnamefont {Miller}}, \bibinfo
  {author} {\bibfnamefont {A.}~\bibnamefont {Boozer}}, \bibinfo {author}
  {\bibfnamefont {T.}~\bibnamefont {Northup}}, \ and\ \bibinfo {author}
  {\bibfnamefont {H.}~\bibnamefont {Kimble}},\ }\href {doi:10.1038/nature03804}
  {\bibfield  {journal} {\bibinfo  {journal} {Nature}\ }\textbf {\bibinfo
  {volume} {436}},\ \bibinfo {pages} {87} (\bibinfo {year} {2005})}\BibitemShut
  {NoStop}%
\bibitem [{\citenamefont {Verger}\ \emph {et~al.}(2006)\citenamefont {Verger},
  \citenamefont {Ciuti},\ and\ \citenamefont {Carusotto}}]{verger06a}%
  \BibitemOpen
  \bibfield  {author} {\bibinfo {author} {\bibfnamefont {A.}~\bibnamefont
  {Verger}}, \bibinfo {author} {\bibfnamefont {C.}~\bibnamefont {Ciuti}}, \
  and\ \bibinfo {author} {\bibfnamefont {I.}~\bibnamefont {Carusotto}},\ }\href
  {10.1103/PhysRevB.73.193306} {\bibfield  {journal} {\bibinfo  {journal}
  {Phys. Rev. B}\ }\textbf {\bibinfo {volume} {73}},\ \bibinfo {pages} {193306}
  (\bibinfo {year} {2006})}\BibitemShut {NoStop}%
\bibitem [{\citenamefont {Faraon}\ \emph {et~al.}(2008)\citenamefont {Faraon},
  \citenamefont {Fushman}, \citenamefont {Englund}, \citenamefont {Stoltz},
  \citenamefont {Petroff},\ and\ \citenamefont {\Vuckovic}}]{faraon08a}%
  \BibitemOpen
  \bibfield  {author} {\bibinfo {author} {\bibfnamefont {A.}~\bibnamefont
  {Faraon}}, \bibinfo {author} {\bibfnamefont {I.}~\bibnamefont {Fushman}},
  \bibinfo {author} {\bibfnamefont {D.}~\bibnamefont {Englund}}, \bibinfo
  {author} {\bibfnamefont {N.}~\bibnamefont {Stoltz}}, \bibinfo {author}
  {\bibfnamefont {P.}~\bibnamefont {Petroff}}, \ and\ \bibinfo {author}
  {\bibfnamefont {J.}~\bibnamefont {\Vuckovic}},\ }\href
  {doi:10.1038/nphys1078} {\bibfield  {journal} {\bibinfo  {journal} {Nature
  Phys.}\ }\textbf {\bibinfo {volume} {4}},\ \bibinfo {pages} {859} (\bibinfo
  {year} {2008})}\BibitemShut {NoStop}%
\bibitem [{\citenamefont {Faraon}\ \emph {et~al.}(2010)\citenamefont {Faraon},
  \citenamefont {Majumdar},\ and\ \citenamefont {\Vuckovic}}]{faraon10a}%
  \BibitemOpen
  \bibfield  {author} {\bibinfo {author} {\bibfnamefont {A.}~\bibnamefont
  {Faraon}}, \bibinfo {author} {\bibfnamefont {A.}~\bibnamefont {Majumdar}}, \
  and\ \bibinfo {author} {\bibfnamefont {J.}~\bibnamefont {\Vuckovic}},\ }\href
  {doi:10.1103/PhysRevA.81.033838} {\bibfield  {journal} {\bibinfo  {journal}
  {Phys. Rev. A}\ }\textbf {\bibinfo {volume} {81}},\ \bibinfo {pages} {033838}
  (\bibinfo {year} {2010})}\BibitemShut {NoStop}%
\bibitem [{\citenamefont {Hoffman}\ \emph {et~al.}(2011)\citenamefont
  {Hoffman}, \citenamefont {Srinivasan}, \citenamefont {Schmidt}, \citenamefont
  {Spietz}, \citenamefont {Aumentado}, \citenamefont {T\"ureci},\ and\
  \citenamefont {Houck}}]{hoffman11a}%
  \BibitemOpen
  \bibfield  {author} {\bibinfo {author} {\bibfnamefont {A.~J.}\ \bibnamefont
  {Hoffman}}, \bibinfo {author} {\bibfnamefont {S.~J.}\ \bibnamefont
  {Srinivasan}}, \bibinfo {author} {\bibfnamefont {S.}~\bibnamefont {Schmidt}},
  \bibinfo {author} {\bibfnamefont {L.}~\bibnamefont {Spietz}}, \bibinfo
  {author} {\bibfnamefont {J.}~\bibnamefont {Aumentado}}, \bibinfo {author}
  {\bibfnamefont {H.~E.}\ \bibnamefont {T\"ureci}}, \ and\ \bibinfo {author}
  {\bibfnamefont {A.~A.}\ \bibnamefont {Houck}},\ }\href
  {doi:10.1103/PhysRevLett.107.053602} {\bibfield  {journal} {\bibinfo
  {journal} {Phys. Rev. Lett.}\ }\textbf {\bibinfo {volume} {107}},\ \bibinfo
  {pages} {053602} (\bibinfo {year} {2011})}\BibitemShut {NoStop}%
\bibitem [{\citenamefont {Rabl}(2011)}]{rabl11a}%
  \BibitemOpen
  \bibfield  {author} {\bibinfo {author} {\bibfnamefont {P.}~\bibnamefont
  {Rabl}},\ }\href {doi:10.1103/PhysRevLett.107.063601} {\bibfield  {journal}
  {\bibinfo  {journal} {Phys. Rev. Lett.}\ }\textbf {\bibinfo {volume} {107}},\
  \bibinfo {pages} {063601} (\bibinfo {year} {2011})}\BibitemShut {NoStop}%
\bibitem [{\citenamefont {Lang}\ \emph {et~al.}(2011)\citenamefont {Lang},
  \citenamefont {Bozyigit}, \citenamefont {Eichler}, \citenamefont {Steffen},
  \citenamefont {Fink}, \citenamefont {{Abdumalikov Jr.}}, \citenamefont
  {Baur}, \citenamefont {Filipp}, \citenamefont {da~Silva}, \citenamefont
  {Blais},\ and\ \citenamefont {Wallraff}}]{lang11a}%
  \BibitemOpen
  \bibfield  {author} {\bibinfo {author} {\bibfnamefont {C.}~\bibnamefont
  {Lang}}, \bibinfo {author} {\bibfnamefont {D.}~\bibnamefont {Bozyigit}},
  \bibinfo {author} {\bibfnamefont {C.}~\bibnamefont {Eichler}}, \bibinfo
  {author} {\bibfnamefont {L.}~\bibnamefont {Steffen}}, \bibinfo {author}
  {\bibfnamefont {J.~M.}\ \bibnamefont {Fink}}, \bibinfo {author}
  {\bibfnamefont {A.~A.}\ \bibnamefont {{Abdumalikov Jr.}}}, \bibinfo {author}
  {\bibfnamefont {M.}~\bibnamefont {Baur}}, \bibinfo {author} {\bibfnamefont
  {S.}~\bibnamefont {Filipp}}, \bibinfo {author} {\bibfnamefont {M.~P.}\
  \bibnamefont {da~Silva}}, \bibinfo {author} {\bibfnamefont {A.}~\bibnamefont
  {Blais}}, \ and\ \bibinfo {author} {\bibfnamefont {A.}~\bibnamefont
  {Wallraff}},\ }\href {doi:10.1103/PhysRevLett.106.243601} {\bibfield
  {journal} {\bibinfo  {journal} {Phys. Rev. Lett.}\ }\textbf {\bibinfo
  {volume} {106}},\ \bibinfo {pages} {243601} (\bibinfo {year}
  {2011})}\BibitemShut {NoStop}%
\bibitem [{\citenamefont {M\"uller}\ \emph {et~al.}(2015)\citenamefont
  {M\"uller}, \citenamefont {Rundquist}, \citenamefont {Fischer}, \citenamefont
  {Sarmiento}, \citenamefont {Lagoudakis}, \citenamefont {Kelaita},
  \citenamefont {{Mu\~noz}}, \citenamefont {del Valle}, \citenamefont
  {Laussy},\ and\ \citenamefont {\Vuckovic}}]{muller15a}%
  \BibitemOpen
  \bibfield  {author} {\bibinfo {author} {\bibfnamefont {K.}~\bibnamefont
  {M\"uller}}, \bibinfo {author} {\bibfnamefont {A.}~\bibnamefont {Rundquist}},
  \bibinfo {author} {\bibfnamefont {K.~A.}\ \bibnamefont {Fischer}}, \bibinfo
  {author} {\bibfnamefont {T.}~\bibnamefont {Sarmiento}}, \bibinfo {author}
  {\bibfnamefont {K.~G.}\ \bibnamefont {Lagoudakis}}, \bibinfo {author}
  {\bibfnamefont {Y.~A.}\ \bibnamefont {Kelaita}}, \bibinfo {author}
  {\bibfnamefont {C.~S.}\ \bibnamefont {{Mu\~noz}}}, \bibinfo {author}
  {\bibfnamefont {E.}~\bibnamefont {del Valle}}, \bibinfo {author}
  {\bibfnamefont {F.~P.}\ \bibnamefont {Laussy}}, \ and\ \bibinfo {author}
  {\bibfnamefont {J.}~\bibnamefont {\Vuckovic}},\ }\href
  {doi:10.1103/PhysRevLett.114.233601} {\bibfield  {journal} {\bibinfo
  {journal} {Phys. Rev. Lett.}\ }\textbf {\bibinfo {volume} {114}},\ \bibinfo
  {pages} {233601} (\bibinfo {year} {2015})}\BibitemShut {NoStop}%
\bibitem [{\citenamefont {Radulaski}\ \emph {et~al.}(2017)\citenamefont
  {Radulaski}, \citenamefont {Fischer}, \citenamefont {Lagoudakis},
  \citenamefont {Zhang},\ and\ \citenamefont {\Vuckovic}}]{radulaski17a}%
  \BibitemOpen
  \bibfield  {author} {\bibinfo {author} {\bibfnamefont {M.}~\bibnamefont
  {Radulaski}}, \bibinfo {author} {\bibfnamefont {K.~A.}\ \bibnamefont
  {Fischer}}, \bibinfo {author} {\bibfnamefont {K.~G.}\ \bibnamefont
  {Lagoudakis}}, \bibinfo {author} {\bibfnamefont {J.~L.}\ \bibnamefont
  {Zhang}}, \ and\ \bibinfo {author} {\bibfnamefont {J.}~\bibnamefont
  {\Vuckovic}},\ }\href {doi:10.1103/PhysRevA.96.011801} {\bibfield  {journal}
  {\bibinfo  {journal} {Phys. Rev. A}\ }\textbf {\bibinfo {volume} {96}},\
  \bibinfo {pages} {011801(R)} (\bibinfo {year} {2017})}\BibitemShut {NoStop}%
\bibitem [{\citenamefont {Deng}\ \emph {et~al.}(2017)\citenamefont {Deng},
  \citenamefont {Li},\ and\ \citenamefont {Qin}}]{deng17a}%
  \BibitemOpen
  \bibfield  {author} {\bibinfo {author} {\bibfnamefont {W.-W.}\ \bibnamefont
  {Deng}}, \bibinfo {author} {\bibfnamefont {G.-X.}\ \bibnamefont {Li}}, \ and\
  \bibinfo {author} {\bibfnamefont {H.}~\bibnamefont {Qin}},\ }\href
  {doi:10.1364/OE.25.006767} {\bibfield  {journal} {\bibinfo  {journal} {Opt.
  Express}\ }\textbf {\bibinfo {volume} {6}},\ \bibinfo {pages} {6767}
  (\bibinfo {year} {2017})}\BibitemShut {NoStop}%
\bibitem [{\citenamefont {Sarma}\ and\ \citenamefont {Sarma}(2017)}]{sarma17a}%
  \BibitemOpen
  \bibfield  {author} {\bibinfo {author} {\bibfnamefont {B.}~\bibnamefont
  {Sarma}}\ and\ \bibinfo {author} {\bibfnamefont {A.~K.}\ \bibnamefont
  {Sarma}},\ }\href {doi:10.1103/PhysRevA.96.053827} {\bibfield  {journal}
  {\bibinfo  {journal} {Phys. Rev. A}\ }\textbf {\bibinfo {volume} {96}},\
  \bibinfo {pages} {053827} (\bibinfo {year} {2017})}\BibitemShut {NoStop}%
\bibitem [{\citenamefont {Ghosh}\ and\ \citenamefont {Liew}(2019)}]{ghosh19a}%
  \BibitemOpen
  \bibfield  {author} {\bibinfo {author} {\bibfnamefont {S.}~\bibnamefont
  {Ghosh}}\ and\ \bibinfo {author} {\bibfnamefont {T.~C.}\ \bibnamefont
  {Liew}},\ }\href {doi:10.1103/PhysRevLett.123.013602} {\bibfield  {journal}
  {\bibinfo  {journal} {Phys. Rev. Lett.}\ }\textbf {\bibinfo {volume} {123}},\
  \bibinfo {pages} {013602} (\bibinfo {year} {2019})}\BibitemShut {NoStop}%
\bibitem [{\citenamefont {Delteil}\ \emph {et~al.}(2019)\citenamefont
  {Delteil}, \citenamefont {Fink}, \citenamefont {Schade}, \citenamefont
  {{H\''ofling}}, \citenamefont {Schneider},\ and\ \citenamefont
  {\Imamoglu}}]{delteil19a}%
  \BibitemOpen
  \bibfield  {author} {\bibinfo {author} {\bibfnamefont {A.}~\bibnamefont
  {Delteil}}, \bibinfo {author} {\bibfnamefont {T.}~\bibnamefont {Fink}},
  \bibinfo {author} {\bibfnamefont {A.}~\bibnamefont {Schade}}, \bibinfo
  {author} {\bibfnamefont {S.}~\bibnamefont {{H\''ofling}}}, \bibinfo {author}
  {\bibfnamefont {C.}~\bibnamefont {Schneider}}, \ and\ \bibinfo {author}
  {\bibfnamefont {A.}~\bibnamefont {\Imamoglu}},\ }\href {\doibase
  doi:10.1038/s41563-019-0282-y} {\bibfield  {journal} {\bibinfo  {journal}
  {Nature Mater.}\ }\textbf {\bibinfo {volume} {18}},\ \bibinfo {pages} {219}
  (\bibinfo {year} {2019})}\BibitemShut {NoStop}%
\bibitem [{\citenamefont {Phillips}\ \emph {et~al.}(2020)\citenamefont
  {Phillips}, \citenamefont {Brash}, \citenamefont {McCutcheon}, \citenamefont
  {Iles-Smith}, \citenamefont {Clarke}, \citenamefont {Royall}, \citenamefont
  {Skolnick},\ and\ \citenamefont {Fox}}]{arXiv_phillips20a}%
  \BibitemOpen
  \bibfield  {author} {\bibinfo {author} {\bibfnamefont {C.~L.}\ \bibnamefont
  {Phillips}}, \bibinfo {author} {\bibfnamefont {A.~J.}\ \bibnamefont {Brash}},
  \bibinfo {author} {\bibfnamefont {D.~P.~S.}\ \bibnamefont {McCutcheon}},
  \bibinfo {author} {\bibfnamefont {J.}~\bibnamefont {Iles-Smith}}, \bibinfo
  {author} {\bibfnamefont {E.}~\bibnamefont {Clarke}}, \bibinfo {author}
  {\bibfnamefont {B.}~\bibnamefont {Royall}}, \bibinfo {author} {\bibfnamefont
  {M.~S.}\ \bibnamefont {Skolnick}}, \ and\ \bibinfo {author} {\bibfnamefont
  {A.~M.}\ \bibnamefont {Fox}},\ }\href@noop {} {\bibfield  {journal} {\bibinfo
   {journal} {arXiv:2002.08192}\ } (\bibinfo {year} {2020})}\BibitemShut
  {NoStop}%
\bibitem [{\citenamefont {Hanschke}\ \emph {et~al.}(2020)\citenamefont
  {Hanschke}, \citenamefont {Schweickert}, \citenamefont {{L\'opez
  Carre\~{n}o}}, \citenamefont {Sch\"oll}, \citenamefont {Zeuner},
  \citenamefont {Lettner}, \citenamefont {{Zubizarreta Casalengua}},
  \citenamefont {Reindl}, \citenamefont {{Covre da Silva}}, \citenamefont
  {Trotta}, \citenamefont {Finley}, \citenamefont {Rastelli}, \citenamefont
  {del Valle}, \citenamefont {Laussy}, \citenamefont {Zwiller}, \citenamefont
  {M\"uller},\ and\ \citenamefont {J\"ons}}]{arXiv_hanschke20a}%
  \BibitemOpen
  \bibfield  {author} {\bibinfo {author} {\bibfnamefont {L.}~\bibnamefont
  {Hanschke}}, \bibinfo {author} {\bibfnamefont {L.}~\bibnamefont
  {Schweickert}}, \bibinfo {author} {\bibfnamefont {J.~C.}\ \bibnamefont
  {{L\'opez Carre\~{n}o}}}, \bibinfo {author} {\bibfnamefont {E.}~\bibnamefont
  {Sch\"oll}}, \bibinfo {author} {\bibfnamefont {K.~D.}\ \bibnamefont
  {Zeuner}}, \bibinfo {author} {\bibfnamefont {T.}~\bibnamefont {Lettner}},
  \bibinfo {author} {\bibfnamefont {E.}~\bibnamefont {{Zubizarreta
  Casalengua}}}, \bibinfo {author} {\bibfnamefont {M.}~\bibnamefont {Reindl}},
  \bibinfo {author} {\bibfnamefont {S.~F.}\ \bibnamefont {{Covre da Silva}}},
  \bibinfo {author} {\bibfnamefont {R.}~\bibnamefont {Trotta}}, \bibinfo
  {author} {\bibfnamefont {J.~J.}\ \bibnamefont {Finley}}, \bibinfo {author}
  {\bibfnamefont {A.}~\bibnamefont {Rastelli}}, \bibinfo {author}
  {\bibfnamefont {E.}~\bibnamefont {del Valle}}, \bibinfo {author}
  {\bibfnamefont {F.~P.}\ \bibnamefont {Laussy}}, \bibinfo {author}
  {\bibfnamefont {V.}~\bibnamefont {Zwiller}}, \bibinfo {author} {\bibfnamefont
  {K.}~\bibnamefont {M\"uller}}, \ and\ \bibinfo {author} {\bibfnamefont
  {K.~D.}\ \bibnamefont {J\"ons}},\ }\href@noop {} {\bibfield  {journal}
  {\bibinfo  {journal} {arXiv:2005.11800}\ } (\bibinfo {year}
  {2020})}\BibitemShut {NoStop}%
\bibitem [{\citenamefont {Visser}\ and\ \citenamefont
  {Nienhuis}(1995)}]{visser95a}%
  \BibitemOpen
  \bibfield  {author} {\bibinfo {author} {\bibfnamefont {P.~M.}\ \bibnamefont
  {Visser}}\ and\ \bibinfo {author} {\bibfnamefont {G.}~\bibnamefont
  {Nienhuis}},\ }\href {doi:10.1103/PhysRevA.52.4727} {\bibfield  {journal}
  {\bibinfo  {journal} {Phys. Rev. A}\ }\textbf {\bibinfo {volume} {52}},\
  \bibinfo {pages} {4727} (\bibinfo {year} {1995})}\BibitemShut {NoStop}%
\bibitem [{\citenamefont {{L\'opez Carre{\~n}o}}\ \emph
  {et~al.}(2019)\citenamefont {{L\'opez Carre{\~n}o}}, \citenamefont
  {Casalengua}, \citenamefont {Laussy},\ and\ \citenamefont {del
  Valle}}]{lopezcarreno19a}%
  \BibitemOpen
  \bibfield  {author} {\bibinfo {author} {\bibfnamefont {J.~C.}\ \bibnamefont
  {{L\'opez Carre{\~n}o}}}, \bibinfo {author} {\bibfnamefont {E.~Z.}\
  \bibnamefont {Casalengua}}, \bibinfo {author} {\bibfnamefont {F.~P.}\
  \bibnamefont {Laussy}}, \ and\ \bibinfo {author} {\bibfnamefont
  {E.}~\bibnamefont {del Valle}},\ }\href {doi:10.1088/1361-6455/aaf68d}
  {\bibfield  {journal} {\bibinfo  {journal} {J. Phys. B.: At. Mol. Opt.
  Phys.}\ }\textbf {\bibinfo {volume} {52}},\ \bibinfo {pages} {035504}
  (\bibinfo {year} {2019})}\BibitemShut {NoStop}%
\bibitem [{\citenamefont {Jaynes}\ and\ \citenamefont
  {Cummings}(1963)}]{jaynes63a}%
  \BibitemOpen
  \bibfield  {author} {\bibinfo {author} {\bibfnamefont {E.}~\bibnamefont
  {Jaynes}}\ and\ \bibinfo {author} {\bibfnamefont {F.}~\bibnamefont
  {Cummings}},\ }\href {doi:10.1109/PROC.1963.1664} {\bibfield  {journal}
  {\bibinfo  {journal} {Proc. IEEE}\ }\textbf {\bibinfo {volume} {51}},\
  \bibinfo {pages} {89} (\bibinfo {year} {1963})}\BibitemShut {NoStop}%
\bibitem [{\citenamefont {Shore}\ and\ \citenamefont
  {Knight}(1993)}]{shore93a}%
  \BibitemOpen
  \bibfield  {author} {\bibinfo {author} {\bibfnamefont {B.~W.}\ \bibnamefont
  {Shore}}\ and\ \bibinfo {author} {\bibfnamefont {P.~L.}\ \bibnamefont
  {Knight}},\ }\href {doi:10.1080/09500349314551321} {\bibfield  {journal}
  {\bibinfo  {journal} {J. Mod. Opt.}\ }\textbf {\bibinfo {volume} {40}},\
  \bibinfo {pages} {1195} (\bibinfo {year} {1993})}\BibitemShut {NoStop}%
\bibitem [{\citenamefont {{Zubizarreta Casalengua}}\ \emph
  {et~al.}(2018)\citenamefont {{Zubizarreta Casalengua}}, \citenamefont
  {{L\'{o}pez Carre\~{n}o}}, \citenamefont {Laussy},\ and\ \citenamefont {del
  Valle}}]{wolfram_casalengua18a}%
  \BibitemOpen
  \bibfield  {author} {\bibinfo {author} {\bibfnamefont {E.}~\bibnamefont
  {{Zubizarreta Casalengua}}}, \bibinfo {author} {\bibfnamefont {J.~C.}\
  \bibnamefont {{L\'{o}pez Carre\~{n}o}}}, \bibinfo {author} {\bibfnamefont
  {F.~P.}\ \bibnamefont {Laussy}}, \ and\ \bibinfo {author} {\bibfnamefont
  {E.}~\bibnamefont {del Valle}},\ }\href@noop {} {\enquote {\bibinfo {title}
  {Polariton and jaynes-cummings blockade},}\ }\bibinfo {howpublished}
  {\url{http://demonstrations.wolfram.com/PolaritonAndJaynesCummingsBlockade/}}
  (\bibinfo {year} {2018})\BibitemShut {NoStop}%
\bibitem [{\citenamefont {del Valle}\ \emph {et~al.}(2009)\citenamefont {del
  Valle}, \citenamefont {Laussy},\ and\ \citenamefont {Tejedor}}]{delvalle09a}%
  \BibitemOpen
  \bibfield  {author} {\bibinfo {author} {\bibfnamefont {E.}~\bibnamefont {del
  Valle}}, \bibinfo {author} {\bibfnamefont {F.~P.}\ \bibnamefont {Laussy}}, \
  and\ \bibinfo {author} {\bibfnamefont {C.}~\bibnamefont {Tejedor}},\ }\href
  {doi:10.1103/PhysRevB.79.235326} {\bibfield  {journal} {\bibinfo  {journal}
  {Phys. Rev. B}\ }\textbf {\bibinfo {volume} {79}},\ \bibinfo {pages} {235326}
  (\bibinfo {year} {2009})}\BibitemShut {NoStop}%
\bibitem [{\citenamefont {Laussy}\ \emph {et~al.}(2012)\citenamefont {Laussy},
  \citenamefont {del Valle}, \citenamefont {Schrapp}, \citenamefont {Laucht},\
  and\ \citenamefont {Finley}}]{laussy12e}%
  \BibitemOpen
  \bibfield  {author} {\bibinfo {author} {\bibfnamefont {F.~P.}\ \bibnamefont
  {Laussy}}, \bibinfo {author} {\bibfnamefont {E.}~\bibnamefont {del Valle}},
  \bibinfo {author} {\bibfnamefont {M.}~\bibnamefont {Schrapp}}, \bibinfo
  {author} {\bibfnamefont {A.}~\bibnamefont {Laucht}}, \ and\ \bibinfo {author}
  {\bibfnamefont {J.~J.}\ \bibnamefont {Finley}},\ }\href
  {doi:10.1117/1.jnp.6.061803} {\bibfield  {journal} {\bibinfo  {journal} {J.
  Nanophoton.}\ }\textbf {\bibinfo {volume} {6}},\ \bibinfo {pages} {061803}
  (\bibinfo {year} {2012})}\BibitemShut {NoStop}%
\bibitem [{\citenamefont {Xu}\ and\ \citenamefont {Li}(2014)}]{xu14a}%
  \BibitemOpen
  \bibfield  {author} {\bibinfo {author} {\bibfnamefont {X.}~\bibnamefont
  {Xu}}\ and\ \bibinfo {author} {\bibfnamefont {Y.}~\bibnamefont {Li}},\ }\href
  {doi:10.1103/PhysRevA.90.043822} {\bibfield  {journal} {\bibinfo  {journal}
  {Phys. Rev. A}\ }\textbf {\bibinfo {volume} {90}},\ \bibinfo {pages} {043822}
  (\bibinfo {year} {2014})}\BibitemShut {NoStop}%
\bibitem [{\citenamefont {Sanvitto}\ \emph {et~al.}(2019)\citenamefont
  {Sanvitto}, \citenamefont {Laussy},\ and\ \citenamefont
  {Gerace}}]{sanvitto19a}%
  \BibitemOpen
  \bibfield  {author} {\bibinfo {author} {\bibfnamefont {D.}~\bibnamefont
  {Sanvitto}}, \bibinfo {author} {\bibfnamefont {F.}~\bibnamefont {Laussy}}, \
  and\ \bibinfo {author} {\bibfnamefont {D.}~\bibnamefont {Gerace}},\ }\href
  {doi:10.1038/s41563-019-0298-3} {\bibfield  {journal} {\bibinfo  {journal}
  {Nature Mater.}\ }\textbf {\bibinfo {volume} {18}},\ \bibinfo {pages} {200}
  (\bibinfo {year} {2019})}\BibitemShut {NoStop}%
\bibitem [{\citenamefont {Mu{\~n}oz-Matutano}\ \emph
  {et~al.}(2019)\citenamefont {Mu{\~n}oz-Matutano}, \citenamefont {Wood},
  \citenamefont {Johnson}, \citenamefont {Asensio}, \citenamefont {Baragiola},
  \citenamefont {Reinhard}, \citenamefont {Lemaitre}, \citenamefont {Bloch},
  \citenamefont {Amo}, \citenamefont {Besga}, \citenamefont {Richard},\ and\
  \citenamefont {Volz}}]{munozmatutano19a}%
  \BibitemOpen
  \bibfield  {author} {\bibinfo {author} {\bibfnamefont {G.}~\bibnamefont
  {Mu{\~n}oz-Matutano}}, \bibinfo {author} {\bibfnamefont {A.}~\bibnamefont
  {Wood}}, \bibinfo {author} {\bibfnamefont {M.}~\bibnamefont {Johnson}},
  \bibinfo {author} {\bibfnamefont {X.~V.}\ \bibnamefont {Asensio}}, \bibinfo
  {author} {\bibfnamefont {B.}~\bibnamefont {Baragiola}}, \bibinfo {author}
  {\bibfnamefont {A.}~\bibnamefont {Reinhard}}, \bibinfo {author}
  {\bibfnamefont {A.}~\bibnamefont {Lemaitre}}, \bibinfo {author}
  {\bibfnamefont {J.}~\bibnamefont {Bloch}}, \bibinfo {author} {\bibfnamefont
  {A.}~\bibnamefont {Amo}}, \bibinfo {author} {\bibfnamefont {B.}~\bibnamefont
  {Besga}}, \bibinfo {author} {\bibfnamefont {M.}~\bibnamefont {Richard}}, \
  and\ \bibinfo {author} {\bibfnamefont {T.}~\bibnamefont {Volz}},\ }\href
  {\doibase doi:10.1038/s41563-019-0281-z} {\bibfield  {journal} {\bibinfo
  {journal} {Nature Mater.}\ }\textbf {\bibinfo {volume} {18}},\ \bibinfo
  {pages} {213} (\bibinfo {year} {2019})}\BibitemShut {NoStop}%
\bibitem [{\citenamefont {Xu}\ and\ \citenamefont {Li}(2013)}]{xu13a}%
  \BibitemOpen
  \bibfield  {author} {\bibinfo {author} {\bibfnamefont {X.-W.}\ \bibnamefont
  {Xu}}\ and\ \bibinfo {author} {\bibfnamefont {Y.-J.}\ \bibnamefont {Li}},\
  }\href {doi:10.1088/0953-4075/46/3/035502} {\bibfield  {journal} {\bibinfo
  {journal} {J. Phys. B.: At. Mol. Opt. Phys.}\ }\textbf {\bibinfo {volume}
  {46}},\ \bibinfo {pages} {035502} (\bibinfo {year} {2013})}\BibitemShut
  {NoStop}%
\bibitem [{\citenamefont {Shen}\ \emph
  {et~al.}(2015{\natexlab{a}})\citenamefont {Shen}, \citenamefont {Zhou},\ and\
  \citenamefont {Yi}}]{shen15a}%
  \BibitemOpen
  \bibfield  {author} {\bibinfo {author} {\bibfnamefont {H.~Z.}\ \bibnamefont
  {Shen}}, \bibinfo {author} {\bibfnamefont {Y.~H.}\ \bibnamefont {Zhou}}, \
  and\ \bibinfo {author} {\bibfnamefont {X.~X.}\ \bibnamefont {Yi}},\ }\href
  {doi:10.1103/PhysRevA.91.063808} {\bibfield  {journal} {\bibinfo  {journal}
  {Phys. Rev. A}\ }\textbf {\bibinfo {volume} {91}},\ \bibinfo {pages} {063808}
  (\bibinfo {year} {2015}{\natexlab{a}})}\BibitemShut {NoStop}%
\bibitem [{\citenamefont {Shen}\ \emph
  {et~al.}(2015{\natexlab{b}})\citenamefont {Shen}, \citenamefont {Zhou},
  \citenamefont {Liu}, \citenamefont {Wang},\ and\ \citenamefont
  {Yi}}]{shen15b}%
  \BibitemOpen
  \bibfield  {author} {\bibinfo {author} {\bibfnamefont {H.~Z.}\ \bibnamefont
  {Shen}}, \bibinfo {author} {\bibfnamefont {Y.~H.}\ \bibnamefont {Zhou}},
  \bibinfo {author} {\bibfnamefont {H.~D.}\ \bibnamefont {Liu}}, \bibinfo
  {author} {\bibfnamefont {G.~C.}\ \bibnamefont {Wang}}, \ and\ \bibinfo
  {author} {\bibfnamefont {X.~X.}\ \bibnamefont {Yi}},\ }\href
  {doi:10.1364/OE.23.032835} {\bibfield  {journal} {\bibinfo  {journal} {Opt.
  Express}\ }\textbf {\bibinfo {volume} {23}},\ \bibinfo {pages} {32835}
  (\bibinfo {year} {2015}{\natexlab{b}})}\BibitemShut {NoStop}%
\bibitem [{\citenamefont {Li}\ \emph {et~al.}(2015)\citenamefont {Li},
  \citenamefont {Yu},\ and\ \citenamefont {Wu}}]{li15c}%
  \BibitemOpen
  \bibfield  {author} {\bibinfo {author} {\bibfnamefont {J.}~\bibnamefont
  {Li}}, \bibinfo {author} {\bibfnamefont {R.}~\bibnamefont {Yu}}, \ and\
  \bibinfo {author} {\bibfnamefont {Y.}~\bibnamefont {Wu}},\ }\href
  {doi:10.1103/PhysRevA.92.053837} {\bibfield  {journal} {\bibinfo  {journal}
  {Phys. Rev. A}\ }\textbf {\bibinfo {volume} {92}},\ \bibinfo {pages} {053837}
  (\bibinfo {year} {2015})}\BibitemShut {NoStop}%
\bibitem [{\citenamefont {Xu}\ \emph {et~al.}(2016)\citenamefont {Xu},
  \citenamefont {Chen},\ and\ \citenamefont {Liu}}]{xu16a}%
  \BibitemOpen
  \bibfield  {author} {\bibinfo {author} {\bibfnamefont {X.-W.}\ \bibnamefont
  {Xu}}, \bibinfo {author} {\bibfnamefont {A.-X.}\ \bibnamefont {Chen}}, \ and\
  \bibinfo {author} {\bibfnamefont {Y.}~\bibnamefont {Liu}},\ }\href
  {doi:10.1103/PhysRevA.94.063853} {\bibfield  {journal} {\bibinfo  {journal}
  {Phys. Rev. A}\ }\textbf {\bibinfo {volume} {94}},\ \bibinfo {pages} {063853}
  (\bibinfo {year} {2016})}\BibitemShut {NoStop}%
\bibitem [{\citenamefont {Wang}\ \emph {et~al.}(2016)\citenamefont {Wang},
  \citenamefont {Miranowicz}, \citenamefont {Li},\ and\ \citenamefont
  {Nori}}]{wang16b}%
  \BibitemOpen
  \bibfield  {author} {\bibinfo {author} {\bibfnamefont {X.}~\bibnamefont
  {Wang}}, \bibinfo {author} {\bibfnamefont {A.}~\bibnamefont {Miranowicz}},
  \bibinfo {author} {\bibfnamefont {H.-R.}\ \bibnamefont {Li}}, \ and\ \bibinfo
  {author} {\bibfnamefont {F.}~\bibnamefont {Nori}},\ }\href
  {doi:10.1103/PhysRevA.93.063861} {\bibfield  {journal} {\bibinfo  {journal}
  {Phys. Rev. A}\ }\textbf {\bibinfo {volume} {93}},\ \bibinfo {pages} {063861}
  (\bibinfo {year} {2016})}\BibitemShut {NoStop}%
\bibitem [{\citenamefont {Zhou}\ and\ \citenamefont {Li}(2016)}]{zhou16a}%
  \BibitemOpen
  \bibfield  {author} {\bibinfo {author} {\bibfnamefont {B.}~\bibnamefont
  {Zhou}}\ and\ \bibinfo {author} {\bibfnamefont {G.}~\bibnamefont {Li}},\
  }\href {doi:10.1103/PhysRevA.94.033809} {\bibfield  {journal} {\bibinfo
  {journal} {Phys. Rev. A}\ }\textbf {\bibinfo {volume} {94}},\ \bibinfo
  {pages} {033809} (\bibinfo {year} {2016})}\BibitemShut {NoStop}%
\bibitem [{\citenamefont {Zhou}\ \emph {et~al.}(2016)\citenamefont {Zhou},
  \citenamefont {Shen}, \citenamefont {Shao},\ and\ \citenamefont
  {Yi}}]{zhou16b}%
  \BibitemOpen
  \bibfield  {author} {\bibinfo {author} {\bibfnamefont {Y.~H.}\ \bibnamefont
  {Zhou}}, \bibinfo {author} {\bibfnamefont {H.~Z.}\ \bibnamefont {Shen}},
  \bibinfo {author} {\bibfnamefont {X.~Q.}\ \bibnamefont {Shao}}, \ and\
  \bibinfo {author} {\bibfnamefont {X.~X.}\ \bibnamefont {Yi}},\ }\href
  {doi:10.1364/OE.24.017332} {\bibfield  {journal} {\bibinfo  {journal} {Opt.
  Express}\ }\textbf {\bibinfo {volume} {24}},\ \bibinfo {pages} {17332}
  (\bibinfo {year} {2016})}\BibitemShut {NoStop}%
\bibitem [{\citenamefont {Liu}\ \emph {et~al.}(2016)\citenamefont {Liu},
  \citenamefont {Wang}, \citenamefont {Liu},\ and\ \citenamefont
  {Nori}}]{liu16a}%
  \BibitemOpen
  \bibfield  {author} {\bibinfo {author} {\bibfnamefont {Y.-L.}\ \bibnamefont
  {Liu}}, \bibinfo {author} {\bibfnamefont {G.-Z.}\ \bibnamefont {Wang}},
  \bibinfo {author} {\bibfnamefont {Y.}~\bibnamefont {Liu}}, \ and\ \bibinfo
  {author} {\bibfnamefont {F.}~\bibnamefont {Nori}},\ }\href
  {doi:10.1103/PhysRevA.93.013856} {\bibfield  {journal} {\bibinfo  {journal}
  {Phys. Rev. A}\ }\textbf {\bibinfo {volume} {93}},\ \bibinfo {pages} {013856}
  (\bibinfo {year} {2016})}\BibitemShut {NoStop}%
\bibitem [{\citenamefont {Kryuchkyan}\ \emph {et~al.}(2016)\citenamefont
  {Kryuchkyan}, \citenamefont {Shahinyan},\ and\ \citenamefont
  {Shelykh}}]{kryuchkyan16a}%
  \BibitemOpen
  \bibfield  {author} {\bibinfo {author} {\bibfnamefont {G.~Y.}\ \bibnamefont
  {Kryuchkyan}}, \bibinfo {author} {\bibfnamefont {A.~R.}\ \bibnamefont
  {Shahinyan}}, \ and\ \bibinfo {author} {\bibfnamefont {I.~A.}\ \bibnamefont
  {Shelykh}},\ }\href {doi:10.1103/PhysRevA.93.043857} {\bibfield  {journal}
  {\bibinfo  {journal} {Phys. Rev. A}\ }\textbf {\bibinfo {volume} {93}},\
  \bibinfo {pages} {043857} (\bibinfo {year} {2016})}\BibitemShut {NoStop}%
\bibitem [{\citenamefont {Zhou}\ \emph {et~al.}(2018)\citenamefont {Zhou},
  \citenamefont {Shen}, \citenamefont {Zhang},\ and\ \citenamefont
  {Yi}}]{zhou18a}%
  \BibitemOpen
  \bibfield  {author} {\bibinfo {author} {\bibfnamefont {Y.~H.}\ \bibnamefont
  {Zhou}}, \bibinfo {author} {\bibfnamefont {H.~Z.}\ \bibnamefont {Shen}},
  \bibinfo {author} {\bibfnamefont {X.~Y.}\ \bibnamefont {Zhang}}, \ and\
  \bibinfo {author} {\bibfnamefont {X.~X.}\ \bibnamefont {Yi}},\ }\href
  {doi:10.1103/PhysRevA.97.043819} {\bibfield  {journal} {\bibinfo  {journal}
  {Phys. Rev. A}\ }\textbf {\bibinfo {volume} {97}},\ \bibinfo {pages} {043819}
  (\bibinfo {year} {2018})}\BibitemShut {NoStop}%
\bibitem [{\citenamefont {Gr\"unwald}(2019)}]{grunwald19a}%
  \BibitemOpen
  \bibfield  {author} {\bibinfo {author} {\bibfnamefont {P.}~\bibnamefont
  {Gr\"unwald}},\ }\href {doi:10.1088/1367-2630/ab3ae0} {\bibfield  {journal}
  {\bibinfo  {journal} {New J. Phys.}\ }\textbf {\bibinfo {volume} {21}},\
  \bibinfo {pages} {093003} (\bibinfo {year} {2019})}\BibitemShut {NoStop}%
\end{thebibliography}%


\begin{thebibliography}{2}%
\makeatletter
\providecommand \@ifxundefined [1]{%
 \@ifx{#1\undefined}
}%
\providecommand \@ifnum [1]{%
 \ifnum #1\expandafter \@firstoftwo
 \else \expandafter \@secondoftwo
 \fi
}%
\providecommand \@ifx [1]{%
 \ifx #1\expandafter \@firstoftwo
 \else \expandafter \@secondoftwo
 \fi
}%
\providecommand \natexlab [1]{#1}%
\providecommand \enquote  [1]{``#1''}%
\providecommand \bibnamefont  [1]{#1}%
\providecommand \bibfnamefont [1]{#1}%
\providecommand \citenamefont [1]{#1}%
\providecommand \href@noop [0]{\@secondoftwo}%
\providecommand \href [0]{\begingroup \@sanitize@url \@href}%
\providecommand \@href[1]{\@@startlink{#1}\@@href}%
\providecommand \@@href[1]{\endgroup#1\@@endlink}%
\providecommand \@sanitize@url [0]{\catcode `\\12\catcode `\$12\catcode
  `\&12\catcode `\#12\catcode `\^12\catcode `\_12\catcode `\%12\relax}%
\providecommand \@@startlink[1]{}%
\providecommand \@@endlink[0]{}%
\providecommand \url  [0]{\begingroup\@sanitize@url \@url }%
\providecommand \@url [1]{\endgroup\@href {#1}{\urlprefix }}%
\providecommand \urlprefix  [0]{URL }%
\providecommand \Eprint [0]{\href }%
\providecommand \doibase [0]{http://dx.doi.org/}%
\providecommand \selectlanguage [0]{\@gobble}%
\providecommand \bibinfo  [0]{\@secondoftwo}%
\providecommand \bibfield  [0]{\@secondoftwo}%
\providecommand \translation [1]{[#1]}%
\providecommand \BibitemOpen [0]{}%
\providecommand \bibitemStop [0]{}%
\providecommand \bibitemNoStop [0]{.\EOS\space}%
\providecommand \EOS [0]{\spacefactor3000\relax}%
\providecommand \BibitemShut  [1]{\csname bibitem#1\endcsname}%
\let\auto@bib@innerbib\@empty
\bibitem [{\citenamefont {Laussy}\ \emph {et~al.}(2009)\citenamefont {Laussy},
  \citenamefont {del Valle},\ and\ \citenamefont {Tejedor}}]{laussy09a}%
  \BibitemOpen
  \bibfield  {author} {\bibinfo {author} {\bibfnamefont {F.~P.}\ \bibnamefont
  {Laussy}}, \bibinfo {author} {\bibfnamefont {E.}~\bibnamefont {del Valle}}, \
  and\ \bibinfo {author} {\bibfnamefont {C.}~\bibnamefont {Tejedor}},\ }\href
  {doi:10.1103/PhysRevB.79.235325} {\bibfield  {journal} {\bibinfo  {journal}
  {Phys. Rev. B}\ }\textbf {\bibinfo {volume} {79}},\ \bibinfo {pages} {235325}
  (\bibinfo {year} {2009})}\BibitemShut {NoStop}%
\bibitem [{\citenamefont {Visser}\ and\ \citenamefont
  {Nienhuis}(1995)}]{visser95a}%
  \BibitemOpen
  \bibfield  {author} {\bibinfo {author} {\bibfnamefont {P.~M.}\ \bibnamefont
  {Visser}}\ and\ \bibinfo {author} {\bibfnamefont {G.}~\bibnamefont
  {Nienhuis}},\ }\href {doi:10.1103/PhysRevA.52.4727} {\bibfield  {journal}
  {\bibinfo  {journal} {Phys. Rev. A}\ }\textbf {\bibinfo {volume} {52}},\
  \bibinfo {pages} {4727} (\bibinfo {year} {1995})}\BibitemShut {NoStop}%
\end{thebibliography}%

\end{document}


\title{Tuning photon statistics with coherent fields\\
Supplementary Material}

\author{Eduardo {Zubizarreta Casalengua}}
\affiliation{Faculty of Science and Engineering,
  University of Wolverhampton, Wulfruna St, Wolverhampton WV1 1LY, UK}

\author{{Juan Camilo} {L\'{o}pez~Carre\~no}}
\affiliation{Faculty of Science and Engineering, University of
  Wolverhampton, Wulfruna St, Wolverhampton WV1 1LY, UK}

\author{Fabrice P. Laussy}
\affiliation{Faculty of Science and Engineering,
  University of Wolverhampton, Wulfruna St, Wolverhampton WV1 1LY, UK}
\affiliation{Russian Quantum Center, Novaya 100, 143025 Skolkovo,
  Moscow Region, Russia}

\author{Elena {del Valle}}
\email{elena.delvalle.reboul@gmail.com}
\affiliation{Faculty of Science and Engineering,
  University of Wolverhampton, Wulfruna St, Wolverhampton WV1 1LY, UK}
\affiliation{Departamento de F\'isica Te\'orica de la Materia
Condensada, Universidad Aut\'onoma de Madrid, 28049 Madrid,
Spain}

\date{\today}

\begin{abstract}
  We describe how the steady states solutions in the main text have
  been obtained in closed-form (Section~I) and detail the wavefunction
  approximation method (Section~II) for the three cases studied in the
  text (A.~Two-level system, B.~Jaynes--Cummings model and
  C.~polaritons). We also provide the exact $\mathcal{I}$
  decompositions for polaritons (Section~3). Equations from the main
  text are prefixed with T, e.g., Eq.~(T\ref{eq:Wed8Apr181920CEST2020}) is
  the mean-field--quantum decomposition of the
  signal~$s=\alpha+\epsilon$.
\end{abstract}
\maketitle

\section{Steady states of light matter coupling at vanishing laser driving}

In this work we need to solve the steady state dynamics of
light-matter interaction in the low coherent driving regime. This can
be done following the method of
Refs.~\cite{laussy09a}. The light field is a cavity
mode~$a$ and the matter field can be either a Two-Level system
(2LS)~$\sigma$ or another bosonic mode~$b$. We solve the dynamics in
terms of a general mean value, a product of any system operator, which
in its most general normally ordered form
reads~$C_{\{m,n,\mu,\nu\}}=\mean{\sigma^{\dagger m}\sigma^n a^{\dagger
    \mu} a^\nu}$ (with~$m$, $n \in\{0,1\}$ and $\mu$,
$\nu\in \mathbb{N}$) if the matter field is a 2LS or
~$C_{\{m,n,\mu,\nu\}}=\mean{b^{\dagger m}b^n a^{\dagger \mu} a^\nu}$
(with~$m$, $n$, $\mu$, $\nu\in \mathbb{N}$) if the matter field is
bosonic. This general element follows the master equation described in
the main text, which can be expressed in a matricial form:
\begin{equation}
\label{eq:TueMay5174356GMT2009}
\partial_t C_{\{m,n,\mu,\nu\}}=\sum_{{m',n',\mu',\nu'}}\mathcal{M}_{\substack{m,n,\mu,\nu\\m',n',\mu',\nu'}}C_{\{m',n',\mu',\nu'\}}\,.
\end{equation}
The regression matrix
elements~$\mathcal{M}_{\substack{m,n,\mu,\nu\\m',n',\mu',\nu'}}$, in
the case of a coupled 2LS, are given by:
\begin{subequations}
	\label{eq:TueDec23114907CET2008}
	\begin{align}
	&\mathcal{M}_{\substack{m,n,\mu,\nu\\m,n,\mu,\nu}}=-\frac{\gamma_a}2(\mu+\nu)-\frac{\gamma_\sigma}2(m+n)
	+ i (\mu - \nu ) \delta_{a} + i (m-n) \Delta_\sigma 
	\end{align}
	\begin{align}
	&\mathcal{M}_{\substack{m,n,\mu,\nu\\1-m,n,\mu,\nu}}=i\Omega_\sigma[m+2n(1-m)]\,,\quad
	&&\mathcal{M}_{\substack{m,n,\mu,\nu\\m,1-n,\mu,\nu}}=-i\Omega_\sigma[n+2m(1-n)]\,\\
	&\mathcal{M}_{\substack{m,n,\mu,\nu\\m,n,\mu-1,\nu}}=i e^{-i \phi} \, \Omega_a\mu\,,\quad
	&&\mathcal{M}_{\substack{m,n,\mu,\nu\\m,n,\mu,\nu-1}}=- i e^{i \phi} \, \Omega_a\nu\,,\\
	&\mathcal{M}_{\substack{m,n,\mu,\nu\\m,1-n,\mu,\nu-1}}=-i g(1-n)\nu\,,\quad 
	&&\mathcal{M}_{\substack{m,n,\mu,\nu\\1-m,n,\mu-1,\nu}}=i g(1-m)\mu\,, \\
	&\mathcal{M}_{\substack{m,n,\mu,\nu\\m,1-n,\mu,\nu+1}}=-i g n \,,\quad
	&&\mathcal{M}_{\substack{m,n,\mu,\nu\\1-m,n,\mu+1,\nu}}=i g m \,, \\
	&\mathcal{M}_{\substack{m,n,\mu,\nu\\1-m,n,\mu,\nu+1}}=2 i n g(1-m) \,,\quad
	&&\mathcal{M}_{\substack{m,n,\mu,\nu\\m,1-n,\mu + 1,\nu}}= -2 i n g(1-m)\,,
	\end{align}
\end{subequations}
and zero everywhere else. In the main text, we discuss first the case
of resonance fluorescence, which corresponds to having only the 2LS
operator~$\sigma$ and no cavity mode~$a$ (taking $g$, $\Omega_a=0$
here). Second, we solve the Jaynes--Cummings model with both cavity
and dot driving with a phase difference between the sources, which
corresponds to setting $\Omega_\sigma / \Omega_a = \chi $
here. Similarly, for the polariton model where the matter field is an
exciton (boson), we have:
\begin{subequations}
	\begin{align}
	&\mathcal{M}_{\substack{m,n,\mu,\nu\\m,n,\mu,\nu}}=-\frac{\gamma_a}2(\mu+\nu)-\frac{\gamma_b}2(m+n)
	+ i (\mu - \nu ) \Delta_{a} + i (m-n) \Delta_b + i \frac{U}{2} \left[m (m-1) - n(n-1) \right],
	\end{align}
	\begin{align}
	&\mathcal{M}_{\substack{m,n,\mu,\nu\\m,n,\mu-1,\nu}}=i e^{-i \phi} \,\Omega_a \mu ,
	&&\mathcal{M}_{\substack{m,n,\mu,\nu\\m,n,\mu,\nu-1}}=-i e^{i \phi} \,\Omega_a \nu \,\\
	&\mathcal{M}_{\substack{m,n,\mu,\nu\\m+1,n,\mu-1,\nu}}= i g \mu\,,\quad
	&&\mathcal{M}_{\substack{m,n,\mu,\nu\\m,n+1,\mu,\nu-1}}=-i g \nu\,,\\
	&\mathcal{M}_{\substack{m,n,\mu,\nu\\m-1,n,\mu+1,\nu}}= i g m\,,\quad
    &&\mathcal{M}_{\substack{m,n,\mu,\nu\\m,n-1,\mu,\nu+1}}=-i g n\,,\\
	&\mathcal{M}_{\substack{m,n,\mu,\nu\\m+1,n+1,\mu,\nu}}=i U (m-n) ,\quad
	\end{align}
\end{subequations}
and, again, the remaining matrix elements are zero.

These equations can be solved numerically, by choosing a high enough
truncation in the number of excitations, in order to obtain the steady
state ($\partial_t C_{\{m,n,\mu,\nu\}}=0$) for any given
pump. However, in the vanishing driving limit
($\Omega_\sigma \rightarrow 0$ or $\Omega_a\rightarrow 0$), one can
obtain analytical solutions. In this case, it is enough to solve
recursively sets of truncated equations. That is, we start with the
lowest order correlators, with only one operator, that we write in a
vectorial form for convenience (using the JC model as an example):
$\mathbf{v}_1=(
\mean{a}~\mean{a^\dagger}~\mean{\sigma}~\mean{\sigma^\dagger
})^\mathrm{T}$. Its equation,
$\partial_t \mathbf{v}_1= M_1 \mathbf{v}_1+A_1+~\mathrm{h.~o.~t.}$,
provides the steady state
value~$\mathbf{v}_1= -M_1^{-1} A_1+~\mathrm{h.~o.~t.}$, to lowest
order in $\Omega_a$ (with ``h. o. t.'' meaning \emph{higher order
  terms}). We proceed in the same way with the two-operator
correlators
$\mathbf{v}_2=( \mean{a^2}~\mean{a^{\dagger 2}}~\mean{a^\dagger
  a}~\mean{\sigma^\dagger\sigma}~\mean{\sigma^\dagger
  a}\,\hdots)^\mathrm{T}$, only, in this case, we also need to include
the steady state value for the one-operator correlators as part of the
independent term in the equation:
$\partial_t \mathbf{v}_2= M_2 \mathbf{v}_2+A_2+X_{21}
\mathbf{v}_1+~\mathrm{h.~o.~t.}$. The steady state
reads~$\mathbf{v}_2= -M_2^{-1} (A_2+X_{21}
\mathbf{v}_1)+~\mathrm{h.~o.~t.}$ with an straightforward
generalization
$\mathbf{v}_N= -M_N^{-1} (A_N+\sum_{j=1}^{N-1}X_{N j}
\mathbf{v}_j)+~\mathrm{h.~o.~t.}$. We aim in particular at obtaining
photon correlators of the type~$\mean{a^{\dagger N} a^N }$ that
follow~$\mean{a^{\dagger N} a^N }\sim (\Omega_a)^{2N} $, to lowest
order in the driving $\Omega_a$. The normalized correlation
functions~$g_a^{(N)}=\mean{a^{\dagger N} a^N }/\mean{a^\dagger a}^N $
are independent of $\Omega_a$ to lowest order, and their computation
requires to solve the $2N$ sets of recurrent equations and to take the
limit~$\lim_{\Omega_a\rightarrow 0}g_a^{(N)}$.

\section{Wavefunction approximation method at vanishing pumping regime} 
\label{sec:Fri28Feb105950GMT2020}

For comprehensiveness, we also describe a popular formalism to
describe photon statistics, namely, the wavefunction
approximation~\cite{visser95a}, which has brought much results in the
literature of unconventional photon blockade given its connection to
interferences.  In the context of this paper, where we assume that the
state of the system is composed by at most two fields, with
annihilation operators~$\xi$ and~$c$, following either fermionic or
bosonic algebra, this can be approximated by a pure state, which in
the Fock state basis reads
\begin{equation}
  \label{eq:WedFeb28112316CET2018}
\ket{\psi} = \sum_{n,m} \mathcal{C}_{nm} \ket{n}_c\ket{m}_\xi \equiv
\sum_{n,m} 
\mathcal{C}_{nm} \ket{n\,,m} \,,
\end{equation}
where~$\mathcal{C}_{nm}$ are the probability amplitude of having~$n$
photons in the field described with operator~$\xi$ and~$m$ photons in
the field described with operator~$c$, and the summation is taken over
the size of the Hilbert space: one for a fermionic field and~$N$ for a
truncated bosonic one. Given that the dynamics of the system is given
by the master equation
\begin{equation}
\label{eq:WedFeb28114318CET2018}
\partial_t \rho = i[\rho,H] + \sum_k (\tilde \Gamma_k/2)
\mathcal{L}_{j_k}\rho\,,
\end{equation}
where~$H$ is the Hamiltonian of the system and we have assumed that
the dissipation is given by ``jump operators''~$j_k$ at
rates~$\tilde\Gamma_k$, the dynamics of the wavefunction is given by
Schr\"odinger's equation
\begin{equation}
  \label{eq:WedFeb28113458CET2018}
  \partial_t \ket{\psi} = - i H_\mathrm{eff}\ket{\psi}
\end{equation}
where~$H_\mathrm{eff}$ is a non-hermitian Hamitonian constructed
as~$ H_\mathrm{eff}=H-i \sum_k \tilde\Gamma_k\, \ud{j_k} j_k$. The
coefficients evolve as:
\begin{equation}
\label{eq:coeffeqs}
\partial_t \, \mathcal{C}_{nm} = -i  \sum_{p,q} \bra{n\,,m}H_\mathrm{eff}
\ket{p\,,q} \mathcal{C}_{pq}\,. 
\end{equation}
In the following sections we make explicit both the effective
Hamiltonians and the differential equations for the coefficients for
all the systems considered in the main text. 

\subsection{Two-level system in the Heitler regime}

The Hamiltonian describing the excitation of a sensor (a cavity) by
the emission of a 2LS, which in turn is driven in the Heitler regime
by a laser, is given by the Hamiltonian
Eq.~(T\ref{eq:Thu31May103357CEST2018}).  To complete the analogy of
beam splitter setting and be consistent with the main text, both
driving and coupling for the sensor has to be defined in terms of
coherent source amplitude $|\beta|$ and BS parameters $\mathrm{T}$ and
$\mathrm{R}$: $\Omega_a \rightarrow i \mathrm{R}|\beta|$,
$g \rightarrow \mathrm{T} g$.
%
%
The system and driving source are not necessarily at resonance
so we define the detuning as $\Delta_{\sigma} = \omega_\sigma - \omega_{\mathrm{L}}$.
The effective Hamiltonian that describes
the dynamics in the wavefunction approximation is
\begin{equation}
  \label{eq:MonMar5160912CET2018}
  H_\mathrm{eff} = H -\frac{i}{2}\left ( \gamma_\sigma
    \ud{\sigma}\sigma + \Gamma \ud{a}a \right)\,,
\end{equation}
where~$H$ is the Hamiltonian in
Eq.~(T\ref{eq:Thu31May103357CEST2018}). Replacing the effective
Hamiltonian in Eq.~(\ref{eq:MonMar5160912CET2018}) on the expression
in Eq.~(\ref{eq:WedFeb28113458CET2018}), we obtain the differential
equations for the coefficients of interest:
\begin{subequations}
  \label{eq:MonMar5161258CET2018}
  \begin{align}
    i \partial_t \mathcal{C}_{01} &= \Omega_\sigma + \mathrm{T}\,g\,
                                    \mathcal{C}_{10} - i
                                    \mathrm{R}|\beta|e^{-i\phi}
                                    \mathcal{C}_{11}+\left(\Delta_{\sigma}-
                                    i\frac{\gamma_\sigma}{2} \right)
                                    \mathcal{C}_{01}\,,\\
    i \partial_t \mathcal{C}_{10} &= i \mathrm{R}
                                    |\beta|e^{i\phi} + \Omega_\sigma
                                    \mathcal{C}_{11}+ 
                                    \mathrm{T}\,g\, \mathcal{C}_{01}          - i  \sqrt{2}\mathrm{R}|\beta|e^{-i\phi}
                                    \mathcal{C}_{20}-i\frac{\Gamma}{2}
                                    \mathcal{C}_{10}\,,\\
    i \partial_t \mathcal{C}_{11} &= \Omega_\sigma \mathcal{C}_{10} + i
                                    \mathrm{R}|\beta| e^{i\phi}
                                    \mathcal{C}_{01}+\sqrt{2}
                                    \mathrm{T}\,g\,\mathcal{C}_{20}
    +\left(\Delta_\sigma-i\frac{\gamma_\sigma+\Gamma}{2}
    \right) \mathcal{C}_{11}\,,\\
    i \partial_t \mathcal{C}_{20} & = \sqrt{2}\mathrm{T}\,g\,\mathcal{C}_{11} +
                                    i \sqrt{2} \mathrm{R}|\beta|e^{i
                                    \phi}\mathcal{C}_{10} - i\Gamma
                                    \mathcal{C}_{20}\,,  
  \end{align}
\end{subequations}
where we have assumed that the driving to the 2LS is low
enough so that the states with three or more excitation can be safely
neglected. Assuming that the coherent field that drives the sensor can
be written as a fraction of the field that drives the two-level
system ($|\beta| = g \frac{\mathrm{T} \Omega_{\sigma}}{\mathrm{R} \gamma_\sigma} \mathcal{F} $), 
and to leading order in the coupling and the driving intensity of the
two-level system, the solution to Eqs.~(\ref{eq:MonMar5161258CET2018})
is
\begin{subequations}
  \label{eq:MonMar5165614CET2018}
  \begin{align}
    \mathcal{C}_{01} &= -\frac{2i\Omega_\sigma}{\gamma_\sigma + 2 i \Delta_\sigma}\,,\\
    \mathcal{C}_{10} &= -\frac{2g\Omega_\sigma \mathrm{T}}{\Gamma}
                       \left(\frac{2}{\gamma_\sigma + 2 i \Delta_\sigma}-\frac{\mathcal{F}e^{i\phi}}{\gamma_\sigma}\right)\,,\\
    \mathcal{C}_{11} &= -\frac{4 i g \Omega_\sigma^2 \mathrm{T}
                       [-2\gamma_\sigma
                       +\mathcal{F}(\gamma_\sigma +\Gamma+ 2 i \Delta_\sigma)e^{i\phi}]}{\Gamma \gamma_\sigma \left(\gamma_\sigma + 2 i \Delta_{\sigma}\right) \left(\gamma_\sigma + \Gamma 
                       + 2 i \Delta_{\sigma}\right) }
                   \,,\\
    \mathcal{C}_{20}&= \frac{2\sqrt{2} g^2\Omega_\sigma^2 \mathrm{T}^2[4 \gamma _{\sigma }^2+\mathcal{F}^2 e^{2 i \phi } \left(\gamma _{\sigma }+2 i \Delta _{\sigma }\right) \left(\gamma _{\sigma }+\Gamma +2 i \Delta _{\sigma }\right)
    -4 \mathcal{F} e^{i \phi } \gamma _{\sigma } \left(\gamma _{\sigma }+\Gamma +2 i \Delta _{\sigma }\right)]}{\gamma_\sigma^2
                      \Gamma^2 \left(\gamma_\sigma + 2 i \Delta_{\sigma}\right) \left(\gamma_\sigma + \Gamma 
                      + 2 i \Delta_{\sigma}\right)}
                     \,.
  \end{align}
\end{subequations}
The population of both the 2LS and the cavity, and the~$\g{2}_a$ can
be obtained from the coefficients in
Eqs.~(\ref{eq:MonMar5165614CET2018}). However, to recover some
information from the unfiltered signal ($\Gamma \rightarrow \infty$),
this limit has to be performed carefully and the previous expressions
need some manipulation first. A new set of coefficients is defined as:
$\mathcal{C}'_{ij} = \left(\frac{\Gamma}{2\mathrm{T} g} \right)^{i+j}
\mathcal{C}_{ij}$ so any explicit dependence of sensor parameters
disappears, resulting in a non-vanishing (finite) solution after the
proper limits are taken.  After the substitution and taking the limit,
the new coefficients are
\begin{subequations}
	\begin{align}
	\mathcal{C}'_{01} &=  -\frac{2i\Omega_\sigma}{\gamma_\sigma + 2 i \Delta_\sigma} \,,\\
	\mathcal{C}'_{10} &= \Omega_{\sigma} \left(\frac{e^{i \phi} \mathcal{F}}{\gamma_\sigma} - \frac{2}{\gamma_\sigma + 2 i \Delta_\sigma}\right) \,,\\
	\mathcal{C}'_{11} &=- \frac{2 i e^{i \phi} \mathcal{F} \Omega_\sigma^2}{\gamma_\sigma
	                    \left(\gamma_\sigma + 2 i \Delta_{\sigma}\right)}\,,\\
	\mathcal{C}'_{20}&= \frac{e^{i \phi} \mathcal{F} \Omega_\sigma^2}{\gamma_\sigma^2
		\left(\gamma_\sigma + 2 i \Delta_{\sigma}\right)} 
	\left[e^{i \phi} \mathcal{F} \left(\gamma_\sigma+ 2 i \Delta_\sigma \right)
	-4 \gamma_\sigma\right] \,.
	\end{align}
\end{subequations}
Now, these solutions provide useful information about the equivalent
filtered signal: $\av{n_a} \approx |\mathcal{C}'_{10}|^2$,
$\mathrm{P}_{20} = |\mathcal{C}'_{20}|^2$ (probability of two-photon
state)
and~$\g{2}_a \approx 2|\mathcal{C}'_{20}|^2/
|\mathcal{C}'_{10}|^4$. The cancellation of the
coefficient~$\mathcal{C}'_{20}$, and therefore of~$\g{2}_a$, yields
the condition on the attenuation factor
\begin{equation}
  \label{eq:TueMar6111957CET2018}
  \mathcal{F} = \frac{4 e^{-i \phi}\gamma_\sigma}{\gamma_\sigma + 2 i \Delta_{\sigma}}\,,
\end{equation}
which can only be satisfied---$\mathcal{F}$~is an attenuation factor,
and thus a real number---when the relative phase between the driving
field and the 2LS coherent contribution is either 0 or $\pi$ (opposite
phase). 
Note, as well, that the cancellation of the
coefficient~$\mathcal{C}_{10}$, and therefore of the population of the
cavity, is obtained
when~$\mathcal{F} = \frac{2 e^{-i \phi}\gamma_\sigma}{\gamma_\sigma +
  2 i \Delta_{\sigma}}$ which is a real number for the same phases for
which Eq.~(\ref{eq:TueMar6111957CET2018}) is a real number.

\subsection{Jaynes--Cummings blockade}
\label{sec:WedFeb28173330CET2018}


The Hamiltonian describing the Jaynes--Cummings model in given in
Eq.~(T\ref{eq:Thu31May103357CEST2018}), and the dynamics is
complemented with a master equation that takes into account the decay
of the 2LS with rate~$\gamma_\sigma$ and of the cavity with
rate~$\gamma_a$. As such, the effective Hamiltonian that described the
dynamics in the wavefunction approximation is
\begin{equation}
  \label{eq:WedFeb28144028CET2018}
  H_\mathrm{eff}  = 
  \left(\Delta_a - i \frac{\gamma_a}{2} \right) \ud{a} a \ +
  \left(\Delta_\sigma - i 
    \frac{\gamma_\sigma}{2} \right) \ud{\sigma} \sigma
 + g (\ud{a} \sigma  + \ud{\sigma} a ) + \Omega_a (e^{i \phi}\ud{a} + e^{-i \phi} a) + \Omega_{\sigma} 
  (\ud{\sigma} + \sigma)\,.
\end{equation}
%
Replacing the Hamiltonian in Eq~(\ref{eq:WedFeb28144028CET2018}) into
Eq.~(\ref{eq:coeffeqs}), we have that the differential equation for
the relevant coefficients are as follows:
\begin{subequations}
  \label{eq:WedFeb28144227CET2018}
  \begin{align}
    i \partial_t \,\mathcal{C}_{10}&= e^{i \phi} \Omega_a  + \left(\Delta_a - i
          \frac{\gamma_a}{2}\right) \mathcal{C}_{10} + g\,
                                   \mathcal{C}_{01} + \Omega_{\sigma} \,
                                     \mathcal{C}_{11} 
    +\sqrt{2} e^{-i \phi} \Omega_a \, \mathcal{C}_{20} \,, \\ 
	i \partial_t \,\mathcal{C}_{01}&= \Omega_{\sigma} +  g \, \mathcal{C}_{10} +
                                         \left(\Delta_\sigma - i 
	\frac{\gamma_\sigma}{2}\right) \mathcal{C}_{01} + e^{-i \phi} \Omega_a \,
                                       \mathcal{C}_{11}\,, \\ 
	i \partial_t \,\mathcal{C}_{11}&=  e^{i \phi} \Omega_a \,  \mathcal{C}_{01} + 
                                         \Omega_{\sigma} \, \mathcal{C}_{10}
    +\left(\Delta_a+\Delta_\sigma - i \frac{\gamma_a + \gamma_\sigma}{2}\right) \mathcal{C}_{11} + \sqrt{2} g \mathcal{C}_{20}\,, \\  
    i \partial_t \,\mathcal{C}_{20}&=  \sqrt{2}  e^{i \phi}  \Omega_a \, \mathcal{C}_{10} + 
                                      \sqrt{2} g \,\mathcal{C}_{11}+
                                      \left(2 \Delta_a - i \gamma_a
                                   \right) \mathcal{C}_{20}\,, 
  \end{align}
\end{subequations}
where we have assumed that the driving is low enough for the states
containing three or more photons to be neglected. The steady-state
solution for the coefficients Eqs.~(\ref{eq:WedFeb28144227CET2018}) is
obtained when the derivatives on the left-hand side of the equations
vanish. Thus, assuming that the coefficient of the vacuum dominates
over all the others, i.e.,~$\mathcal{C}_{00}\approx 1$, and to leading
order in the driving of the cavity, the coefficients are
\begin{subequations}
  \label{eq:WedFeb28145309CET2018}
  \begin{align}
    \mathcal{C}_{10}  &=  \Omega_a \, \frac{2 e^{i \phi} \left( 2 \Delta_\sigma 
                        - i \gamma _{\sigma } \right) - 4 \chi g}{4 g^2+\left(\gamma _a+2 i
                        \Delta_a\right) \left(\gamma _{\sigma }+2 i \Delta_\sigma \right)}, \\ 
    \mathcal{C}_{01} &= \Omega_a \, \frac{2 \chi \left( 2 \Delta_a
                       - i 
                       \gamma _a \right) - 4 e^{i \phi} g}{4 g^2+\left(\gamma _a+2 i
                       \Delta_a\right) \left(\gamma _{\sigma }+2 i \Delta_\sigma \right)}, \\ 
    \mathcal{C}_{11} & = 4 \Omega_a^2 \, \frac{[2 e^{i \phi} g - \chi (2 \Delta_a - i \gamma_a)]
                       [2 g \chi + i e^{i \phi} (\tilde{\gamma}_{11} + 2 i \tilde{\Delta}_{11})]}
                       { [4 g^2+\left(\gamma _a+2 i
                       \Delta_a\right) \left(\gamma _{\sigma }+2 i \Delta_\sigma \right)]
                       [4 g^2+\left(\gamma _a+2 i
                       \Delta_a\right) (\tilde{\gamma}_{11} +2 i \tilde{\Delta}_{11})]},\\   
    \mathcal{C}_{20}  &= \sqrt{8} \Omega _a^2 \,
                        \frac{4 g^2 \chi^2
                        + 4 i e^{i \phi} g \chi (\tilde{\gamma}_{11} +2 i \tilde{\Delta}_{11})
                        +e^{i 2 \phi} [4 g^2 - (\gamma_\sigma + 2 i \Delta_\sigma)
                        (\tilde{\gamma}_{11} +2 i \tilde{\Delta}_{11})] }
                        { [4 g^2+\left(\gamma _a+2 i
                        \Delta_a\right) \left(\gamma _{\sigma }+2 i \Delta_\sigma \right)]
                        [4 g^2+\left(\gamma _a+2 i
\Delta_a\right) (\tilde{\gamma}_{11} +2 i \tilde{\Delta}_{11})]}, 
  \end{align}
\end{subequations}
where $\Delta_c = \omega_c - \omega_\mathrm{L}$ (for $c = a, \sigma$),
$\chi = \Omega_{\sigma}/ \Omega_a$ is the ratio between dot and cavity
driving and $\tilde{\Delta}_{ij} = i \Delta_a + j \Delta_{\sigma}$
(the same notation for $\tilde{\gamma}_{ij}$). The population of both
the 2LS and the cavity, and the~$\g{2}_a$ can be obtained from the
coefficients in Eqs.~(\ref{eq:WedFeb28145309CET2018})
as~$ n_a = |\mathcal{C}_{10}|^2$,
$\mean{n_\sigma}= |\mathcal{C}_{01}|^2$
and~$\g{2}_a = 2|\mathcal{C}_{20}|^2/ |\mathcal{C}_{10}|^4$,
respectively; which coincide with the expressions given in
Eq.~(T\ref{eq:Tue29May182623CEST2018}) and
Eq.~(T\ref{eq:Tue29May184632CEST2018}).



\subsection{Exciton-polaritons blockade}
\label{sec:FriMar2181711CET2018}

The Hamiltonian describing exciton-polaritons is given in
Eq.~(T\ref{eq:Mon5Jun145532BST2017}), and the dynamics is complemented
with a master equation that takes into account the decay rate of the
cavity with rate~$\gamma_a$ and of the excitons with
rate~$\gamma_b$. Thus, the effective Hamiltonian that described the
dynamics in the wavefunction approximation is
\begin{equation}
  \label{eq:FriMar2182643CET2018}
H_{\mathrm{eff}}
= \left(\Delta_a 
- i \frac{\gamma_a}{2}\right) \ud{a} a + \left(\Delta_{b}- i
\frac{\gamma_b}{2}\right) \ud{b} b+ g \left(\ud{a} b 
+ \ud{b} a \right) + \frac{U}{2}\ud{b} \ud{b} b b
{}+ \Omega_a \left(e^{i\phi} \ud{a} + e^{-i \phi} a\right) + \Omega_b (\ud{b} + b)\,.
\end{equation}
Replacing the Hamiltonian Eq.~(\ref{eq:FriMar2182643CET2018}) in
Eq.~(\ref{eq:coeffeqs}), we find that the differential equations for
the relevant coefficients are
%
\begin{subequations}
  \label{eq:polaritonSteadyState_wavefunc}
  \begin{align}
	i \partial_t\, \mathcal{C}_{10} &= e^{i\phi} \Omega_a + \left(\Delta_a - i
          \frac{\gamma_a}{2}\right) \mathcal{C}_{10} + g \ \mathcal{C}_{01} +
                                          \Omega_b \, \mathcal{C}_{11} 
    +
          \sqrt{2} e^{-i\phi} \Omega_a \, \mathcal{C}_{20}\,, \\
	i \partial_t \,\mathcal{C}_{01} &= \Omega_b +  g \ \mathcal{C}_{10} +
                                        \left(\Delta_\sigma - i 
	\frac{\gamma_\sigma}{2}\right) \mathcal{C}_{01} + e^{-i \phi} \Omega_a \,
                                          \mathcal{C}_{11} 
    + \sqrt{2} \Omega_b \, \mathcal{C}_{02} \,, \\ 
	i \partial_t \,\mathcal{C}_{11} &= e^{i \phi} \Omega_a \, \mathcal{C}_{01} +
	                                  \Omega_b \, \mathcal{C}_{01} 
    +\left(\Delta_a +\Delta_\sigma -
                                      i 
	\frac{\gamma_a + \gamma_\sigma}{2}\right) \mathcal{C}_{11} +
                                        \sqrt{2} g 
          \left( \mathcal{C}_{20} + \mathcal{C}_{02} \right)\,, \\ 
	i \partial_t \,\mathcal{C}_{20} &= \sqrt{2} e^{i \phi} \Omega_a \,
                                        \mathcal{C}_{10} +  
	\sqrt{2} g \, \mathcal{C}_{11}+
	\left(2 \Delta_a - i \gamma_a \right) \mathcal{C}_{20}\,, \\
	i \partial_t\, \mathcal{C}_{02} &= \sqrt{2} \Omega_b \, \mathcal{C}_{01} + \sqrt{2} g \,
                                        \mathcal{C}_{11}+ 
	\left(2 \Delta_b + U - i \gamma_b \right) \mathcal{C}_{02}\,,
	\end{align}
\end{subequations}
%
where we have assumed that the driving is low enough for the states
with three or more photons to be neglected. The steady-state solution
of the coefficients in Eqs.~(\ref{eq:polaritonSteadyState_wavefunc}) is
obtained when the derivatives in the left-hand side of the equation
vanish.  Thus, assuming that the coefficient of the vacuum dominates
over all the others, i.e.,~$\mathcal{C}_{00}\approx 1$, and to leading
order in the driving of the cavity, the coefficients are
\begin{subequations}
\label{eq:FriMar2190022CET2018}
	\begin{align}
	 \mathcal{C}_{10}  &= 2 \Omega_a \, \frac{e^{i \phi} (2 \Delta_b - i \gamma_b) - 2 g \chi}
	 {4 g^2+(\gamma _a + 2 i \Delta_{a}) (\gamma_b + 2 i \Delta_{b})}\,, \\ 
	\mathcal{C}_{01} &=2 \Omega_a \, \frac{\chi (2 \Delta_a - i \gamma_a) - 2 e^{i \phi} g }
	{4 g^2+(\gamma _a + 2 i \Delta_{a}) (\gamma_b + 2 i \Delta_{b})}\,, \\ 
	 \mathcal{C}_{11}  &= 4 \Omega_a^2 \, [-2 i e^{i \phi} g + (\gamma_a + 2 i \Delta_a)]
	 [e^{i \phi} (U + 2 \Delta_b - i \gamma_b)(\tilde{\gamma}_{11} + 2 i \tilde{\Delta}_{11})
	 - i \chi g (U + 2 \tilde{\Delta}_{11} - i \tilde{\gamma}_{11})] 
	/\mathcal{N} \,,  \\
		\mathcal{C}_{20} & =
	\begin{aligned}[t]
     & i \sqrt{8} \Omega_a^2 \, \lbrace 4 \chi^2 g^2 (U + \tilde{\Delta}_{11} - i
	 \tilde{\gamma}_{11})+ 4 i e^{i \phi} \chi g (U + 2 \Delta_b - i \gamma_b)
	 (U + \tilde{\Delta}_{11} - i \tilde{\gamma}_{11}) + {}\\ 
	 & \kern5cm{}+ e^{i 2 \phi}
	 [4 g^2 U - (\gamma_b + 2 i \Delta_b)(\tilde{\gamma}_{11} + 2 i \tilde{\Delta}_{11})
	 (U + \Delta_b - i \gamma_b)]\rbrace / 
	\ \mathcal{N},
    \end{aligned} 
	\\
	 \mathcal{C}_{02}& = i \sqrt{8} \Omega_a^2 \, (\tilde{\gamma}_{11} + 2 i \tilde{\Delta}_{11})
	 (2 e^{i \phi} + i \chi \gamma_a - 2 \chi \Delta_a)^2 / 
	\ \mathcal{N}\,,	 
	\end{align}
\end{subequations}	
where we have used
\begin{equation}
\mathcal{N}= [4 g^2 + (\gamma_a + i \Delta_a)(\gamma_b + \Delta_b)] 
[(\gamma_a + i \Delta_a) (\tilde{\gamma}_{11} + i \tilde{\Delta}_{11})
(U + 2 \Delta_b - i \gamma_b) + 4 g^2 (U + 2 \tilde{\Delta}_{11} - i \tilde{\gamma}_{11})]\,,
\end{equation}
and $\chi$, $\tilde{\gamma}_{ij}$ and $\tilde{\Delta}_{ij}$ share the
same definition as described above changing $\sigma$ by $b$. The
population of both the cavity and the excitons, and the~$\g{2}_{a,b}$
can be obtained from the coefficients in
Eqs.~(\ref{eq:FriMar2190022CET2018}) as~$ n_a = |\mathcal{C}_{10}|^2$,
$\mean{n_b}= |\mathcal{C}_{01}|^2$,
$\g{2}_a = 2|\mathcal{C}_{20}|^2/ |\mathcal{C}_{10}|^4$
and~$\g{2}_b = 2|\mathcal{C}_{02}|^2/ |\mathcal{C}_{01}|^4$,
respectively; which coincide with the expressions given in
Eq.~(T\ref{eq:Thu31May102314CEST2018}).

\section{$\mathcal{I}_k$ coefficients for the polariton statistics}

The $\mathcal{I}_k$ coefficients which give the decomposition of
Eq.~(T\ref{eq:JCg2}) in the case of polaritons can also be given in
closed form, although they become extremely bulky (compare with
Eqs.~(T\ref{eq:FriFeb23150914CET2018}) for the two-level system and
Eqs.~(T\ref{eq:Mon4May142336CEST2020}) for the Jaynes--Cummings
model):
\begin{subequations}
\label{eq:Wed22Apr182633CEST2020}
\begin{align}
\mathcal{I}_0 = & 256 U^2 \, g^8 /
 f_2\left(g, \Delta_{a,b}, \gamma_{a,b}\right)\,,\\
\mathcal{I}_1 = & 0,\\
\begin{split}
\mathcal{I}_2= & 
-32 g^4 U \bigg[ \gamma_b^2 \Big(U \gamma_a \left[\gamma_a + \gamma_b\right]
 + 2 \gamma_b \Delta_a \left[2 \gamma_a + \gamma_b \right]
  - 4 U \Delta_a^2 + 4 g^2 \left[U+2 \Delta_a \right]\Big) + \\
& 2 \gamma_b \Delta_b \Big( 3 \gamma_a^2 \gamma_b + 4 g^2 \left[2 \gamma_a + 3 
\gamma_b\right] - 6 \gamma_b \Delta_a \left[U + 2 \Delta_a\right] + 4 \gamma_a
\left[\gamma_b^2 - 2 U \Delta_a \right]\Big)
- 4 \Delta_b^2 \Big(U \gamma_a \left[\gamma_a + 3 \gamma_b \right]+ 12 
\gamma_b \Delta_a \left[\gamma_a + \gamma_b\right] - \\ &
4 U \Delta_a^2 + 4 g^2 \left[U + 2 \Delta_a\right]\Big) -
8 \Delta_b^3 \Big(4 g^2+ \gamma_a^2 + 4 \gamma_a \gamma_b - 2 \Delta_a
\left[ U + \Delta_a\right]\Big) + 32 \Delta_a \Delta_b^4 
 \bigg] \biggm/ f_2\left(g, \Delta_{a,b},\gamma_{a,b}\right)\,,
\end{split} 
\end{align}
\end{subequations}
where the auxiliary function $f_2$ has the following form:
\begin{multline}
f_2 \left(g, \Delta_{a,b},\gamma_{a,b}\right) = 
\Big[\gamma _b^2+4 \Delta_b^2\Big]^2 
\bigg[ \left(\gamma_a^2 + 4 \Delta_a^2\right) \left(\left(\gamma_a+\gamma_b\right)^2 
+ 4 \left(\Delta_a + \Delta_b\right)^2\right)
\left(\gamma_b^2 + \left(U + 2\Delta_b\right)^2\right) + \\
 16 g^4 \left(\left(\gamma_a + \gamma_b\right)^2 + \left(U + 2 \left[ \Delta_a + \Delta_b\right]\right)^2\right)
+ 8 g^2 \Big(U^2 \left(\gamma_a \left[\gamma_a + \gamma_b\right]-
4 \Delta_a \left[\Delta_a + \Delta_b\right] \right) + \\
\big(\gamma_a \gamma_b - 4 \Delta_a \Delta_b\big)
\big(\left[\gamma_a + \gamma_b\right]^2 + 4 \left[\Delta_a + \Delta_b\right]^2\big)
- 2 U \left(\gamma_a^2 \left[\Delta_a -\Delta_b\right]- 2 \gamma_a \gamma_b \Delta_b
+ 4 \Delta_a \left[\Delta_a + \Delta_b\right] \left[\Delta_a + 2 \Delta_b\right] \right)
\Big)
\bigg]\,.
\end{multline}
As in the other cases, the squeezing contribution~$\mathcal{I}_2$ is
significantly more complex than the other terms.

\bibliography{Sci,Books,pb,arXiv}